\documentclass[journal]{IEEEtran}

\usepackage{amssymb}
\usepackage{lipsum}
\usepackage{amsthm}
\usepackage{subfigure}
\usepackage{gensymb}
\usepackage{textcomp}
\usepackage{multirow}
\usepackage{makecell}
\usepackage{soul,framed} 
\usepackage{url}
\usepackage[hidelinks]{hyperref}
\usepackage[table]{xcolor}
\usepackage{cite}
\usepackage{amsmath,amsfonts}
\usepackage{graphicx}
\usepackage{dcolumn}
\usepackage{bm}
\usepackage{pict2e}
\usepackage{epstopdf}
\usepackage{arydshln}

\usepackage{enumitem}
\usepackage[mathscr]{euscript}
\usepackage{algorithmic}
\usepackage{algorithm}
\usepackage{array}
\usepackage{textcomp}
\usepackage{verbatim}
\usepackage{graphicx}
\usepackage{cite}
\usepackage{stfloats}
\usepackage{pifont}
\usepackage{colortbl}
\usepackage{pgf}

\definecolor{latGreen}{RGB}{183, 239, 197}
\definecolor{latOrange}{RGB}{250, 229, 136}
\definecolor{latRed}{RGB}{244,109,67}

\newcommand{\ccl}[1]{%
  \begingroup
    \def\val{#1}%
    \ifnum\pdfstrcmp{\val}{Out.}=0
      \cellcolor{latRed!40}Out.%
    \else
      \pgfmathparse{#1}%
      \ifnum\pgfmathresult<51
        \cellcolor{latGreen!}#1%
      \else
        \pgfmathsetmacro{\percent}{100-(\pgfmathresult/500*100)}%
        \edef\colmix{\noexpand\cellcolor{latGreen!\percent!latOrange!}}%
        \colmix #1%
      \fi
    \fi
  \endgroup
}

\newcommand{\cct}[1]{%
  \begingroup
    \def\val{#1}%
    \ifnum\pdfstrcmp{\val}{Out.}=0
      \cellcolor{latRed!40}Out.%
    \else
        \pgfmathparse{#1}%
        \pgfmathsetmacro{\percent}{(\pgfmathresult/5*100)}%
        \edef\colmix{\noexpand\cellcolor{latGreen!\percent!latOrange!}}%
        \colmix #1%
    \fi
  \endgroup
}

\definecolor{gold}{RGB}{255, 215, 0}
\definecolor{silver}{RGB}{192, 192, 192}
\definecolor{bronze}{RGB}{205, 127, 50}

\definecolor{bluerevision}{RGB}{0,112,192}

\makeatletter
\def\hlinewd#1{%
\noalign{\ifnum0=`}\fi\hrule \@height #1 \futurelet
\reserved@a\@xhline}
\makeatother

\makeatletter
 \let\old@ps@headings\ps@headings
 \let\old@ps@IEEEtitlepagestyle\ps@IEEEtitlepagestyle
 \def\confheader#1{%
 \def\ps@headings{%
 \old@ps@headings%
 \def\@oddhead{\strut\hfill#1\hfill\strut}%
 \def\@evenhead{\strut\hfill#1\hfill\strut}%
 }%
 \def\ps@IEEEtitlepagestyle{%
 \old@ps@IEEEtitlepagestyle%
 \def\@oddhead{\strut\hfill#1\hfill\strut}%
 \def\@evenhead{\strut\hfill#1\hfill\strut}%
 }%
 \ps@headings%
 }
 \makeatother
\confheader{%
This work has been accepted for publication in \textit{IEEE Open J. Commun. Soc.} DOI: 10.1109/OJCOMS.2026.3682548}

\begin{document}

\title{Towards Reliable Connectivity: Measurement-Driven Assessment\\of Starlink and OneWeb Non-Terrestrial\\and 5G Terrestrial Networks}

\author{Alejandro Ramírez-Arroyo, O. S. Peñaherrera-Pulla, Preben Mogensen

\thanks{\textit{Corresponding author: Alejandro Ramírez-Arroyo.}}

\thanks{Alejandro Ramírez-Arroyo and Preben Mogensen are with the Department of Electronic Systems, Aalborg University (AAU), 9220 Aalborg, Denmark (e-mail: araar@es.aau.dk, pm@es.aau.dk).}

\thanks{O. S. Peñaherrera-Pulla is with the Telecommunication Research Institute (TELMA), E.T.S. Ingeniería de Telecomunicación, Universidad de Málaga, 29010 Málaga, Spain (e-mail: sppulla@ic.uma.es).}}

\markboth{Ramírez-Arroyo \MakeLowercase{\textit{et al.}}: Measurement-Driven Assessment of Starlink and OneWeb Non-Terrestrial and 5G Terrestrial Networks}%
{Ramírez-Arroyo \MakeLowercase{\textit{et al.}}: Measurement-Driven Assessment of Starlink and OneWeb Non-Terrestrial and 5G Terrestrial Networks}

\maketitle

\begin{abstract}
The emergence of commercial satellite communications networks, such as Starlink and OneWeb, has significantly transformed the communications landscape over the last years. As a complement to terrestrial cellular networks, non-terrestrial systems enable coverage extension and reliability enhancement beyond the limits of conventional infrastructure. Currently, the high reliance on terrestrial networks exposes communications to vulnerabilities in the event of terrestrial infrastructure failures, e.g., due to natural disasters. Therefore, this work proposes the joint evaluation of Key Performance Indicators (KPIs) for two non-terrestrial satellite networks (Starlink and OneWeb) and two terrestrial cellular networks to assess the current performance of these technologies across three different environments: (i)~urban, (ii)~suburban, and (iii)~forest scenarios. Additionally, multi-connectivity techniques are explored to determine the benefits in connectivity when two technologies are used simultaneously. For instance, the outage probability of Starlink and OneWeb in urban areas is reduced from approximately 12-21\% to 2\% when both solutions are employed together. Finally, the joint analysis of KPIs in both terrestrial and non-terrestrial networks demonstrates that their integration enhances coverage, improves performance, and increases reliability, which highlights the benefits of combining satellite and terrestrial systems in the analyzed environments.
\end{abstract}

\begin{IEEEkeywords}
5G networks, latency, satellite networks, multi-connectivity, throughput, urban, vehicular.
\end{IEEEkeywords}

\section{Introduction}

\IEEEPARstart{S}{ince} the end of the last century, the development of terrestrial wireless mobile communications, from the First Generation (1G) to the fifth (5G), has brought about a profound change in the habits of society~\cite{5G_society}. Thus, this technology has become the \textit{de facto} standard for wireless Internet connectivity in wide-area cellular coverage scenarios~\cite{5G_defacto}. This phenomenon is particularly pronounced in urban areas, where high population density and significant market potential drive Mobile Network Operators (MNOs) to concentrate their resources~\cite{5G_urban}. All of the above factors lead to a strong dependence on 4G/5G networks for Internet connectivity~\cite{5G_ericsson}.

Despite the positive aspects derived from the homogeneity of Terrestrial Network (TN) infrastructure, these terrestrial deployments can be vulnerable in the event of emergencies or natural disasters, such as earthquakes, tsunamis, or sabotage in hybrid conflicts~\cite{5G_disaster_1, 5G_disaster_2}. This vulnerability arises due to the terrestrial deployment of network elements, such as base stations or power and backhaul infrastructure, which may be susceptible to failures and prolonged service outages~\cite{5G_disaster_3}. Thus, an end user may experience interrupted coverage if the radio access part of the network is unusable. For the above reasons, one of the main research lines in the deployment of future mobile generations is related to Non-Terrestrial Networks (NTN)~\cite{NTN_SOTA_1, NTN_SOTA_2, NTN_SOTA_3}. Since the access network is deployed in space, these networks are more robust against failures, as connectivity relies on space-based platforms that are largely immune to disruptions at ground level~\cite{NTN_P2P}. In addition, growth in terms of terrestrial infrastructure has slowed, as new generations generally use the same physical masts as in previous generations. Therefore, standardization bodies such as the 3rd Generation Partnership Project (3GPP) are focusing on the fact that 6G communications should not rely exclusively on terrestrial infrastructure~\cite{3GPP_NTN_1, 3GPP_NTN_2, 3GPP_NTN_3}.

In view of the aforementioned reasons, there has been a notable deployment of satellite systems in Low Earth Orbits (LEO) over the last decade~\cite{LEO_1}. The popularity of LEO orbits is mainly due to the (i)~relative ease of deployment in orbits between 500~km and 1200~km, (ii)~lower latency, and (iii)~improved global coverage even at high latitude regions, compared to Geostationary Earth Orbits (GEO)~\cite{LEO_2}. Some commercial examples of satellite constellations deployed in LEO orbits are Starlink, OneWeb, and Iridium~\cite{LEO_examples_1, LEO_examples_2}. Despite the advantages in terms of resilience and global coverage compared to terrestrial networks, several challenges need to be addressed for NTN communications~\cite{challenges_1}. Focusing on the radio layer and network performance, the large range of the radio link between the user terminal and the satellite compromises the link budget in Non-Line-of-Sight (NLoS) scenarios, leading to connectivity failures~\cite{link_budget_1}. This fact is particularly noticeable in urban environments, where the probability of losing visibility of the satellite increases significantly due to the presence of buildings on the horizon. On the contrary, terrestrial networks are more robust under NLoS conditions, due to their dense infrastructure close to the end user, which helps mitigate radio propagation issues and enables more reliable signal transmission in complex environments~\cite{TN_NLoS_1, TN_NLoS_2, TN_NLoS_3}.

Given the respective benefits and costs of each communications system, the state-of-the-art includes multiple studies that quantify the performance and the quality of TN and NTN under different assumptions, using simulations as well as real-world measurements. Focusing on terrestrial networks, there is extensive work related to the simulation of KPIs in the radio and network layers for 5G networks. For instance, Kang \textit{et al.} present an extensive simulation analysis of the feasibility of the upper mid-band for use in cellular networks~\cite{SOTA_THE_5G_1}. Iranmanesh \textit{et al.} study capacity improvement in 5G networks through mobile base stations located on drones~\cite{SOTA_THE_5G_2}. Vielhaus \textit{et al.} propose a system-level simulator, which allows realistic network deployments to be reproduced in simulation~\cite{SOTA_THE_5G_3}. Both theoretical advances and the development of simulators are fundamental to the practical deployment of cellular networks. From an experimental point of view, studies such as those by Aerts \textit{et~al.}~\cite{SOTA_PRA_5G_1} and Chiaraviglio \textit{et al.}~\cite{SOTA_PRA_5G_2} conduct empirical studies of the physical layer to determine the levels of electromagnetic fields generated by 5G deployments. Additionally, Mohebi~\textit{et~al.}~\cite{SOTA_PRA_5G_3} report network performance in terms of latency and throughput for commercial 5G networks. Concerning non-terrestrial networks, multiple efforts have been made in relation to stochastic theoretical modeling and simulation of satellite communication systems, such as the work of Jung \textit{et~al.}~\cite{SOTA_THE_SAT_1} and Abdelsadek \textit{et~al.}~\cite{SOTA_THE_SAT_2}. Focusing on experimental work, empirical performance evaluations of commercial global constellations have also recently begun to be explored, such as the examples in the work of Laniewski~\textit{et~al.}~\cite{SOTA_PRA_SAT_1, SOTA_PRA_SAT_2}. In addition, Pan \textit{et~al.}~\cite{SOTA_PAN_1,SOTA_PAN_2} have thoroughly characterized Starlink by analyzing its routing behavior, network architecture, and backbone infrastructure, as well as performance metrics acquired from terminals distributed worldwide. Similarly, Zhao \textit{et~al.}~\cite{SOTA_PAN_3} provide a cross-layer analysis of OneWeb ground architecture and the impact of traffic routing through the satellite network on network KPIs. Recent state-of-the-art has also evaluated the performance of NTN networks under different traffic patterns, focusing on the impact of transport-layer protocols. For instance, Careau~\textit{et~al.}~\cite{SOTA_TCPUDP_1} have performed a comparative analysis of TCP and UDP traffic over the Starlink network. Similarly, García~\textit{et~al.}~\cite{SOTA_TCPUDP_2} have studied several congestion control variants over Starlink, showing that protocol selection can strongly affect network performance. Note that due to the recent deployment of these networks, the experimental state-of-the-art related to NTN networks remains relatively scarce compared to the extensive state-of-the-art on terrestrial networks. Taking a step further and combining the fields described above, we have conducted preliminary studies jointly analyzing the performance of cellular networks and Starlink in rural areas~\cite{PW_1} and high-mobility environments along highways~\cite{PW_2}.

In this paper, we propose a joint empirical analysis of the performance of up to four commercial communications systems around the city of Copenhagen (Denmark): two major Danish MNOs, and two NTN satellite systems, Starlink and OneWeb. Thus, the aim is to assess the feasibility of their deployment across multiple environments that are prone to exhibit heterogeneous and potentially challenging radio propagation conditions due to radio obstruction, NLoS conditions, and limited satellite visibility, such as urban, suburban/residential, and rural/forest environments. The evaluation focuses on Key Performance Parameters (KPIs) related to data rate and latency in both directions, downlink and uplink. This unidirectional measurement approach enables the independent characterization of uplink and downlink paths, revealing latency and performance asymmetries that cannot be captured by conventional round-trip measurements. While optimal performance of the terrestrial network is expected in such urban scenarios, the study shows a comparison between well-established terrestrial systems and emerging non-terrestrial alternatives. In addition, the study provides an estimate of the expected performance if multiple combinations of the aforementioned systems were integrated, covering TN–TN, TN–NTN, and NTN–NTN multi-connectivity configurations. For instance, the analysis of exclusively satellite multi-connectivity enables the assessment of reliability gains achievable through satellite diversity. Finally, the assessment focuses both on mobility conditions, where the communications system is located in a vehicle that is constantly in motion, and on static conditions, where the equipment is fixed in a specific location. Overall, the main contributions of this work are:

\begin{itemize}

    \item Development of a mobility measurement campaign for the joint evaluation of TN 5G, and NTN Starlink and Oneweb commercial networks in urban, suburban, and forest scenarios under challenging radio propagation conditions due to radio obstruction, NLoS conditions, and limited satellite visibility.
    
    \item Unidirectional statistical analysis of end-to-end KPIs that impact end-user performance, such as throughput and one-way delay latency metrics in both uplink and downlink, enabling the identification of link performance asymmetries that cannot be captured through round-trip measurements.

    \item Evaluation of network performance improvement in the case of integrating multiple technologies through multi-connectivity, including terrestrial-only (TN–TN), hybrid terrestrial and non-terrestrial (TN–NTN), and exclusively non-terrestrial (NTN–NTN) multi-connectivity configurations.
    
    \item Assessment of the robustness of commercial satellite connectivity under NLoS conditions in both mobile and static scenarios, representative of challenging real-world deployment scenarios.

    \item Development of a set of recommendations for the use of satellite networks in emergencies or natural disasters based on the conclusions of the empirical analysis of commercial networks.
    
\end{itemize}

In summary, this work, carried out in collaboration with the Center of Emergency Communication (CFB, Denmark), aims to be useful as an empirical analysis in terms of expected and achievable performance for mobile network and satellite operators, as well as shed light on the use of emerging communication technologies in real-world scenarios for emergency centers and systems. Note that, compared to our previous works~\cite{PW_1, PW_2}, this study is novel in terms of (i) the number of interfaces studied simultaneously and the joint analysis of OneWeb and Starlink LEO networks, (ii) the three scenarios encompassing heterogeneous environments, ranging from high-density urban to low-density rural regions where physical characteristics significantly influence performance, (iii) the independent study of uplink/downlink links in terms of latency, thus enabling analysis of possible asymmetries in the bidirectionality of the link, and (iv) the experimental analysis of the impact of NLoS probability in urban areas. 

The document is organized as follows. Sections~II~and~III illustrate the proposed measurement campaign and setup, respectively. They include details about the choice of each scenario and its characteristics, as well as the different technologies used. Section IV presents the statistical network performance for each technology individually. Section V extends the previous section by considering the integration of connectivity technologies through multi-connectivity. Section VI conducts an empirical analysis of how NLoS conditions affect communications between user terminals and satellites in urban environments. Finally, conclusions are drawn in Section VII.

\begin{figure}[!b]
	\centering
	\includegraphics[width= 1\columnwidth]{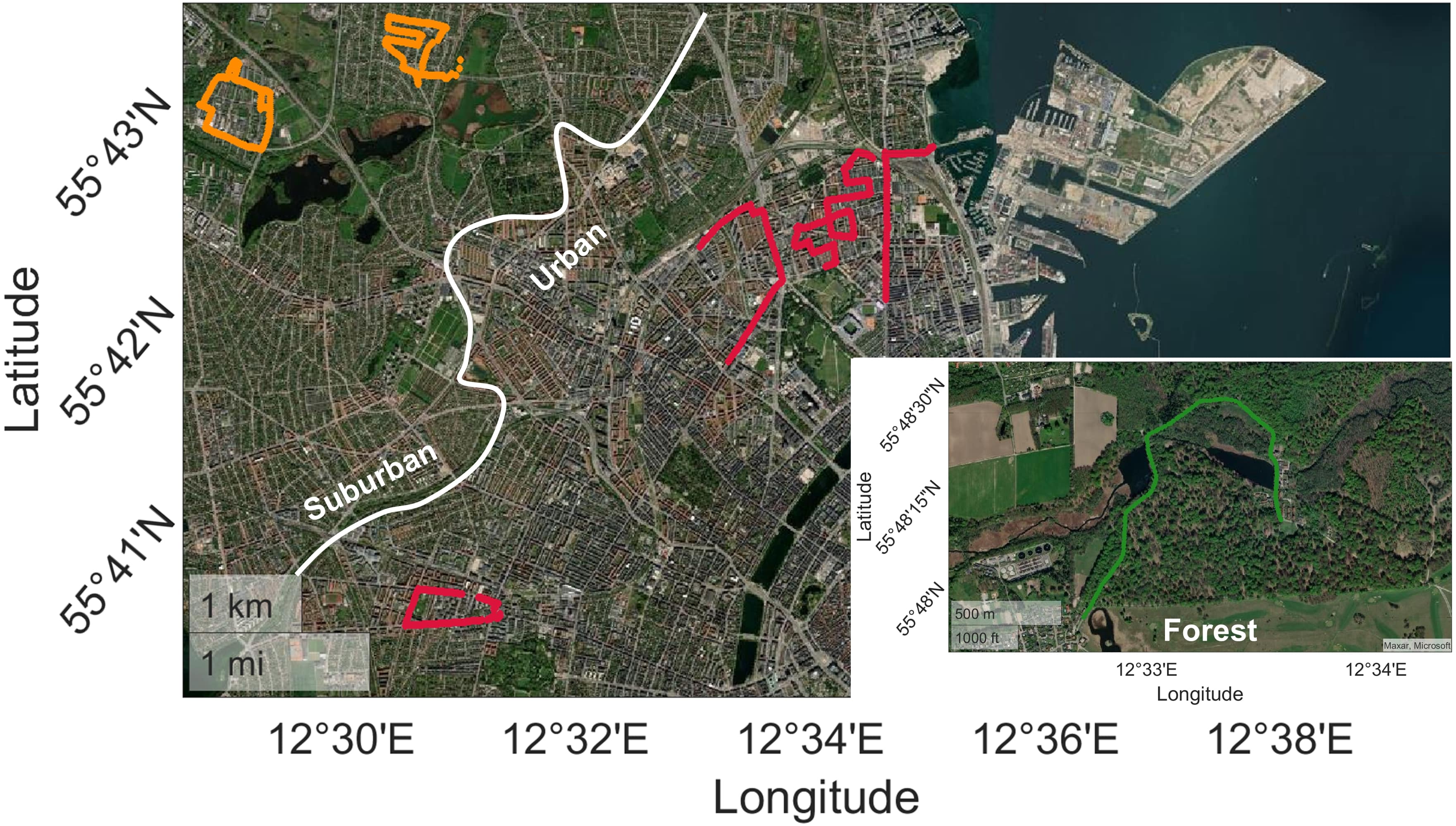}
	\caption{Satellite map of the area under study. The three regions studied and the drive test areas are shown in different colors: urban area (red), suburban area (orange), forest area (green).} 
	\label{fig1}
\end{figure}

\section{Measurement Campaign and Scenario}

This study presents a measurement campaign carried out under mobility conditions to evaluate the network performance of terrestrial and non-terrestrial networks across different environments in the city of Copenhagen, Denmark. In particular, the following measurement scenarios are considered: (i) the downtown of the city of Copenhagen, referred to as \textit{urban}, (ii) the outskirts of the city corresponding to residential areas, referred to as \textit{suburban}, and (iii) a wooded rural area northeast of the city, referred to as \textit{forest}. Fig.~\ref{fig1} shows a satellite view of Copenhagen, along with the division of the different regions considered and the geolocated traces of the scenario studied. The measurement equipment has been mounted on various vehicles (see Section III), and in-motion measurements have been conducted at driving speeds ranging from 0~km/h to 50~km/h due to local traffic regulations. Additionally, several measurements have been acquired at static locations in the urban area to study the probability of satellite link blockage (see Section VI). Overall, around 7.7~million network traffic samples, which will be used to evaluate several network KPIs, have been acquired for TN and NTN across three different scenarios throughout three days.

This campaign simultaneously evaluates two cellular connectivity solutions, corresponding to Denmark's two main 5G cellular operators, as well as the non-terrestrial networks Starlink and OneWeb. At the time of performing the measurement campaign, both mobile network operators have a cellular network of hundreds of base stations within the region under study~\cite{mastedatabasen}. With regard to satellite operators, Starlink has approximately 8000 satellites in LEO orbits at the time of measurement between 550~km and 700~km~\cite{number_starlink}, while OneWeb has approximately 648 satellites deployed in LEO orbits at an altitude of 1200~km~\cite{number_oneweb}. Regarding inter-satellite links (ISLs), Starlink has introduced ISL capabilities in its newer generations of satellites, while OneWeb does not currently support ISLs. In the case of Starlink, the vast majority of satellites are deployed in orbits with an inclination of 53º since the largest proportion of the global population is located below these latitudes. In addition, new polar orbits have been deployed to provide coverage at higher latitudes. In the case of the city of Copenhagen, located at a latitude of 55º, this deployment is sufficient to ensure visibility with multiple satellites simultaneously. In the case of OneWeb, the constellation orbits exclusively in polar orbits around~87º, which means better coverage at high latitudes compared to the equator. For the latitude at which the study area is located, this deployment is sufficient to ensure continuous coverage if there are no obstructions on the horizon. Note that the results shown throughout the work are specific to the 55º Copenhagen high latitude scenario, and although both constellations provide coverage at this latitude, their relative performance might differ in terms of the latitude. In particular, OneWeb’s relative performance is expected to improve with increasing latitude due to its near-polar orbital configuration, while Starlink benefits from higher satellite density at lower latitudes closer to the equator. Regarding the network topology, the scientific community has carried out extensive work in mapping the terrestrial infrastructure for NTN networks, i.e., the Ground Stations or Satellite Network Portals (SNPs) and Points of Presence (PoPs). For instance, the works of Wang~\textit{et al.}~\cite{NT_Starlink} and Zhao \textit{et al.}~\cite{SOTA_PAN_3} have identified this infrastructure for Starlink and OneWeb, respectively. Considering our geographical position in Copenhagen, the Starlink ground stations in Aerzen (Germany), Frankfurt (Germany) and Hoofddorp (Netherlands), and the Starlink PoP in Frankfurt (Germany) are noteworthy due to their geographical proximity. For OneWeb, the nearest ground stations are geographically sparser in Sintra (Portugal), Palermo (Italy), Stara Zagora (Bulgaria) and Piteå (Sweden), while the PoP closest to our location is in Amsterdam (Netherlands). Therefore, the larger distance between the ground infrastructure and our location indicates higher expected end-to-end latency in OneWeb compared to Starlink. Note also that, from a global perspective, the development of ground infrastructure is more mature for Starlink, with 49~PoPs compared to OneWeb’s 29~PoPs, both as of 2025~\cite{NT_Starlink, SOTA_PAN_3}. Hence, it is more likely that a Starlink PoP is geographically closer to the user’s location given a random location. In terms of signal structure, Starlink uses Orthogonal Frequency Division Multiplexing (OFDM)~\cite{SS_Starlink}, while OneWeb implements MultiFrequency Time Division Multiple Access (MF-TDMA) Single Carrier (SC)~\cite{SS_OneWeb}.

\begin{figure}[!t]
	\centering
	\subfigure[]{\includegraphics[width=1\columnwidth]{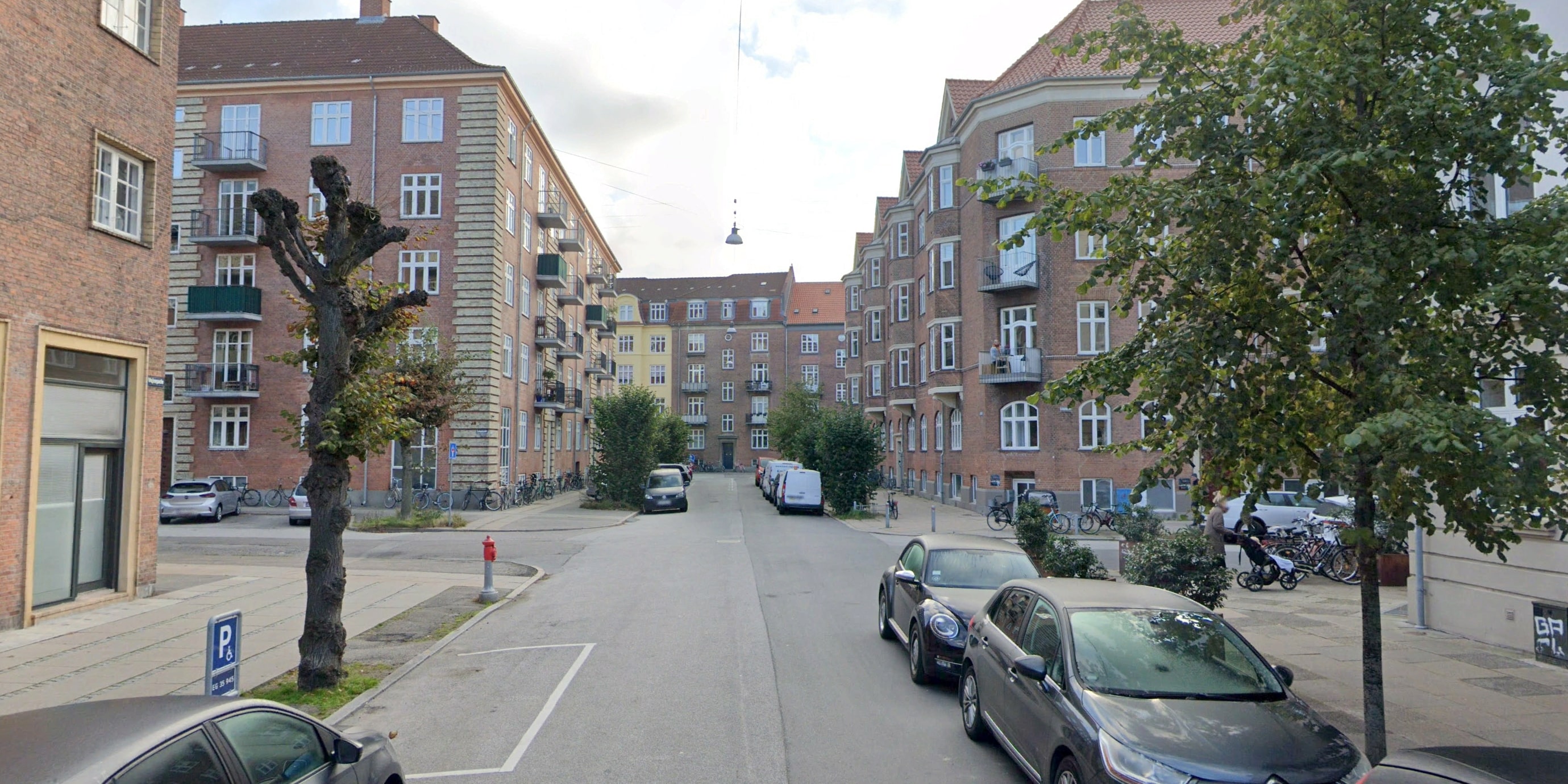}
	} 
    \subfigure[]{\includegraphics[width= 1\columnwidth]{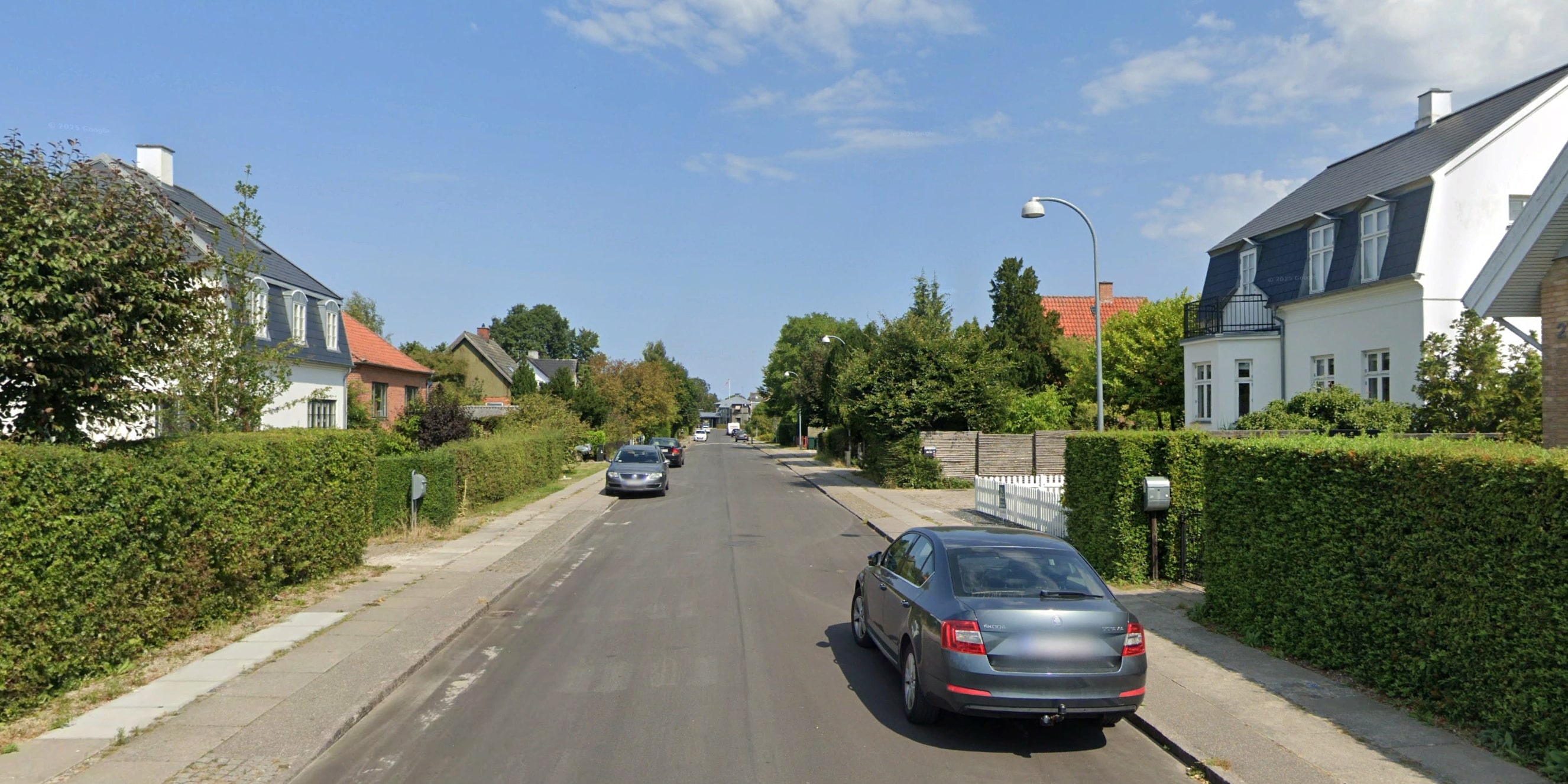}
	} 
    \subfigure[]{\includegraphics[width= 1\columnwidth]{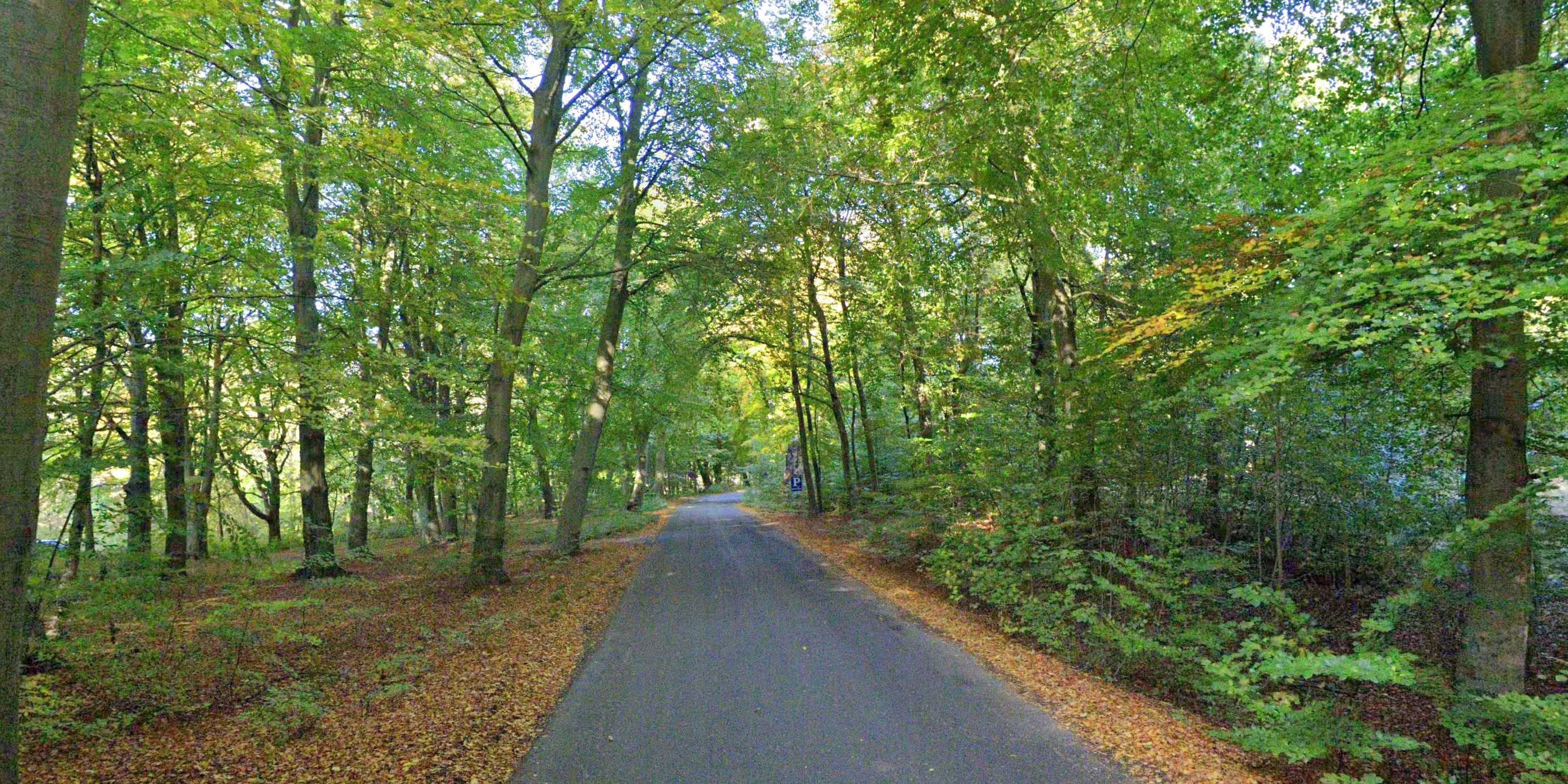}
	}
	\caption{Representative photographs corresponding to each of the scenarios studied: (a) urban area, (b) suburban area, and (c) forest area. Images extracted from Google Street View.} 
	\label{fig2}
\end{figure}

Focusing on the physical characteristics of each scenario, Figs.~\ref{fig2}(a)-(c) show representative photographs of the environments corresponding to various locations along the routes covered. The urban environment is characterized by avenues and streets lined with buildings of four to five stories high. Due to the high population density in this area, MNOs have a highly densified network of base stations. According to statistics from the Danish Agency for Digital Government, there are approximately 15 base stations per km$^2$ in this area, indicating a high density of cell sites~\cite{mastedatabasen}. In the context of Starlink and OneWeb constellations, although the coverage is global, this scenario is prone to signal blockage at low elevation angles due to the presence of buildings, requiring the satellite to establish connectivity in regions close to the zenith. On the other hand, the suburban scenario is characterized by a landscape consisting of single-family homes and one to two-story buildings. Due to the lower population density compared to urban scenarios and more favorable radio propagation conditions, the number of base stations in this area drops to 3-5 base stations per~km$^2$~\cite{mastedatabasen}. However, the landscape is more favorable for satellite constellations given the lower number of obstacles on the horizon, increasing the probability of LoS between the ground equipment and the satellite. The forest scenario corresponds to a narrow road surrounded by dense vegetation, where tall trees are found in both sides of the route. Tree crowns, branches, and leaves introduce significant obstruction of the sky, resulting in a predominantly blocked zenith and limited satellite visibility. In most sections, the dense vegetation causes strong attenuation and frequent blockage of satellite links, particularly at low and medium elevation angles. Limited visibility is occasionally available through small gaps in the foliage. In some short segments, partial visibility at very high elevation angles ($>70\degree$) is observed. However, these conditions are transient and typically persist only for a few seconds due to mobility. Regarding the cellular network conditions, the area under study is located approximately one kilometer from the closest population center, with only one base station from each operator within a two-kilometer radius~\cite{mastedatabasen}.

In summary, these three scenarios have been selected based on their physical characteristics, which directly impact the performance of terrestrial and non-terrestrial networks. The interaction between these scenarios and the feasibility of providing end-user connectivity are discussed in the following sections.

\begin{figure}[!b]
	\centering
	{\includegraphics[width=1\columnwidth]{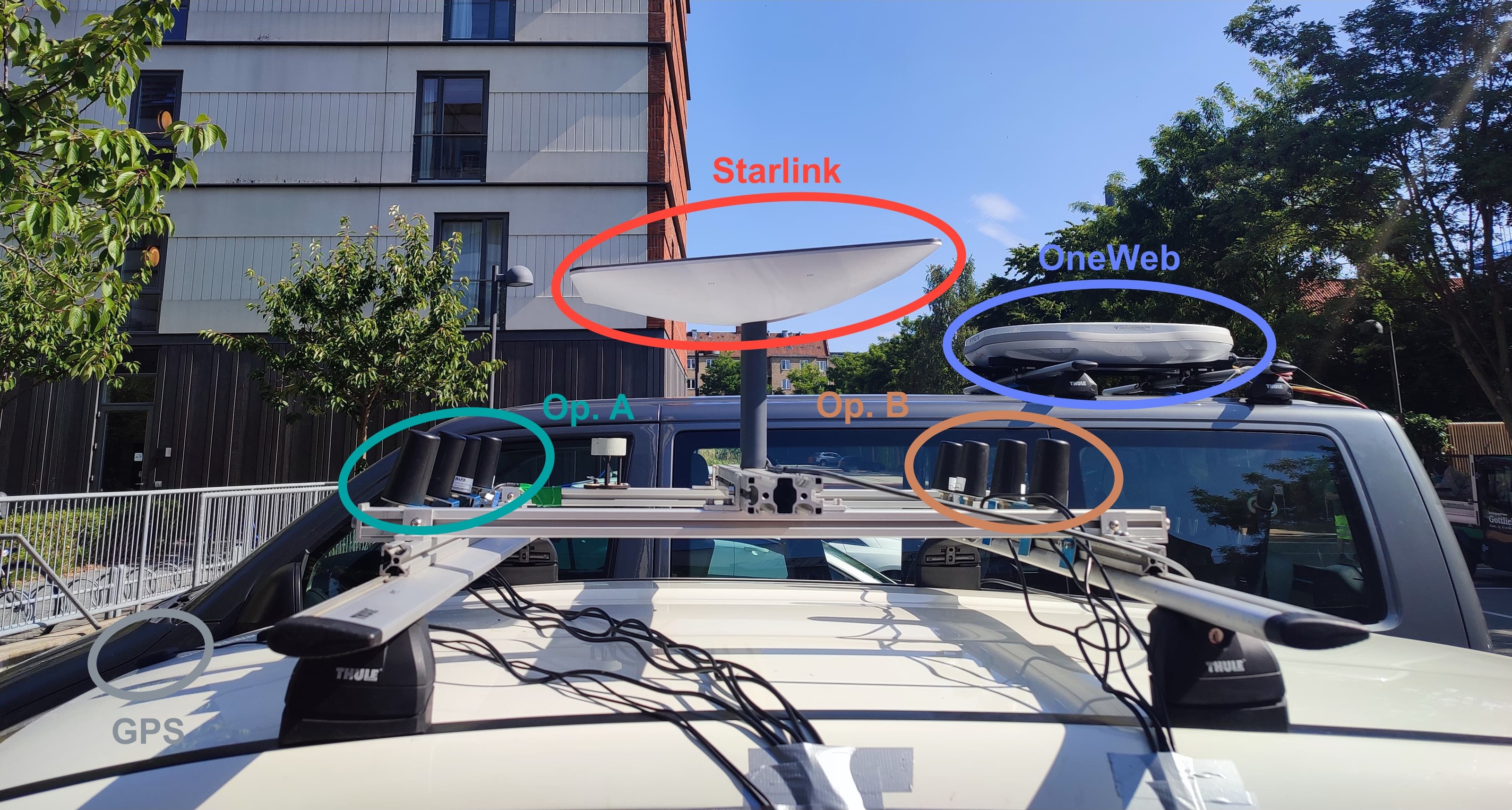}
	} 
	\caption{User equipment for the two 5G operators, Starlink and OneWeb.} 
	\label{fig3}
\end{figure}

\section{Measurement Setup}

\begin{figure*}[!t]
	\centering
	\includegraphics[width= 0.85\textwidth]{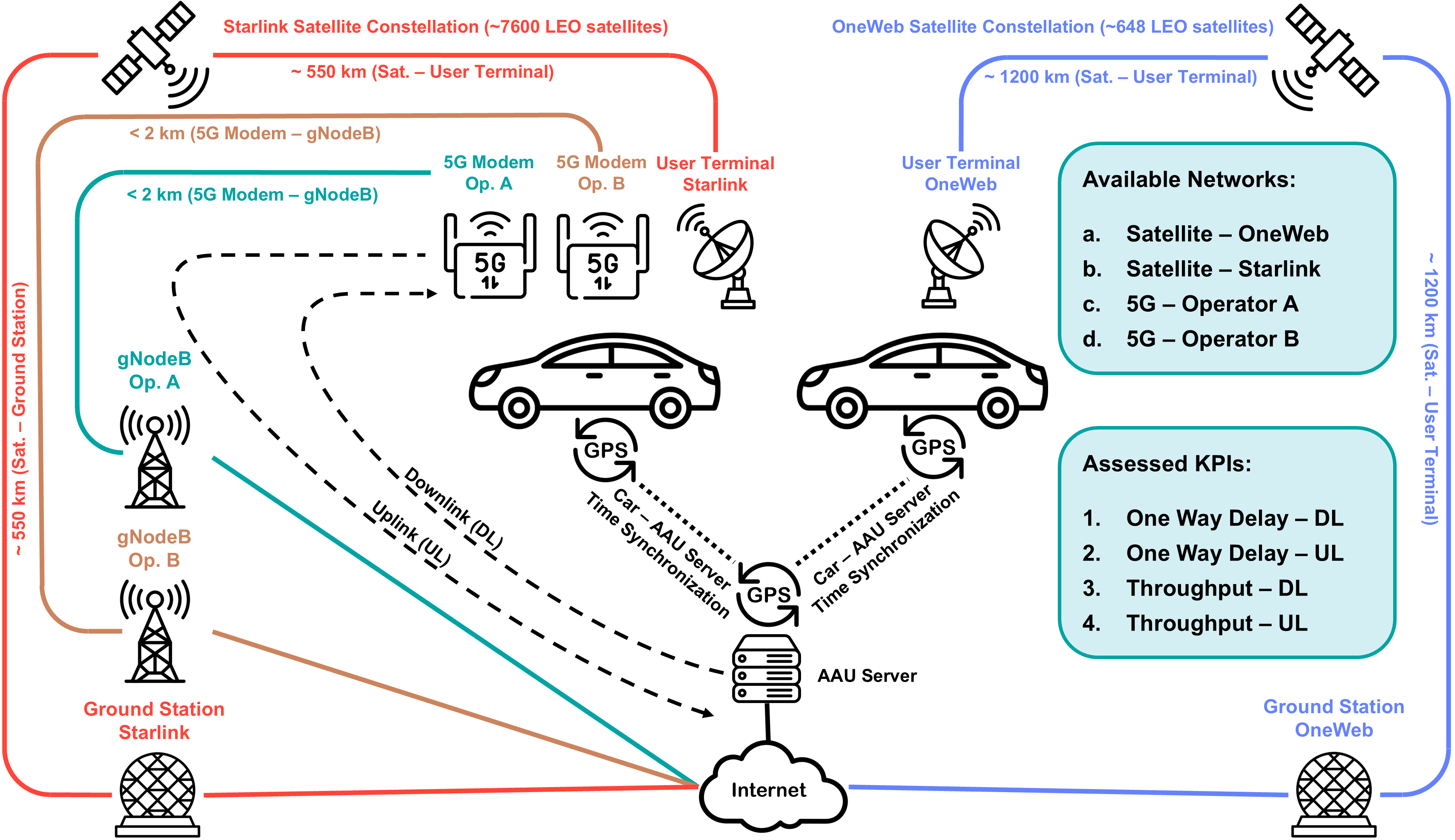}
	\caption{Connectivity scheme between the user equipment and the AAU server.} 
	\label{fig4}
\end{figure*}

For the simultaneous evaluation of the two cellular networks and the two satellite networks, a dedicated setup is required. In particular, given the mobility conditions of the measurement campaign, the user equipment needs to be mounted on a vehicle. Specifically, two vehicles are used due to space limitations. The first one is equipped with the user equipment for the cellular and Starlink networks, and the second one is equipped with the user equipment for OneWeb. Fig.~\ref{fig3} shows the user equipment for the four networks involved, mounted on the two vehicles driven. During the measurement campaign, both vehicles were driven together at a short distance to preserve similar physical conditions among the four connectivity interfaces. Regarding the cellular connectivity framework, the experimental setup comprises two Intel NUC (Next Unit of Computing) devices. Each device is equipped with two onboard Gigabit Ethernet interfaces and expanded with two USB 3.1 Gigabit Ethernet adapters. These NUCs operate as traffic generators, hosting a UDP client that synchronizes with a remote UDP server located at the AAU premises. To establish connectivity, two Teltonika RUTX50 Industrial 5G routers were deployed to provide high-performance and independent network interfaces to one NUC. Routers are configured to prioritize the operator's 5G network, automatically falling back to 4G when 5G coverage is not available. Each router connects to four omnidirectional antennas with 5 dBi gain (model AOA-4G-5W-SMM) depicted in Fig.~\ref{fig3}. As for the Starlink setup, it employs a Starlink Standard Actuated Gen 2 antenna (51.3~$\times$~30.3~cm). Due to mobility during the measurement campaign, this antenna was configured to maintain a fixed orientation toward the sky, avoiding mechanical reconfiguration. A Starlink Gen 2 router serves as the gateway, establishing IP connectivity and linking the satellite backhaul to the NUC via an Ethernet interface. Regarding the OneWeb setup, a Kymeta Hawk U8 terminal (89.5~$\times$~89.5~cm) is employed. This device features a built-in Ethernet interface that connects directly to the OneWeb backhaul, thereby providing the NUC with satellite internet access. Differences between the antennas arise from the distinct market segments targeted by the two satellite operators, with OneWeb user terminal focusing on enterprise and government levels, and Starlink user terminal being designed primarily for the consumer market. The effect of using these two different antennas is discussed in detail in the following sections. To ensure precise time alignment, a GPS-based NTP source distributes a Pulse Per Second (PPS)-driven time reference via Ethernet for each NUC, thus enabling microsecond accuracy time synchronization. In addition to all of the above, each vehicle is equipped with batteries and DC/AC inverters power electronic components, laptops to monitor measurements in real time, and GPS antennas to keep track of the location. 

\begin{figure*}[!b]
	\centering
	\subfigure[{\footnotesize One-Way Delay}]{\includegraphics[width=0.7\textwidth]{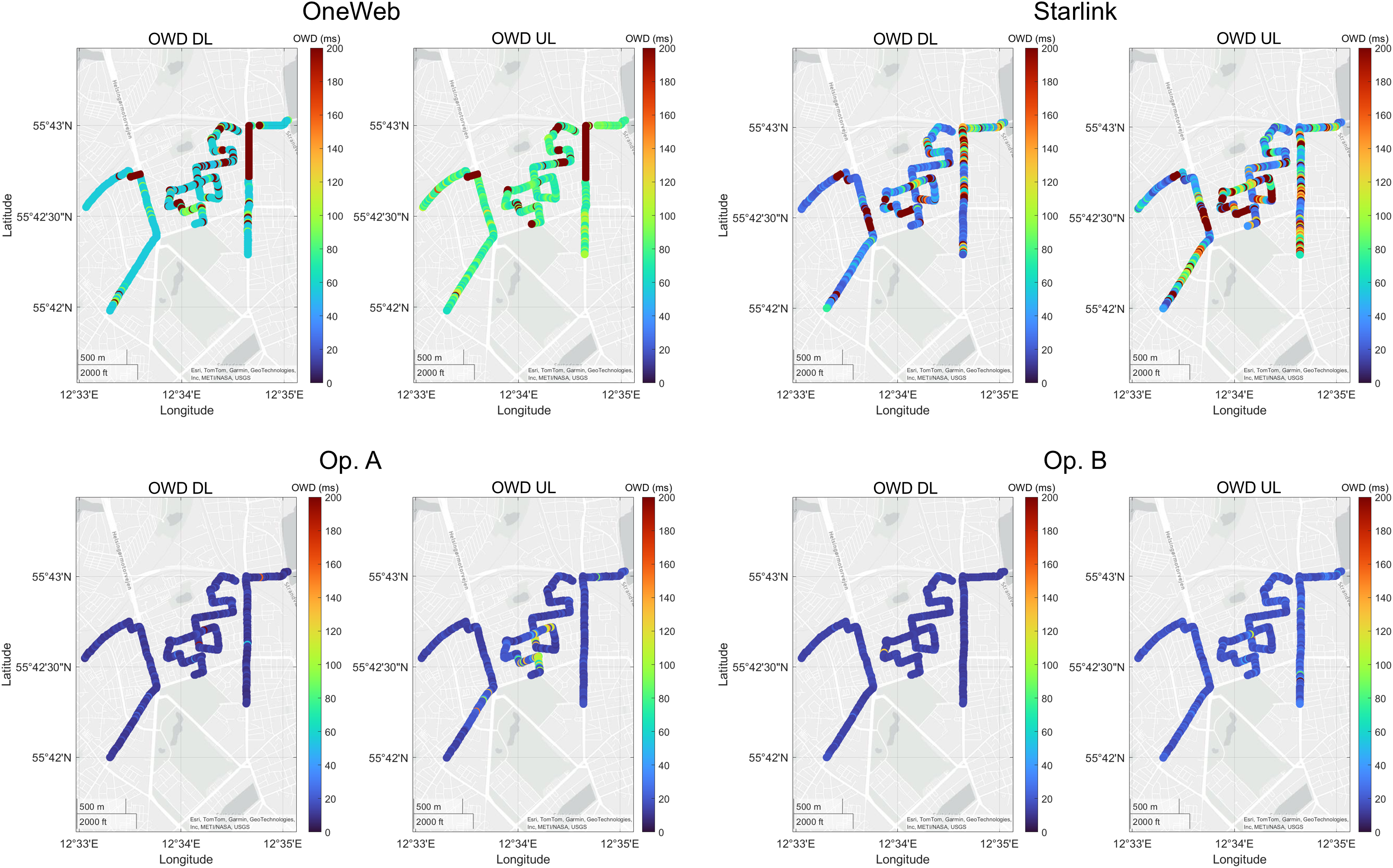}
	} 
    \subfigure[{\footnotesize Throughput}]{\includegraphics[width=0.7\textwidth]{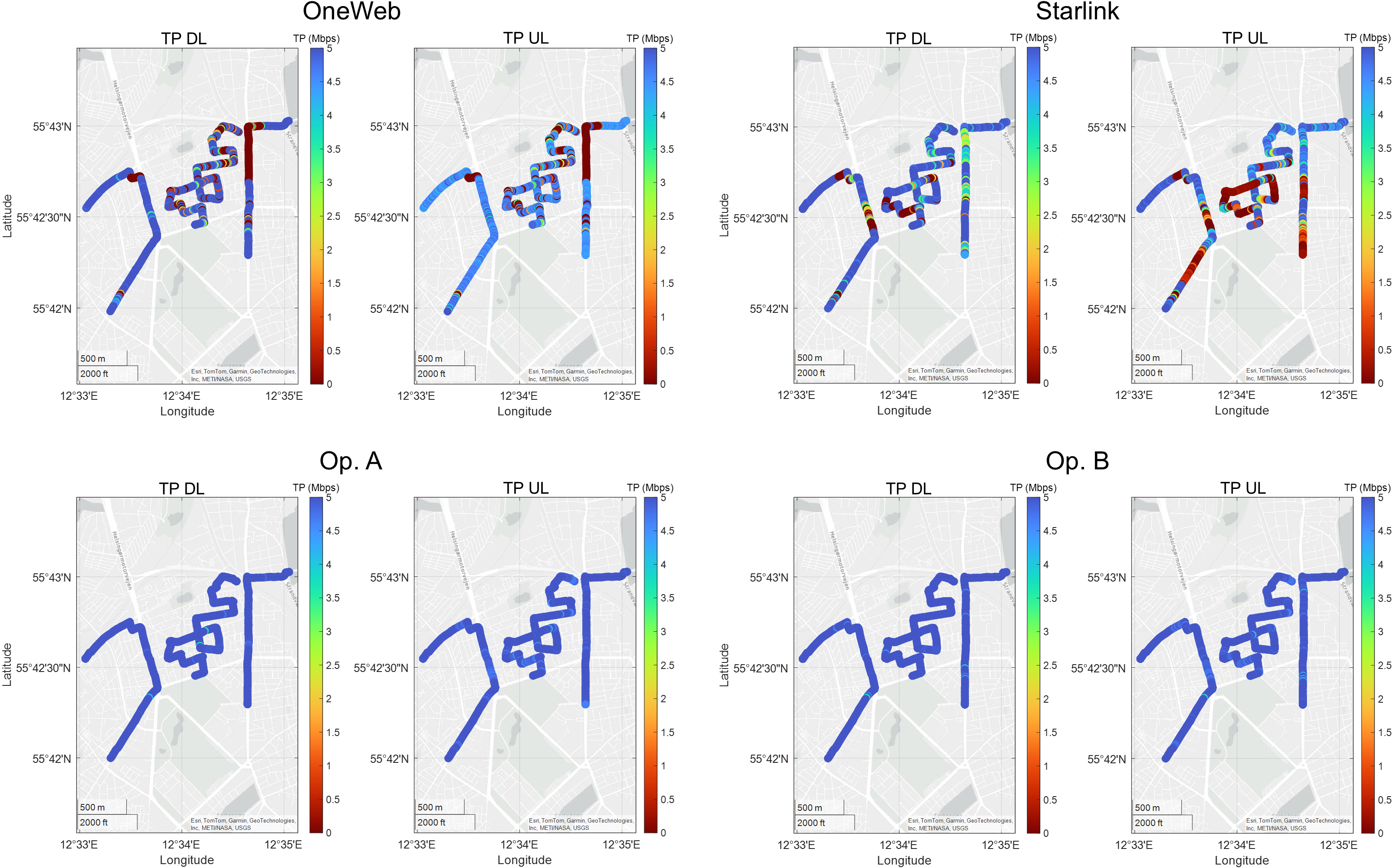}
	} 
	\caption{Geolocated samples of (a) downlink and uplink one-way delay, and (b) downlink and uplink throughput for in-motion measurements in the urban area. The colormap for the one-way delay KPI has been limited to an upper bound of 200 ms for visualization purposes.} 
	\label{fig5}
\end{figure*}

On the server side (AAU server), physically located in the city of Aalborg (Denmark), 220 km from Copenhagen, UDP packets are sent (downlink) and received (uplink) to/from the user equipment mounted in the vehicle to evaluate network performance. This server also features time synchronization via PPS for accurate and synchronized measurement between the parties involved. Fig.~\ref{fig4} shows a conceptual diagram of the connectivity between the user equipment and the server, and the routing through the radio access network towards the Internet for each of the solutions. In particular, 5G modems route traffic through a gNodeB typically located less than 2~km away. In the case of satellite connectivity, the satellite user terminal transmits the data to the satellite (550~km~-~1200~km), which acts as a node forwarding this data to a ground station, either directly or via inter-satellite links, which retransmits it to the AAU server. Note that the server has simultaneous synchronization with each of the vehicles, allowing for simultaneous and synchronized evaluation of the four available interfaces. This is essential for evaluating the multi-connectivity solutions that will be detailed later in Section V.

Regarding the measurement procedures, this setup evaluates four Key Performance Indicators: (i) downlink one-way delay (OWD DL), (ii) uplink one-way delay (OWD UL), (iii)~downlink throughput (TP DL), and (iv) uplink throughput~(TP UL). For this purpose, UDP packets are transmitted in both directions at 2 ms intervals, thus measuring the one-way delay as the difference between the time of sending and receiving on the server and client (DL), respectively, and vice versa (UL). Therefore, the OWD evaluates the end-to-end delay experienced by the user in both uplink (OWD UL) and downlink (OWD DL), accounting for both the physical propagation delay and network-induced delays, including routing, queuing, and handover effects. The short interval between packets ensures constant monitoring of the link. Additionally, the UDP packet size is adjusted so that the traffic corresponds to 5/5 Mbps DL/UL, thus establishing a target data rate of 5 Mbps in both directions. This target data rate has been chosen since it is the throughput required in the CFB for high-resolution real-time video transmission in emergency situations. Therefore, this target is representative of a broad class of applications relying on continuous uplink or downlink video streaming. Note that these values are not intended to evaluate the networks under maximum capacity or heavy traffic conditions, but rather to assess the reliability of network connectivity across the three considered measurement scenarios shown in Section II. In short, this setup allows the measurement of 4~KPIs/interface for the two cellular operators, Starlink and OneWeb, thus enabling the simultaneous and synchronized measurement of a total of 16 KPIs.

\section{Single-Connectivity Evaluation}

This section presents an individual assessment of each of the connectivity solutions in the three scenarios presented through statistical analysis of the different KPIs and evaluation of the samples based on geolocation.

\subsection{GPS Traces}

\begin{figure}[!t]
	\centering
	\includegraphics[width= 1\columnwidth]{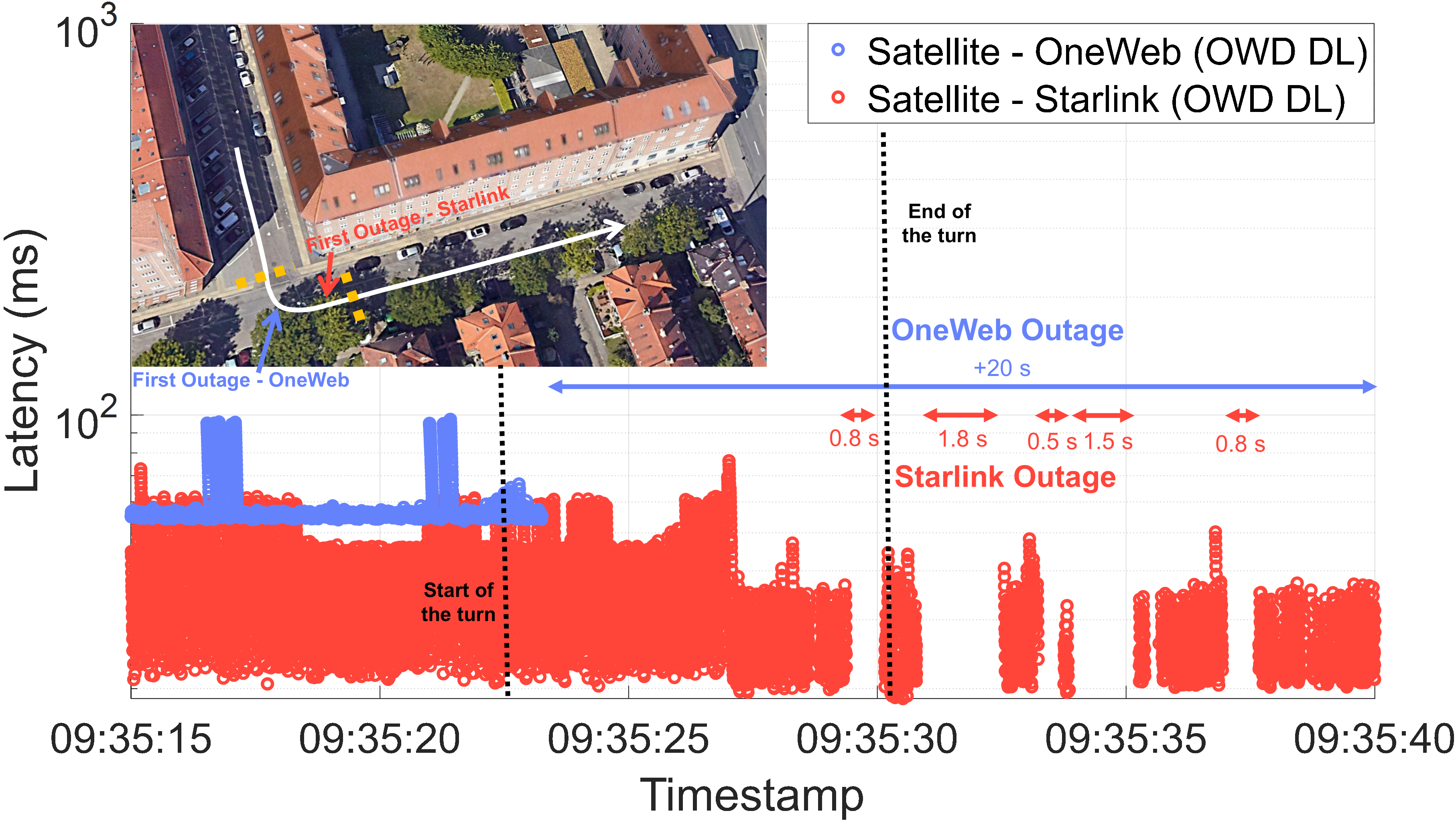}
	\caption{Time trace of a 90-degree turn at an intersection in the urban area for satellite operators OneWeb (\scalebox{0.8}{\textcolor[rgb]{0.3961,0.5098,0.9922}{\ding{108}}}) and Starlink (\scalebox{0.8}{\textcolor[rgb]{1,0.2706,0.2275}{\ding{108}}}).} 
	\label{fig6}
\end{figure}

As a first experiment, connectivity in the urban environment is assessed keeping track of the measurements geolocation. Figs.~\ref{fig5}(a) and~\ref{fig5}(b) show the latency and throughput values, respectively, for each of the terrestrial and non-terrestrial connectivity solutions.

Analyzing the latency values for Op. A and Op. B in Fig.~\ref{fig5}(a), it can be seen that the OWD is stable with uniform values between 15 and 25 ms throughout the different locations. This fact is especially remarkable in the case of the downlink. For the uplink OWD, although the values remain stable, there are several instances where latency increases to values between 100 and 120 ms (see Op. A). In any case, the low latency values obtained for terrestrial networks are to be expected given the redundancy of base stations in this area, where there are multiple cells covering the user terminal that can function as backup even if the main cell is not operating correctly or is congested. In the case of the uplink, the link budget can play a fundamental role in the occasional increase in latency, since the transmission power in the user equipment is certainly more restricted than the transmission power in the base station.

Regarding the satellite operators, there are multiple coverage interruptions along the route where both uplink and downlink OWD exceed several seconds. Given the characteristics of the route, the buildings on the different avenues and streets act as sources of LoS obstruction, causing these interruptions. In particular, after analyzing the time traces of the measurement campaign for the urban environment, it was observed that the main coverage interruptions occur when the vehicle turns at an intersection. The turning of the vehicle causes a change in the position of the satellites on the horizon from the user terminal perspective, so that during turns at intersections or curves, LoS between the satellite and the ground equipment is momentarily obstructed. As an illustrative example, Fig.~\ref{fig6} shows the downlink OWD time trace corresponding to a 90-degree turn in the urban area. While prior to the heading change, both connectivity options maintain a stable latency below 100 ms, after initiating the turn and changing the orientation of our vehicle, connectivity is completely lost with OneWeb for more than 20~seconds and intermittently with Starlink at intervals of around 1~second. This suggests that the change in the vehicle's orientation during the turn changes the spatial distribution of the satellites from the user equipment perspective, which eventually causes a momentary loss of visibility with some of the satellites, resulting in a loss of connectivity for several seconds. Although this Fig.~\ref{fig6} shows a specific example, this behavior has been observed repeatedly throughout the measurements conducted in the urban area. In some cases, OneWeb is the most affected constellation, as shown in Fig.~\ref{fig6}, while in other cases, Starlink is the most affected. This lack of correlation between the instants at which the visibility of each constellation is impaired due to the independence of the orbits of the satellites in each constellation will be exploited to increase the reliability of satellite networks throughout Section V.

In the case of throughput analysis, similar behavior to that shown by latency is found. Cellular operators show stable values around the targeted 5/5 Mbps in both downlink and uplink due to the high density of available cells. In the case of satellite operators, constant connection drops are observed, with throughput instantly falling from 5~Mbps to 0~Mbps, suggesting a sudden blockage of the LoS. It is still significant that these drops tend to occur after turns at intersections.

In summary, this subsection has shown a qualitative analysis of the superiority of cellular links over satellite links in urban environments, mainly due to challenging radio propagation conditions for satellite operators and extensive cellular deployment due to high population density.

\subsection{One Way Delay Statistics}

\begin{figure}[!t]
	\centering
	\subfigure[]{\includegraphics[width=1\columnwidth]{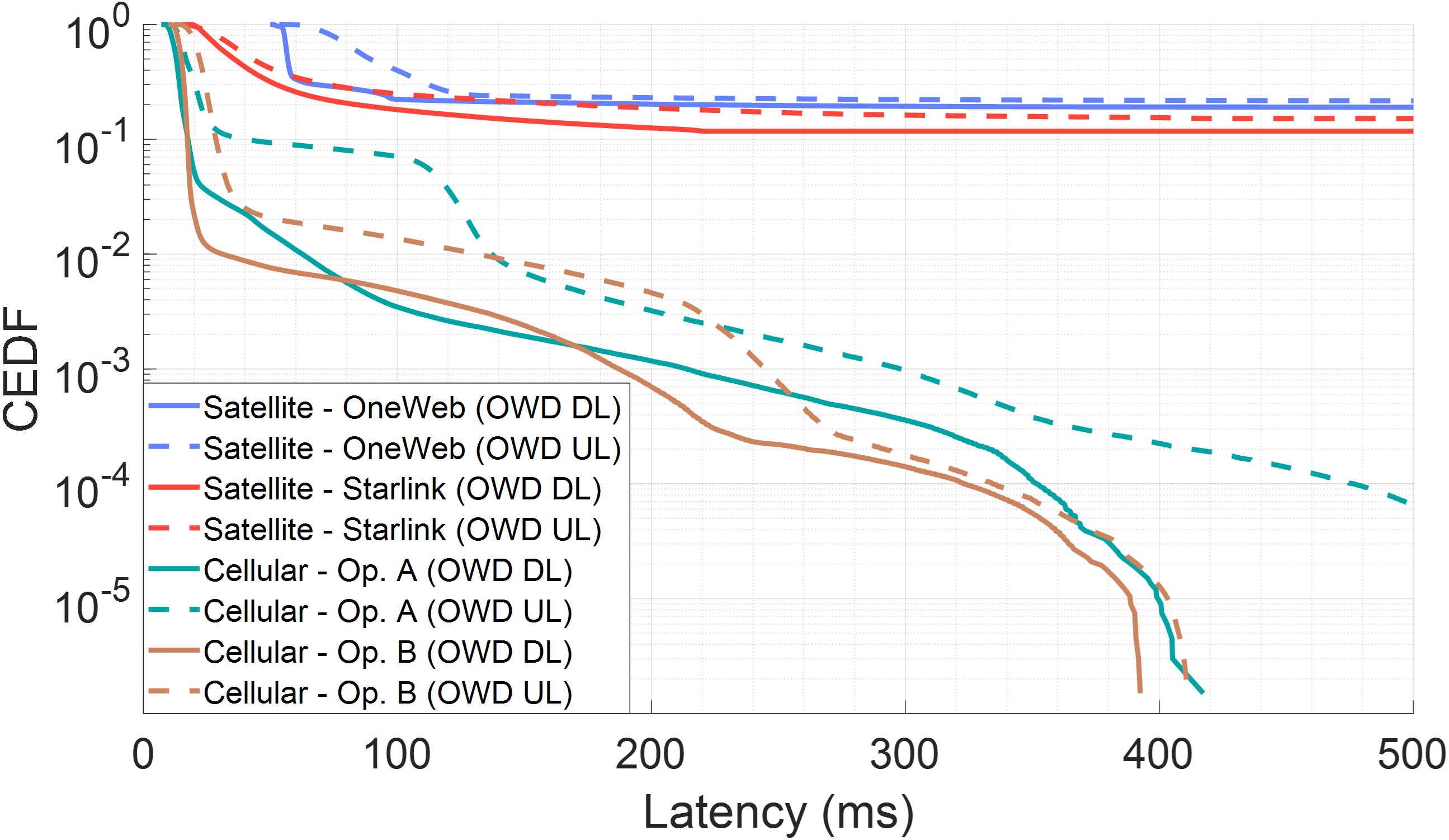}
	} 
    \subfigure[]{\includegraphics[width=1\columnwidth]{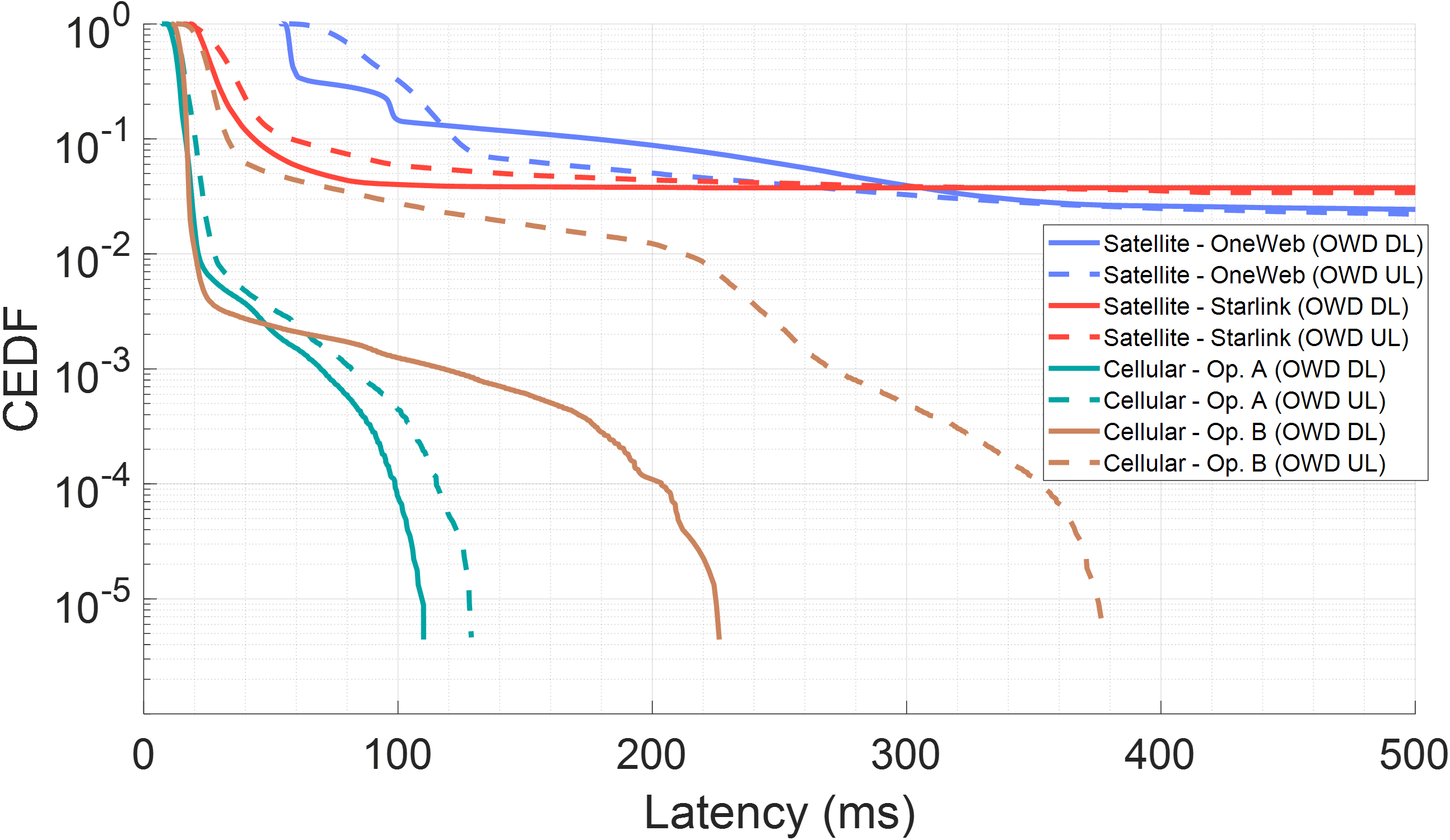}
	}
	\subfigure[]{\includegraphics[width=1\columnwidth]{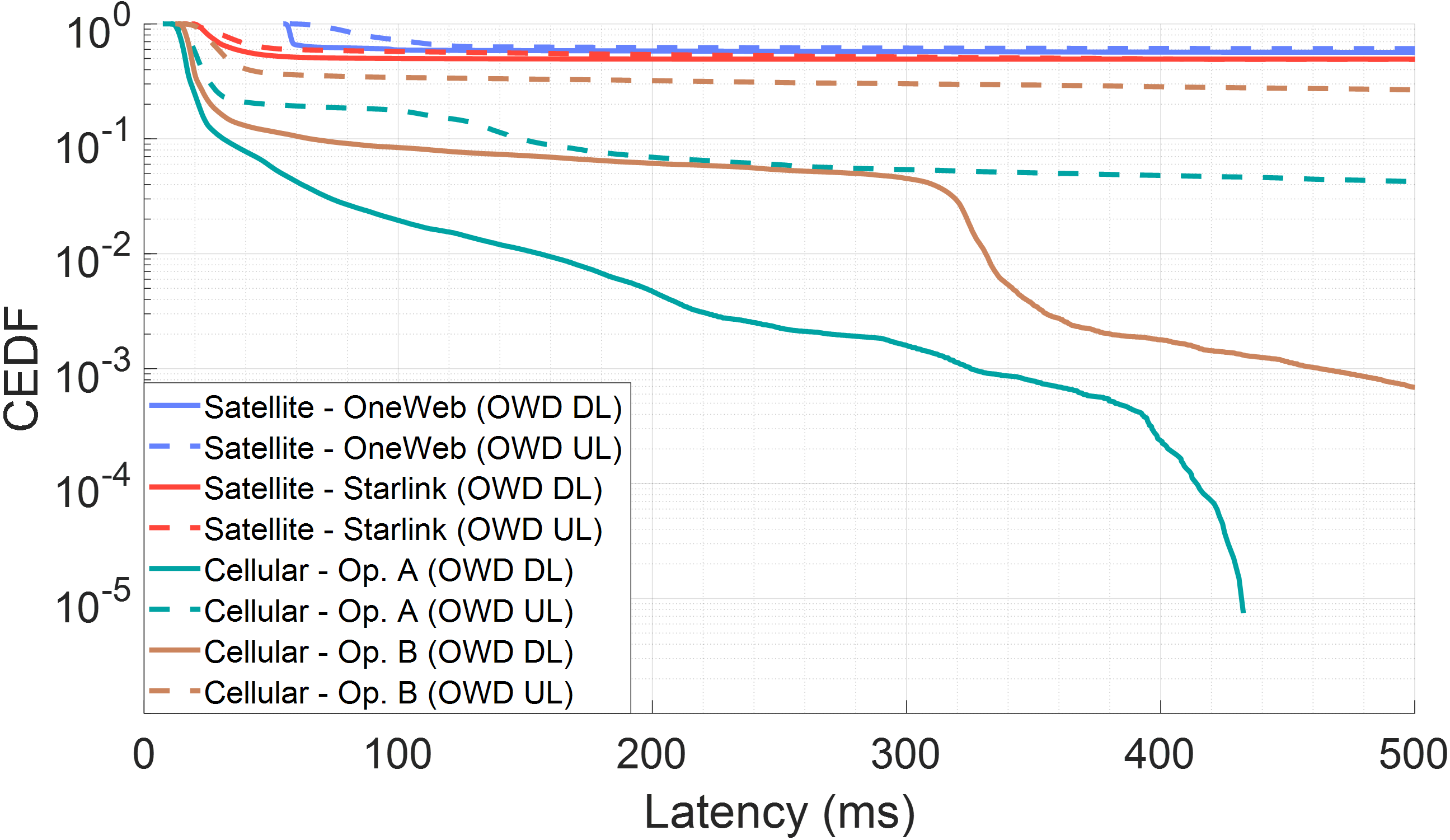}
	}
    \caption{Complementary empirical distribution function for one-way delay in downlink (solid line) and uplink (dashed line) for satellite operators OneWeb (\scalebox{0.8}{\textcolor[rgb]{0.3961,0.5098,0.9922}{\ding{108}}}) and Starlink (\scalebox{0.8}{\textcolor[rgb]{1,0.2706,0.2275}{\ding{108}}}), and cellular operators Op. A (\scalebox{0.8}{\textcolor[rgb]{0,0.6392,0.6392}{\ding{108}}}) and Op. B (\scalebox{0.8}{\textcolor[rgb]{0.7961,0.5176,0.3647}{\ding{108}}}) in (a) urban, (b) suburban, and (c) forest environments.} 
	\label{fig7}
\end{figure}

To perform a quantitative analysis of the four interfaces in terms of latency, a statistical analysis is performed on the 7.7 million network traffic samples captured throughout the three scenarios described. As explained in Section II, the one-way delay analysis is based on packets corresponding to bidirectional UDP traffic generated at 2 ms intervals on each of the available interfaces. Figs.~\ref{fig7}(a)-(c) show the one-way delay Complementary Empirical Distribution Function (CEDF) for every interface in urban, suburban and rural scenarios, respectively. Additionally, Table~\ref{tab:percentiles_OWD} presents OWD values for several percentiles across the different technologies and scenarios.

Focusing on the urban environment, it is observed that cellular connectivity clearly outperforms satellite connectivity. In terms of median values, cellular operators have a median OWD of around 15-20 ms, both downlink and uplink, compared to 40/40 ms (DL/UL) for Starlink, and 60/90 ms (DL/UL) for OneWeb. These results are mainly due to the network architecture (see Fig.~\ref{fig4}) of each technology, where cellular operators tend to have a base station within hundreds of meters, while satellite connectivity requires packet forwarding through the satellite and the ground station located hundreds of kilometers apart. Additionally, OneWeb's higher average latency compared to Starlink is noteworthy, which could be partially attributed to the higher altitude of the constellation in OneWeb (1,200~km) compared to Starlink (550~km). In terms of network reliability and availability, cellular operators also outperform satellite solutions due to the high density of deployed base stations, with one-way delay values constrained to below 500 ms for network reliability larger than 99.99\%. This contrasts sharply with the performance of satellite communications, where OneWeb is unable to meet this threshold up to 19\%/21\% (DL/UL) of the time, and Starlink does not fulfill it with a probability of 12\%/17\% (DL/UL). Even with the sharp performance degradation for satellite technology, Starlink performs significantly better than OneWeb in terms of coverage loss, which can be attributed to the larger number of satellites deployed in the constellation, thus contributing to minimize shadow zones with NLoS. Finally, it is also worth noting the asymmetry found between downlink and uplink, where the uplink OWD tends to be statistically higher than the downlink OWD. This can be attributed to several factors, such as the limited transmission power of user equipment, and more constrained carrier bandwidth allocation compared to the downlink.

\renewcommand{\arraystretch}{0.9}

\begin{table}[!t]
\centering
\caption{One-way delay (ms) percentiles (downlink/uplink) for OneWeb, Starlink, Op. A, and Op. B in urban, suburban, and forest scenarios}
\label{tab:percentiles_OWD}
\resizebox{\columnwidth}{!}{%
\begin{tabular}{|c|c|c||c|c|c|c|c|c|}
\hline\hline
Scenario                                     & Link                                     & Technology & Min & P50 & P90  & P99  & P99.9 & P99.99 \\ \hline\hline
\multirow{8}{*}{\rotatebox{50}{Urban}}       & \multirow{4}{*}{DL}                      & OneWeb     & \ccl{50}  & \ccl{57}  & \ccl{Out.} & \ccl{Out.} & \ccl{Out.}  & \ccl{Out.} \\ \cline{3-9} 
                                             &                                          & Starlink   & \ccl{17}  & \ccl{36}  & \ccl{Out.} & \ccl{Out.} & \ccl{Out.}  & \ccl{Out.} \\ \cline{3-9} 
                                             &                                          & Op. A      & \ccl{7}   & \ccl{13}  & \ccl{18}   & \ccl{63}   & \ccl{214}   & \ccl{352}  \\ \cline{3-9} 
                                             &                                          & Op. B      & \ccl{10}  & \ccl{15}  & \ccl{17}   & \ccl{31}   & \ccl{187}   & \ccl{323}  \\ \cline{2-9} 
                                             & \multirow{4}{*}{UL}                      & OneWeb     & \ccl{54}  & \ccl{89}  & \ccl{Out.} & \ccl{Out.} & \ccl{Out.}  & \ccl{Out.} \\ \cline{3-9} 
                                             &                                          & Starlink   & \ccl{15}  & \ccl{43}  & \ccl{Out.} & \ccl{Out.} & \ccl{Out.}  & \ccl{Out.} \\ \cline{3-9} 
                                             &                                          & Op. A      & \ccl{7}   & \ccl{17}  & \ccl{40}   & \ccl{137}  & \ccl{298}   & \ccl{476}  \\ \cline{3-9} 
                                             &                                          & Op. B      & \ccl{12}  & \ccl{22}  & \ccl{29}   & \ccl{131}  & \ccl{245}   & \ccl{334}  \\ \hline\hline
\multirow{8}{*}{\rotatebox{50}{Suburban}}    & \multirow{4}{*}{DL}                      & OneWeb     & \ccl{54}  & \ccl{58}  & \ccl{177}  & \ccl{Out.} & \ccl{Out.}  & \ccl{Out.} \\ \cline{3-9} 
                                             &                                          & Starlink   & \ccl{18}  & \ccl{26}  & \ccl{43}   & \ccl{Out.} & \ccl{Out.}  & \ccl{Out.} \\ \cline{3-9} 
                                             &                                          & Op. A      & \ccl{9}   & \ccl{13}  & \ccl{17}   & \ccl{21}   & \ccl{70}    & \ccl{99}   \\ \cline{3-9} 
                                             &                                          & Op. B      & \ccl{11}  & \ccl{15}  & \ccl{17}   & \ccl{20}   & \ccl{118}   & \ccl{204}  \\ \cline{2-9} 
                                             & \multirow{4}{*}{UL}                      & OneWeb     & \ccl{57}  & \ccl{88}  & \ccl{122}  & \ccl{Out.} & \ccl{Out.}  & \ccl{Out.}  \\ \cline{3-9} 
                                             &                                          & Starlink   & \ccl{16}  & \ccl{32}  & \ccl{58}   & \ccl{Out.} & \ccl{Out.}  & \ccl{Out.}  \\ \cline{3-9} 
                                             &                                          & Op. A      & \ccl{7}   & \ccl{15}  & \ccl{20}   & \ccl{28}   & \ccl{82}    & \ccl{115}    \\ \cline{3-9} 
                                             &                                          & Op. B      & \ccl{13}  & \ccl{24}  & \ccl{33}   & \ccl{212}  & \ccl{270}   & \ccl{353}    \\ \hline\hline
\multirow{8}{*}{\rotatebox{50}{Forest}}       & \multirow{4}{*}{DL}                     & OneWeb     & \ccl{55}  & \ccl{Out.}& \ccl{Out.} & \ccl{Out.} & \ccl{Out.}  & \ccl{Out.}   \\ \cline{3-9} 
                                             &                                          & Starlink   & \ccl{19}  & \ccl{99}  & \ccl{Out.} & \ccl{Out.} & \ccl{Out.}  & \ccl{Out.}   \\ \cline{3-9} 
                                             &                                          & Op. A      & \ccl{11}  & \ccl{16}  & \ccl{31}   & \ccl{156}  & \ccl{325}   & \ccl{414}    \\ \cline{3-9} 
                                             &                                          & Op. B      & \ccl{12}  & \ccl{19}  & \ccl{65}   & \ccl{331}  & \ccl{462}   & \ccl{Out.}   \\ \cline{2-9} 
                                             & \multirow{4}{*}{UL}                      & OneWeb     & \ccl{57}  & \ccl{Out.}& \ccl{Out.} & \ccl{Out.} & \ccl{Out.}  & \ccl{Out.}    \\ \cline{3-9} 
                                             &                                          & Starlink   & \ccl{15}  & \ccl{376} & \ccl{Out.} & \ccl{Out.} & \ccl{Out.}  & \ccl{Out.}    \\ \cline{3-9} 
                                             &                                          & Op. A      & \ccl{7}   & \ccl{21}  & \ccl{148}  & \ccl{Out.} & \ccl{Out.}  & \ccl{Out.}    \\ \cline{3-9} 
                                             &                                          & Op. B      & \ccl{13}  & \ccl{33}  & \ccl{Out.} & \ccl{Out.} & \ccl{Out.}  & \ccl{Out.}    \\ \hline\hline
\end{tabular}%
}
\end{table}

In the suburban scenario (see Fig.~\ref{fig7}(b)), there is a generalized statistical improvement in terrestrial and non-terrestrial solutions. While the median values are comparable between the urban and suburban scenarios, there is an improvement in network reliability reflected in the improved CEDF tail distribution. Specifically, the downlink OWD is reduced from values of 352 ms (Op.~A) and 323 ms (Op.~B) in the urban environment to 99 ms (Op.~A) and 204 ms (Op.~B) in the suburban environment for 99.99\% reliability. Regarding OneWeb and Starlink, the loss of coverage probability decreases from 19\%/21\% (urban DL/UL) and 12\%/17\% (urban DL/UL) to 2\%/2\% (suburban DL/UL) and 3\%/3\% (suburban DL/UL), respectively. Taking a closer look at Fig.~\ref{fig7}(b), the OWD DL CEDF for OneWeb in the suburban scenario exhibits a bimodal behavior characterized by two latency clusters at around 55-60~ms and 95-100~ms, respectively. This behavior is consistent with handover dynamics between different ground stations, as previously reported in~\cite{SOTA_PAN_3}. Similar bimodal latency patterns have been associated with changes in gateway routing, where traffic is dynamically forwarded through geographically distributed ground stations. In our measurements, this behavior can be attributed to the connection of OneWeb satellites to different European ground stations (see Section~II), which leads to different latency clusters. Note that these patterns are not as noticeable in urban and rural environments, as they are dominated by the NLoS condition and signal blockage, which intrinsically modifies the latency distributions, generating long tails without a clear bimodal behavior. Overall, the previous results indicate an improvement in radio conditions that enhances connectivity across all four interfaces compared to the urban scenario. In the case of cellular networks, the improvement in LoS conditions and lower building density enable improved network KPIs despite the lower number of base stations per~km$^2$. In the case of satellite networks, the lower number of obstacles on the horizon also leads to a notable improvement in LoS conditions and, consequently, connectivity. It is noteworthy that when the horizon is mostly clear, as it is the case for the suburban environment compared to the urban scenario, the performance between the two satellite networks tends to become more homogeneous. Therefore, the redundancy provided by the larger number of satellites in Starlink is no longer as relevant.

\renewcommand{\arraystretch}{1}

\begin{table*}[!t]
\centering
\caption{Best connectivity ranking based on three one-way delay thresholds (100 ms, 300 ms, and 500 ms) for several technologies and scenarios.\\1st (\scalebox{0.8}{\textcolor[rgb]{1.0000,0.8431,0}{\ding{108}}}), 2nd (\scalebox{0.8}{\textcolor[rgb]{0.7529,0.7529,0.7529}{\ding{108}}}), and 3rd (\scalebox{0.8}{\textcolor[rgb]{0.8039,0.4980,0.1961}{\ding{108}}}) best connectivity solutions have been highlighted.}
\label{tab:statistics_OWD}
\resizebox{\textwidth}{!}{%
\begin{tabular}{|c|||cccccc|||cccccc|||cccccc|}
\hline \hline
\textit{Scenario} &
  \multicolumn{6}{c|||}{Urban} &
  \multicolumn{6}{c|||}{Suburban} &
  \multicolumn{6}{c|}{Forest} \\ \hline
\textit{One Way Delay} &
  \multicolumn{3}{c||}{OWD DL} &
  \multicolumn{3}{c|||}{OWD UL} &
  \multicolumn{3}{c||}{OWD DL} &
  \multicolumn{3}{c|||}{OWD UL} &
  \multicolumn{3}{c||}{OWD DL} &
  \multicolumn{3}{c|}{OWD UL} \\ \hline
$P(\textrm{OWD DL/UL}) < X \textrm{ ms}$ &
  \multicolumn{1}{c|}{100} &
  \multicolumn{1}{c|}{300} &
  \multicolumn{1}{c||}{500} &
  \multicolumn{1}{c|}{100} &
  \multicolumn{1}{c|}{300} &
  500 &
  \multicolumn{1}{c|}{100} &
  \multicolumn{1}{c|}{300} &
  \multicolumn{1}{c||}{500} &
  \multicolumn{1}{c|}{100} &
  \multicolumn{1}{c|}{300} &
  500 &
  \multicolumn{1}{c|}{100} &
  \multicolumn{1}{c|}{300} &
  \multicolumn{1}{c||}{500} &
  \multicolumn{1}{c|}{100} &
  \multicolumn{1}{c|}{300} &
  500 \\ \hline \hline
OneWeb &
  \multicolumn{1}{c|}{77.71} &
  \multicolumn{1}{c|}{80.64} &
  \multicolumn{1}{c||}{81.00} &
  \multicolumn{1}{c|}{60.29} &
  \multicolumn{1}{c|}{77.83} &
  78.37 &
  \multicolumn{1}{c|}{85.41} &
  \multicolumn{1}{c|}{96.07} &
  \multicolumn{1}{c||}{\cellcolor{bronze}97.56} &
  \multicolumn{1}{c|}{67.73} &
  \multicolumn{1}{c|}{\cellcolor{bronze}96.73} &
  \cellcolor{bronze}97.79 &
  \multicolumn{1}{c|}{40.67} &
  \multicolumn{1}{c|}{42.71} &
  \multicolumn{1}{c||}{43.47} &
  \multicolumn{1}{c|}{27.78} &
  \multicolumn{1}{c|}{38.44} &
  39.12 \\ \hline
Starlink &
  \multicolumn{1}{c|}{\cellcolor{bronze}81.89} &
  \multicolumn{1}{c|}{\cellcolor{bronze}88.24} &
  \multicolumn{1}{c||}{\cellcolor{bronze}88.24} &
  \multicolumn{1}{c|}{\cellcolor{bronze}75.12} &
  \multicolumn{1}{c|}{\cellcolor{bronze}83.76} &
  \cellcolor{bronze}84.83 &
  \multicolumn{1}{c|}{\cellcolor{bronze}95.99} &
  \multicolumn{1}{c|}{\cellcolor{bronze}96.24} &
  \multicolumn{1}{c||}{96.24} &
  \multicolumn{1}{c|}{\cellcolor{bronze}94.08} &
  \multicolumn{1}{c|}{96.14} &
  96.59 &
  \multicolumn{1}{c|}{\cellcolor{bronze}50.00} &
  \multicolumn{1}{c|}{\cellcolor{bronze}50.65} &
  \multicolumn{1}{c||}{\cellcolor{bronze}50.65} &
  \multicolumn{1}{c|}{\cellcolor{bronze}42.53} &
  \multicolumn{1}{c|}{\cellcolor{bronze}48.51} &
  \cellcolor{bronze}50.88 \\ \hline
Op. A &
  \multicolumn{1}{c|}{\cellcolor{gold}99.65} &
  \multicolumn{1}{c|}{\cellcolor{silver}99.96} &
  \multicolumn{1}{c||}{\cellcolor{silver}99.999} &
  \multicolumn{1}{c|}{\cellcolor{silver}92.92} &
  \multicolumn{1}{c|}{\cellcolor{silver}99.90} &
  \cellcolor{silver}99.994 &
  \multicolumn{1}{c|}{\cellcolor{gold}99.992} &
  \multicolumn{1}{c|}{\cellcolor{gold}99.999} &
  \multicolumn{1}{c||}{\cellcolor{gold}99.999} &
  \multicolumn{1}{c|}{\cellcolor{gold}99.96} &
  \multicolumn{1}{c|}{\cellcolor{gold}99.999} &
  \cellcolor{gold}99.999 &
  \multicolumn{1}{c|}{\cellcolor{silver}80.43} &
  \multicolumn{1}{c|}{\cellcolor{gold}99.84} &
  \multicolumn{1}{c||}{\cellcolor{gold}99.999} &
  \multicolumn{1}{c|}{\cellcolor{gold}82.36} &
  \multicolumn{1}{c|}{\cellcolor{gold}94.60} &
  \cellcolor{gold}95.75 \\ \hline
Op. B &
  \multicolumn{1}{c|}{\cellcolor{silver}99.52} &
  \multicolumn{1}{c|}{\cellcolor{gold}99.98} &
  \multicolumn{1}{c||}{\cellcolor{gold}99.999} &
  \multicolumn{1}{c|}{\cellcolor{gold}99.86} &
  \multicolumn{1}{c|}{\cellcolor{gold}99.98} &
  \cellcolor{gold}99.999 &
  \multicolumn{1}{c|}{\cellcolor{silver}99.88} &
  \multicolumn{1}{c|}{\cellcolor{silver}99.999} &
  \multicolumn{1}{c||}{\cellcolor{silver}99.999} &
  \multicolumn{1}{c|}{\cellcolor{silver}97.22} &
  \multicolumn{1}{c|}{\cellcolor{silver}99.95} &
  \cellcolor{silver}99.999 &
  \multicolumn{1}{c|}{\cellcolor{gold}91.61} &
  \multicolumn{1}{c|}{\cellcolor{silver}95.48} &
  \multicolumn{1}{c||}{\cellcolor{silver}99.93} &
  \multicolumn{1}{c|}{\cellcolor{silver}65.83} &
  \multicolumn{1}{c|}{\cellcolor{silver}69.86} &
  \cellcolor{silver}73.31 \\ \hline\hline
\end{tabular}%
}
\end{table*}

Finally, the forest scenario is the most challenging among all tested connectivity solutions, showing the worst results among all the scenarios studied. For cellular operators, it is possible to establish downlink connectivity, although with significant degradation in the distribution tails, e.g., 414 ms one-way delay (Op.~A) and loss of connectivity (Op.~B) at the P99.99 percentile. This behavior worsens even further in the uplink, with latencies exceeding 500 ms in 4\% (Op.~A) and 26\% (Op.~B) of the samples, respectively. These results are consistent with the previous discussion related to the one-way delay asymmetry due to the limitation of the uplink link budget, whose influence is even more noticeable in scenarios with heavy signal attenuation and few base stations, such as in the forest. These constraints suggest that under conditions in which we can establish connectivity, the system typically operates close to or beyond its uplink capacity when attempting to sustain the target 5~Mbps UL throughput (see Section III). This may lead to more conservative modulation and coding schemes at the MAC layers, increased HARQ retransmissions, and scheduling and buffering delays, which translate into significantly increased uplink one-way delay and, therefore, OWD asymmetry. In the satellite case, this scenario is predominantly affected by the NLoS condition due to the presence of vegetation along the horizon (see Fig.~\ref{fig2}(c)). Therefore, the probability of complete coverage loss is up to 56\%/60\% (DL/UL) in OneWeb and up to 49\%/49\% (DL/UL) in Starlink. In this case, Starlink's larger constellation is once again relevant for improving visibility compared to OneWeb.

\begin{figure}[!t]
	\centering
	\subfigure[]{\includegraphics[width=1\columnwidth]{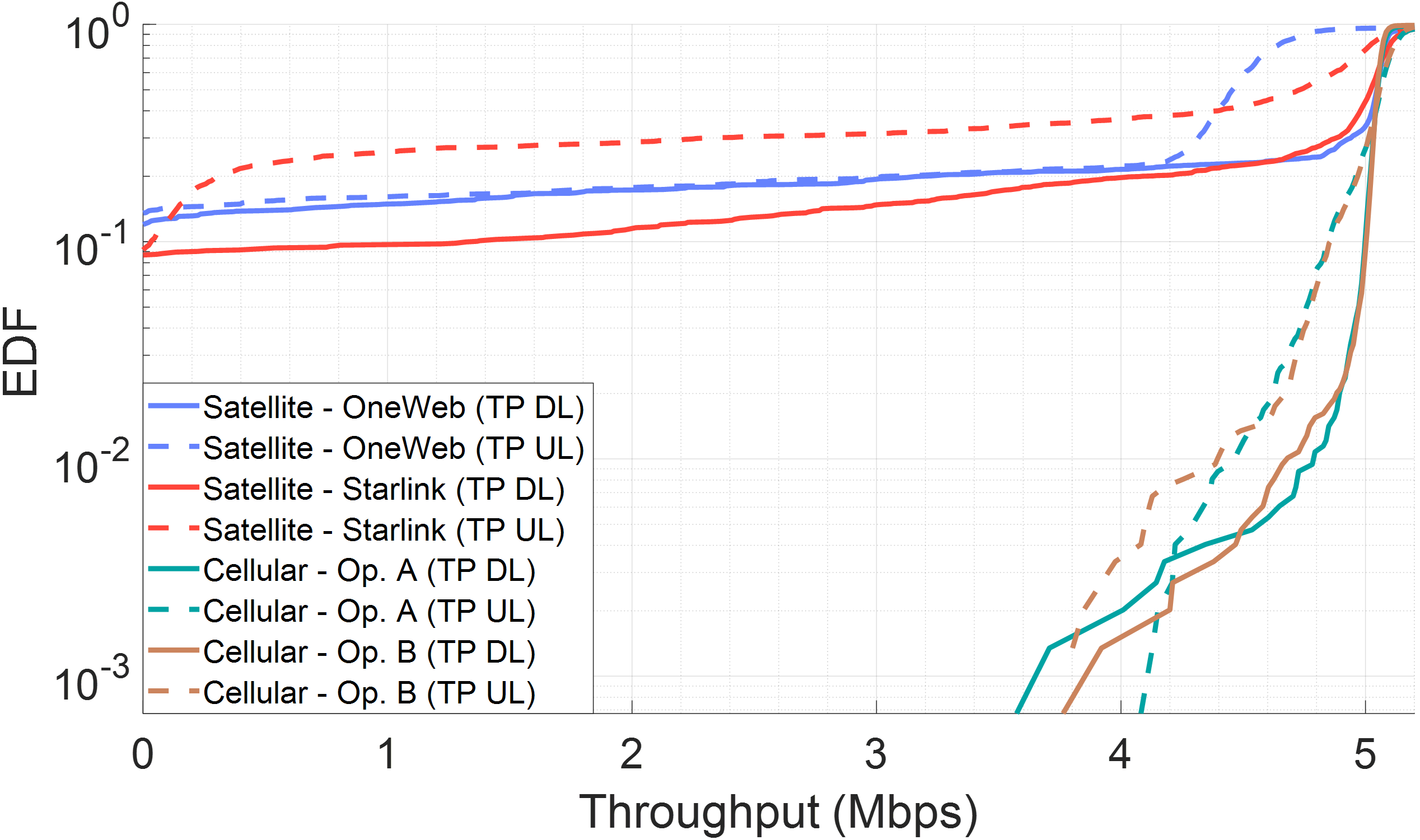}
	} 
    \subfigure[]{\includegraphics[width=1\columnwidth]{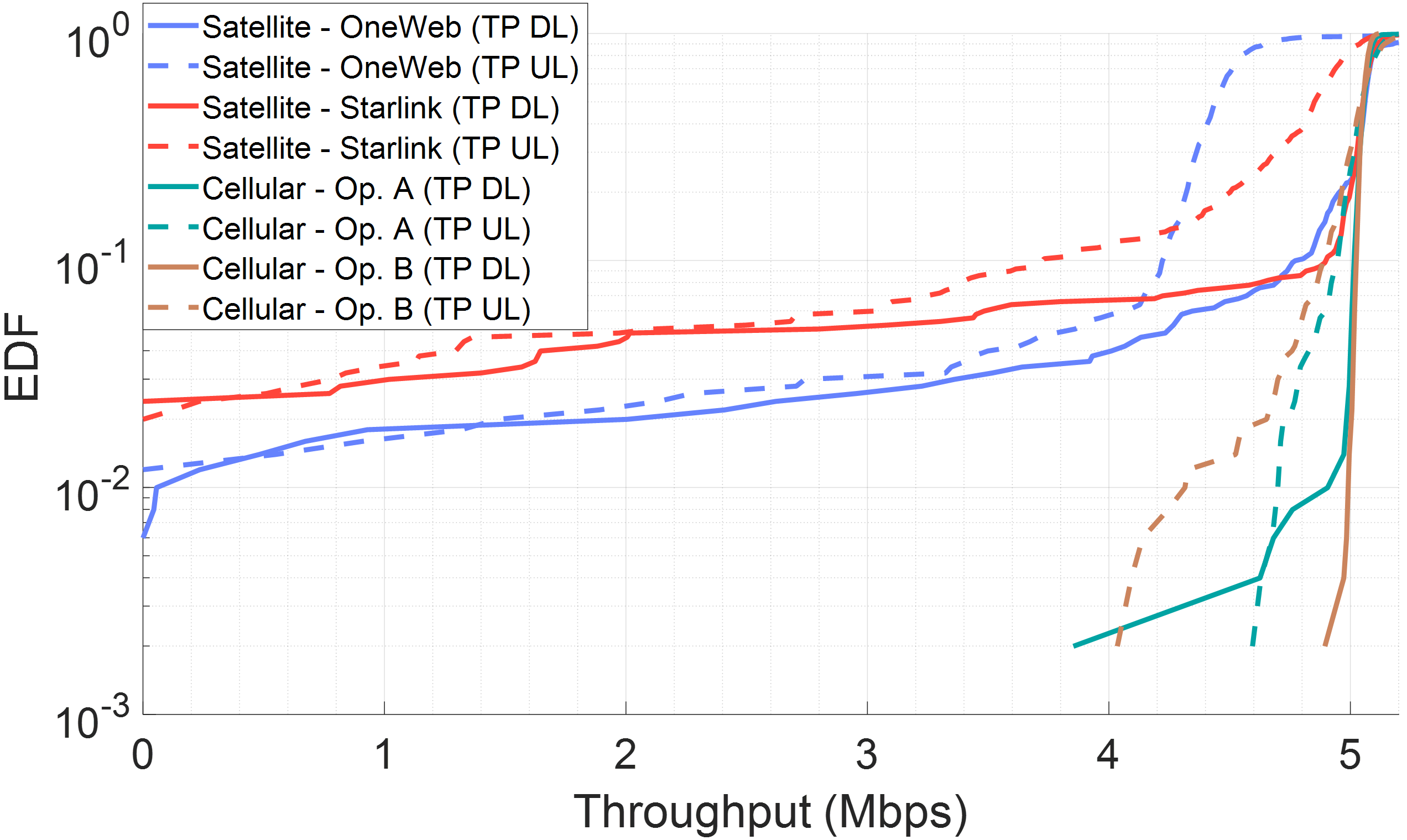}
	}
	\subfigure[]{\includegraphics[width=1\columnwidth]{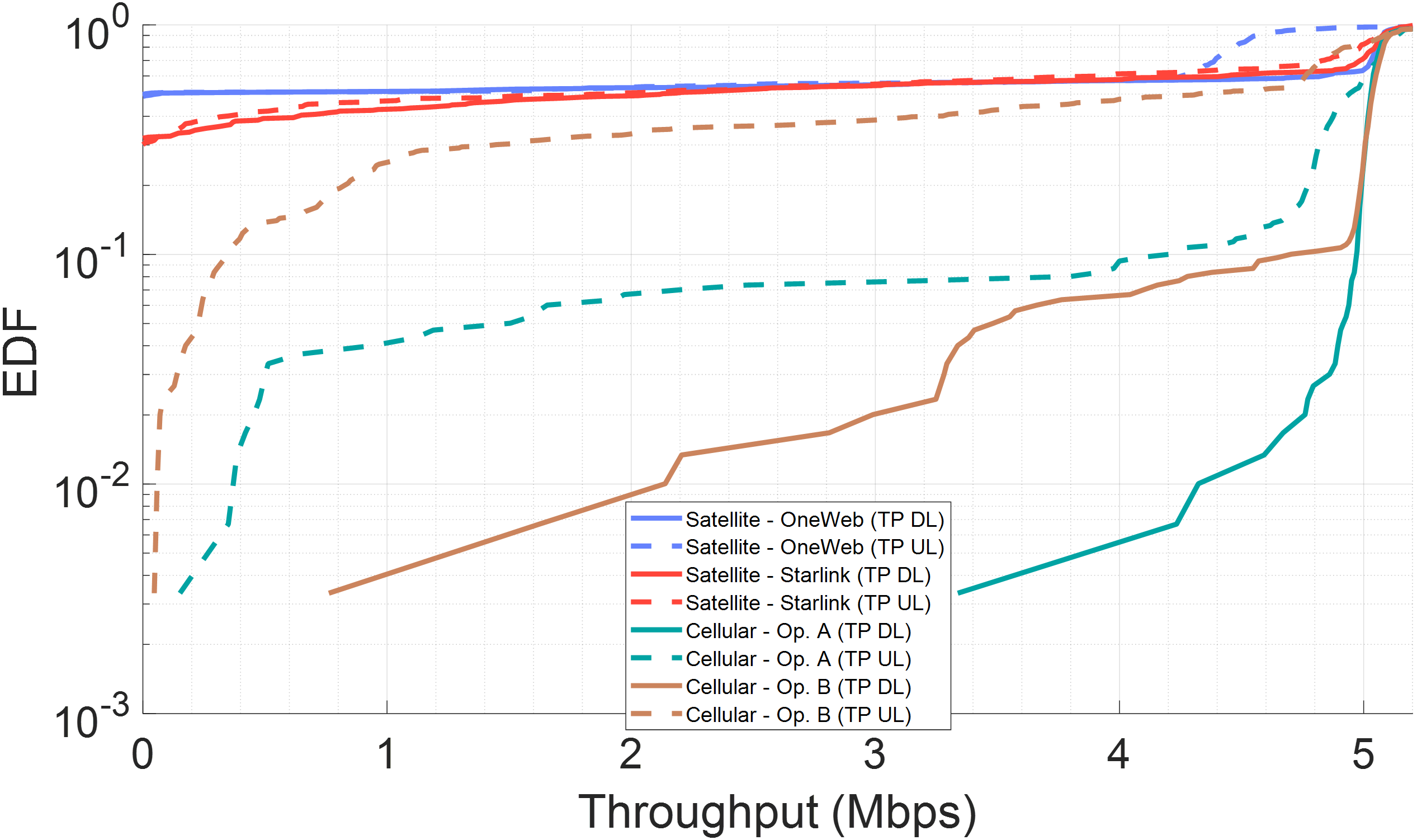}
	}
    \caption{Empirical distribution function for throughput in downlink (solid line) and uplink (dashed line) for satellite operators OneWeb (\scalebox{0.8}{\textcolor[rgb]{0.3961,0.5098,0.9922}{\ding{108}}}) and Starlink (\scalebox{0.8}{\textcolor[rgb]{1,0.2706,0.2275}{\ding{108}}}), and cellular operators Op. A (\scalebox{0.8}{\textcolor[rgb]{0,0.6392,0.6392}{\ding{108}}}) and Op. B (\scalebox{0.8}{\textcolor[rgb]{0.7961,0.5176,0.3647}{\ding{108}}}) in (a) urban, (b) suburban, and (c) forest environments.} 
	\label{fig8}
\end{figure}

To determine the best connectivity solution at a certain instance, Table~\ref{tab:statistics_OWD} shows the probability that the one-way delay will meet several thresholds (100~ms, 300~ms and 500~ms) for different scenarios and technologies\footnote{Table~\ref{tab:statistics_OWD} presents all results with two significant decimals, except for probabilities larger than 99.99\%, for which a third significant decimal is provided.}. Based on these probabilities, a ranking of the best technologies is determined for each scenario, link (DL/UL), and threshold. As previously anticipated based on the CEDF analysis, both cellular solutions are better than satellite options in all cases. Overall, cellular connectivity can be provided with OWD latencies lower than 100 ms with reliability larger than or close to 99\% of cases, except in rural areas where the lack of infrastructure significantly degrades performance. When it comes to satellite connectivity, Starlink is often the better solution over OneWeb for these thresholds in urban and forested areas due to its larger satellite constellation, which counteracts the high probability of NLoS. In the suburban scenario, both technologies show similar performance for 300 ms and 500 ms thresholds, with \mbox{P(OWD DL/UL) $<$ 300/500 ms} around 96\%-97\%. However, when constraining OWD to 100 ms, Starlink achieves \mbox{P(OWD DL/UL) $<$ 100 ms} of 95.99\%/94.08\% (DL/UL) compared to 85.41\%/67.73\% (DL/UL) for OneWeb. These results suggest that Starlink is better suited for low-latency applications due to several factors, including the constellation's lower altitude and its denser network of ground stations, which supports more efficient traffic routing.

Concerning the comparison between the hardware equipment used for each satellite constellation, i.e., Kymeta Hawk U8 (OneWeb) vs. Standard Actuated Gen 2 (Starlink), as indicated in Section III, the larger OneWeb antenna is focused on enterprise market share, while the compact Starlink antenna is focused on the mass market consumer. Due to the different sizes, and assuming the same efficiency and frequency band, the difference in antenna gain $\Delta G$ can be approximated from the factor between areas as $10log_{10}(A_{Starlink}/A_{OneWeb}) = 10log_{10}(0.155\,\,\textrm{m}^2/0.801\,\,\textrm{m}^2) = -7.1\,\,\textrm{dB}$. To determine whether this difference is significant for the results, the time traces and the CEDFs from the measurement campaign have been examined. In scenarios prone to signal blocking, such as urban and forest areas, latency behavior tends to exhibit a binary pattern: either the network operates under regular conditions with occasional latency spikes due to, for instance, periodic reconfiguration of satellite network links~\cite{Reconfiguration_Starlink}, or prolonged coverage interruptions are found with OWDs exceeding 500 ms and sustained packet losses indicating continuous loss of connectivity. Under these conditions, where the signal is prone to NLoS propagation in the satellite frequency bands, the difference in antenna gain becomes irrelevant due to the high attenuation of tens of decibels that occurs when obstacles obstruct the link. In these scenarios, Starlink performance is superior, which can be attributed to the higher satellite density that mitigates the probability of signal blockage (see Table~\ref{tab:statistics_OWD}). In scenarios with excellent visibility where NLoS conditions are not as relevant, such as suburban environments, the performance of both constellations converges in the high latency regime at very similar values. According to the ranking in Table~\ref{tab:statistics_OWD}, OneWeb ensures an OWD of less than 500 ms with a probability 1.2\%-1.3\% higher than Starlink. In this scenario, where visibility for both constellations is favorable and similar performance is obtained, this slight difference in favor of OneWeb may be linked to the user terminal design difference.  

\subsection{Throughput Statistics}

This subsection analyzes the data rate achieved on each of the interfaces throughout each scenario. Following the setup described in Section II, a bidirectional 5/5 Mbps target throughput is established. Figs.~\ref{fig8}(a)-(c) and Table~\ref{tab:percentiles_TP} show the throughput empirical distribution function (EDF) and percentiles, respectively, for each of the scenarios and operators analyzed. Note that the results presented correspond to the instantaneous throughput averaged at one-second intervals.

\renewcommand{\arraystretch}{0.9}

\begin{table}[!t]
\centering
\caption{Throughput (Mbps) percentiles (downlink/uplink) for OneWeb, Starlink, Op. A, and Op. B in urban, suburban, and forest scenarios}
\label{tab:percentiles_TP}
\resizebox{\columnwidth}{!}{%
\begin{tabular}{|c|c|c||c|c|c|c|c|c|}
\hline\hline
Scenario                                     & Link                                     & Technology & Min         & P0.5          & P1          & P5           & P10          & P50       \\ \hline\hline
\multirow{8}{*}{\rotatebox{50}{Urban}}       & \multirow{4}{*}{DL}                      & OneWeb     & \cct{Out.}  & \cct{Out.}    & \cct{Out.}  & \cct{Out.}   & \cct{Out.}   & \cct{5.0} \\ \cline{3-9} 
                                             &                                          & Starlink   & \cct{Out.}  & \cct{Out.}    & \cct{Out.}  & \cct{Out.}   & \cct{1.4}    & \cct{5.0} \\ \cline{3-9} 
                                             &                                          & Op. A      & \cct{3.6}   & \cct{4.6}     & \cct{4.8}   & \cct{5.0}    & \cct{5.0}    & \cct{5.0} \\ \cline{3-9} 
                                             &                                          & Op. B      & \cct{3.8}   & \cct{4.5}     & \cct{4.7}   & \cct{5.0}    & \cct{5.0}    & \cct{5.0} \\ \cline{2-9} 
                                             & \multirow{4}{*}{UL}                      & OneWeb     & \cct{Out.}  & \cct{Out.}    & \cct{Out.}  & \cct{Out.}   & \cct{Out.}   & \cct{4.5} \\ \cline{3-9} 
                                             &                                          & Starlink   & \cct{Out.}  & \cct{Out.}    & \cct{Out.}  & \cct{Out.}   & \cct{0.1}    & \cct{4.7} \\ \cline{3-9} 
                                             &                                          & Op. A      & \cct{4.1}   & \cct{4.3}     & \cct{4.5}   & \cct{4.8}    & \cct{4.9}    & \cct{5.0} \\ \cline{3-9} 
                                             &                                          & Op. B      & \cct{3.8}   & \cct{4.1}     & \cct{4.4}   & \cct{4.8}    & \cct{4.9}    & \cct{5.0} \\ \hline\hline
\multirow{8}{*}{\rotatebox{50}{Suburban}}    & \multirow{4}{*}{DL}                      & OneWeb     & \cct{Out.}  & \cct{Out.}    & \cct{0.1}   & \cct{4.3}    & \cct{4.8}    & \cct{5.0} \\ \cline{3-9} 
                                             &                                          & Starlink   & \cct{Out.}  & \cct{Out.}    & \cct{Out.}  & \cct{2.8}    & \cct{4.9}    & \cct{5.0} \\ \cline{3-9} 
                                             &                                          & Op. A      & \cct{3.9}   & \cct{4.7}     & \cct{4.9}   & \cct{5.0}    & \cct{5.0}    & \cct{5.0} \\ \cline{3-9} 
                                             &                                          & Op. B      & \cct{4.9}   & \cct{5.0}     & \cct{5.0}   & \cct{5.0}    & \cct{5.0}    & \cct{5.0} \\ \cline{2-9} 
                                             & \multirow{4}{*}{UL}                      & OneWeb     & \cct{Out.}  & \cct{Out.}    & \cct{Out.}  & \cct{3.9}    & \cct{4.2}    & \cct{4.4} \\ \cline{3-9} 
                                             &                                          & Starlink   & \cct{Out.}  & \cct{Out.}    & \cct{Out.}  & \cct{2.1}    & \cct{3.7}    & \cct{4.9} \\ \cline{3-9} 
                                             &                                          & Op. A      & \cct{4.6}   & \cct{4.7}     & \cct{4.7}   & \cct{4.9}    & \cct{4.9}    & \cct{5.0} \\ \cline{3-9} 
                                             &                                          & Op. B      & \cct{4.0}   & \cct{4.1}     & \cct{4.3}   & \cct{4.8}    & \cct{4.9}    & \cct{5.0} \\ \hline\hline
\multirow{8}{*}{\rotatebox{50}{Forest}}       & \multirow{4}{*}{DL}                     & OneWeb     & \cct{Out.}  & \cct{Out.}    & \cct{Out.}  & \cct{Out.}   & \cct{Out.}   & \cct{0.1} \\ \cline{3-9} 
                                             &                                          & Starlink   & \cct{Out.}  & \cct{Out.}    & \cct{Out.}  & \cct{Out.}   & \cct{Out.}   & \cct{2.1} \\ \cline{3-9} 
                                             &                                          & Op. A      & \cct{3.3}   & \cct{3.9}     & \cct{4.3}   & \cct{4.9}    & \cct{5.0}    & \cct{5.0}    \\ \cline{3-9} 
                                             &                                          & Op. B      & \cct{0.8}   & \cct{1.3}     & \cct{2.1}   & \cct{3.5}    & \cct{4.7}    & \cct{5.0}   \\ \cline{2-9} 
                                             & \multirow{4}{*}{UL}                      & OneWeb     & \cct{Out.}  & \cct{Out.}    & \cct{Out.}  & \cct{Out.}   & \cct{Out.}   & \cct{0.1}    \\ \cline{3-9} 
                                             &                                          & Starlink   & \cct{Out.}  & \cct{Out.}    & \cct{Out.}  & \cct{Out.}   & \cct{Out.}   & \cct{2.0}    \\ \cline{3-9} 
                                             &                                          & Op. A      & \cct{0.2}   & \cct{0.3}     & \cct{0.4}   & \cct{1.5}    & \cct{4.2}    & \cct{4.9}    \\ \cline{3-9} 
                                             &                                          & Op. B      & \cct{0.1}   & \cct{0.1}     & \cct{0.1}   & \cct{0.2}    & \cct{0.3}    & \cct{4.3}   \\ \hline\hline
\end{tabular}%
}
\end{table}

In urban areas, cellular operators show stable performance with median DL/UL throughput values of 5/5 Mbps, i.e., the required target throughput. Even in the distribution tails, the minimum throughput values found throughout the campaign are around 3.5-4 Mbps, indicating network stability at \mbox{1-second} intervals, thus consistent with the results previously shown for OWD. In the case of satellite operators, the median values are around 5 Mbps, indicating good network availability for at least 50\% of the samples. However, towards the distribution tails, there is a sharp degradation in service, resulting in null coverage for 9\%/9\% (DL/UL) of the samples in Starlink and 12\%/14\% (DL/UL) of the samples in OneWeb. It is remarkable the horizontality of the EDF curves when throughput degradation occurs. This indicates that satellite connectivity is either capable of providing the target throughput or directly degrades to out-of-coverage scenarios, which is consistent with the physical event of LoS visibility loss in the link.

\begin{table*}[!t]
\centering
\caption{Best connectivity ranking based on three throughput thresholds (0 Mbps, 2 Mbps, and 4 Mbps) for several technologies and scenarios.\\1st (\scalebox{0.8}{\textcolor[rgb]{1.0000,0.8431,0}{\ding{108}}}), 2nd (\scalebox{0.8}{\textcolor[rgb]{0.7529,0.7529,0.7529}{\ding{108}}}), and 3rd (\scalebox{0.8}{\textcolor[rgb]{0.8039,0.4980,0.1961}{\ding{108}}}) best connectivity solutions have been highlighted.}
\label{tab:statistics_TP}
\resizebox{\textwidth}{!}{%
\begin{tabular}{|c|||cccccc|||cccccc|||cccccc|}
\hline \hline
\textit{Scenario} &
  \multicolumn{6}{c|||}{Urban} &
  \multicolumn{6}{c|||}{Suburban} &
  \multicolumn{6}{c|}{Forest} \\ \hline
\textit{Throughput} &
  \multicolumn{3}{c||}{TP DL} &
  \multicolumn{3}{c|||}{TP UL} &
  \multicolumn{3}{c||}{TP DL} &
  \multicolumn{3}{c|||}{TP UL} &
  \multicolumn{3}{c||}{TP DL} &
  \multicolumn{3}{c|}{TP UL} \\ \hline
$P(\textrm{TP DL/UL}) > X \textrm{ Mbps}$ &
  \multicolumn{1}{c|}{0} &
  \multicolumn{1}{c|}{2} &
  \multicolumn{1}{c||}{4} &
  \multicolumn{1}{c|}{0} &
  \multicolumn{1}{c|}{2} &
  4 &
  \multicolumn{1}{c|}{0} &
  \multicolumn{1}{c|}{2} &
  \multicolumn{1}{c||}{4} &
  \multicolumn{1}{c|}{0} &
  \multicolumn{1}{c|}{2} &
  4 &
  \multicolumn{1}{c|}{0} &
  \multicolumn{1}{c|}{2} &
  \multicolumn{1}{c||}{4} &
  \multicolumn{1}{c|}{0} &
  \multicolumn{1}{c|}{2} &
  4 \\ \hline \hline
OneWeb &
  \multicolumn{1}{c|}{88.0} &
  \multicolumn{1}{c|}{82.8} &
  \multicolumn{1}{c||}{78.5} &
  \multicolumn{1}{c|}{86.5} &
  \multicolumn{1}{c|}{\cellcolor{bronze}82.2} &
  \cellcolor{bronze}77.7 &
  \multicolumn{1}{c|}{\cellcolor{bronze}99.4} &
  \multicolumn{1}{c|}{\cellcolor{bronze}98.0} &
  \multicolumn{1}{c||}{\cellcolor{bronze}96.0} &
  \multicolumn{1}{c|}{\cellcolor{bronze}98.8} &
  \multicolumn{1}{c|}{\cellcolor{bronze}97.8} &
  \cellcolor{bronze}94.2 &
  \multicolumn{1}{c|}{51.2} &
  \multicolumn{1}{c|}{46.8} &
  \multicolumn{1}{c||}{\cellcolor{bronze}42.8} &
  \multicolumn{1}{c|}{50.2} &
  \multicolumn{1}{c|}{46.5} &
  \cellcolor{bronze}42.8 \\ \hline
Starlink &
  \multicolumn{1}{c|}{\cellcolor{bronze}91.3} &
  \multicolumn{1}{c|}{\cellcolor{bronze}88.5} &
  \multicolumn{1}{c||}{\cellcolor{bronze}80.3} &
  \multicolumn{1}{c|}{\cellcolor{bronze}90.8} &
  \multicolumn{1}{c|}{71.2} &
  63.2 &
  \multicolumn{1}{c|}{97.6} &
  \multicolumn{1}{c|}{95.4} &
  \multicolumn{1}{c||}{93.4} &
  \multicolumn{1}{c|}{98.0} &
  \multicolumn{1}{c|}{95.2} &
  88.4 &
  \multicolumn{1}{c|}{\cellcolor{bronze}67.9} &
  \multicolumn{1}{c|}{\cellcolor{bronze}50.8} &
  \multicolumn{1}{c||}{42.1} &
  \multicolumn{1}{c|}{\cellcolor{bronze}69.9} &
  \multicolumn{1}{c|}{\cellcolor{bronze}49.5} &
  38.8 \\ \hline
Op. A &
  \multicolumn{1}{c|}{\cellcolor{silver}99.9} &
  \multicolumn{1}{c|}{\cellcolor{silver}99.9} &
  \multicolumn{1}{c||}{\cellcolor{silver}99.8} &
  \multicolumn{1}{c|}{\cellcolor{gold}99.9} &
  \multicolumn{1}{c|}{\cellcolor{gold}99.9} &
  \cellcolor{gold}99.9 &
  \multicolumn{1}{c|}{\cellcolor{silver}99.9} &
  \multicolumn{1}{c|}{\cellcolor{silver}99.9} &
  \multicolumn{1}{c||}{\cellcolor{silver}99.8} &
  \multicolumn{1}{c|}{\cellcolor{gold}99.9} &
  \multicolumn{1}{c|}{\cellcolor{gold}99.9} &
  \cellcolor{gold}99.9 &
  \multicolumn{1}{c|}{\cellcolor{gold}99.9} &
  \multicolumn{1}{c|}{\cellcolor{gold}99.9} &
  \multicolumn{1}{c||}{\cellcolor{gold}99.3} &
  \multicolumn{1}{c|}{\cellcolor{gold}99.9} &
  \multicolumn{1}{c|}{\cellcolor{gold}93.3} &
  \cellcolor{gold}90.6 \\ \hline
Op. B &
  \multicolumn{1}{c|}{\cellcolor{gold}99.9} &
  \multicolumn{1}{c|}{\cellcolor{gold}99.9} &
  \multicolumn{1}{c||}{\cellcolor{gold}99.9} &
  \multicolumn{1}{c|}{\cellcolor{silver}99.9} &
  \multicolumn{1}{c|}{\cellcolor{silver}99.9} &
  \cellcolor{silver}99.7 &
  \multicolumn{1}{c|}{\cellcolor{gold}99.9} &
  \multicolumn{1}{c|}{\cellcolor{gold}99.9} &
  \multicolumn{1}{c||}{\cellcolor{gold}99.9} &
  \multicolumn{1}{c|}{\cellcolor{silver}99.9} &
  \multicolumn{1}{c|}{\cellcolor{silver}99.9} &
  \cellcolor{silver}99.9 &
  \multicolumn{1}{c|}{\cellcolor{silver}99.9} &
  \multicolumn{1}{c|}{\cellcolor{silver}99.0} &
  \multicolumn{1}{c||}{\cellcolor{silver}93.3} &
  \multicolumn{1}{c|}{\cellcolor{silver}99.9} &
  \multicolumn{1}{c|}{\cellcolor{silver}66.6} &
  \cellcolor{silver}52.5 \\ \hline \hline
\end{tabular}%
}
\end{table*}

In suburban areas, cellular operators show very similar behavior to that in urban areas, with median values of 5~Mbps and minimum values of around 4~Mbps. However, there is an improvement in coverage outage for satellite operators. In particular, this probability decreases to 2\%/2\% (DL/UL) for Starlink and 0.6\%/1\% (DL/UL) for OneWeb. Numerically, this fact is reflected in Table 3, where the outage appears around the P10 percentile for the urban area and the P1 percentile for the suburban area, denoting an improvement in terms of reliability of an order of magnitude. As discussed in Section~IV.B, this improvement is attributable to better radio propagation conditions in suburban environments compared to urban ones.

In the forest scenario, cellular operators are able to offer values close or equal to 5 Mbps for median values. Notwithstanding, connectivity begins to degrade for percentiles below P50. Although connectivity is never completely lost, the throughput rate drops to values close to 0 Mbps. This is especially noticeable in the uplink, where there is a significant asymmetry in the available throughput compared to the downlink. For satellite connectivity, even at median values, throughput suffers a sharp degradation, with values reaching 2.1/2.0~Mbps (DL/UL) for Starlink and 0.1/0.1~Mbps (DL/UL) for OneWeb for the P50 percentile. Starlink's better throughput could be due to a larger number of visible satellites and rerouting capacity during the driven route. Despite this, both satellite solutions are very prone to losing connectivity, with probabilities of 32\%/30\% (DL/UL) for Starlink and 49\%/50\% (DL/UL) for OneWeb. 

Finally, Table~\ref{tab:statistics_TP} presents the ranking of the best connectivity solution for all three scenarios based on three possible thresholds: 0 Mbps (i.e., null connectivity), 2 Mbps, and 4 Mbps. As a general note, cellular solutions tend to meet all three thresholds with probabilities of around 99.9\%, except in the forest scenario, where the limited infrastructure available degrades connectivity, especially in Op. B. Focusing on satellite operators, Starlink tends to be the best option in the urban scenario for downlink. In terms of uplink, Starlink is the better option (90.8\% Starlink vs. 86.5\% OneWeb) for establishing connectivity in the link, i.e., \mbox{P(TP UL $>$ 0 Mbps)}. Nonetheless, when higher uplink throughput is required, e.g., 2 Mbps (71.2\% Starlink vs. 82.2\% OneWeb) or 4 Mbps (63.2\% Starlink vs. 77.7\% OneWeb), OneWeb offers better results. This may result from the larger antenna size in OneWeb (see Section~III and~IV.B), thus allowing for more beneficial radio propagation conditions, i.e., antenna gain and transmission power in the uplink. These results support the discussion in the previous Section~IV.B, where it was indicated that Starlink is generally a better option in terms of connectivity capacity in blockage-prone environments due to its higher satellite density. However, under the assumption of LoS between ground equipment and satellites, the performance of both satellite solutions tends to become more homogeneous, although with a slight performance advantage for OneWeb due to its larger antenna size. This hypothesis is also supported by the results in suburban and forest environments. In particular, in suburban environments where excellent LoS conditions are expected, the performance of both solutions is very similar, with OneWeb having a slight advantage over Starlink. However, in forest environments, where the probability of NLoS is exceptionally high, Starlink tends to be a better solution.

\begin{figure*}[!t]
	\centering
	\includegraphics[width= 1\textwidth]{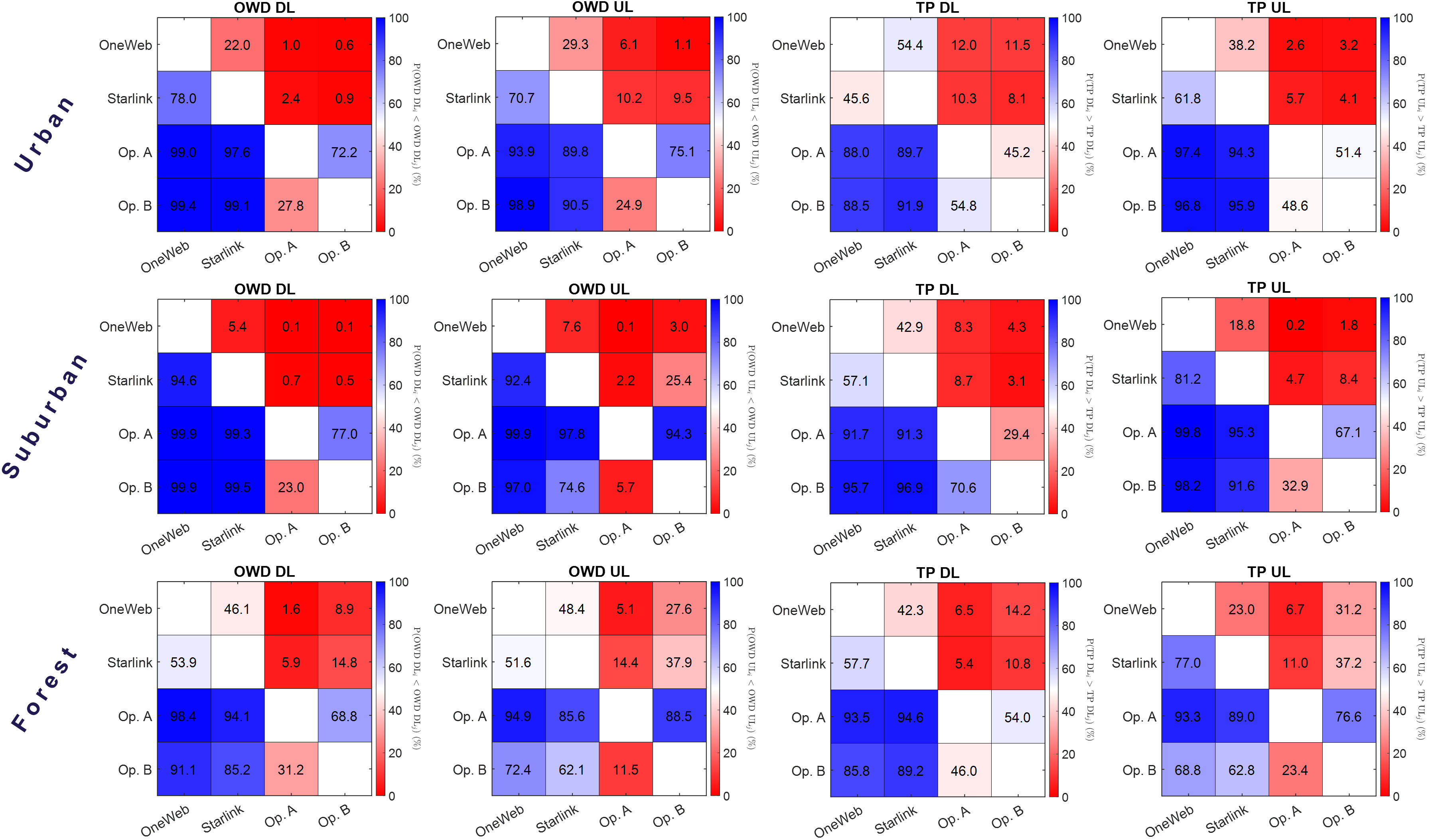}
	\caption{Direct technology-to-technology comparison of one-way delay (DL/UL) and throughput (DL/UL) KPIs for the three scenarios analyzed.} 
	\label{fig9}
\end{figure*}

\subsection{Technology Comparison and Discussion}

Based on the results presented throughout this Section, it is clear that in a direct comparison between TN and NTN solutions, the former are better based on the KPIs analyzed statistically. This is due to a number of specific factors, such as the fact that in the scenarios under study, cellular networks are particularly well developed and have an infrastructure that has been established over decades. In contrast, non-terrestrial networks are still in a phase of expansion and do not have the same density or global coverage as terrestrial networks. Despite this, the previous results indicate that satellite networks can be a good alternative as backups in emergencies, with connectivity of around 90\% in urban areas and 98-99\% in suburban areas. Going a step beyond the statistical study of the previous subsections, Fig.~\ref{fig9} shows a direct comparison of performance between technologies. This analysis is possible due to the temporal synchronization between UDP packets sent through different interfaces, thus allowing us to assess which interface offers lower latency or higher throughput at a given moment. Let $\mathbf{A} \in \mathbb{R}^{4 \times 4}$ be a matrix whose rows $i$ and columns $j$ represent the OneWeb, Starlink, Op.~A, and Op.~B technologies, respectively. The element $A_{i,j}$ determines the probability that the OWD of the technology in row $i$ is lower than that of the technology in row $j$, i.e., \mbox{$P(\textrm{OWD DL/UL}_i <\textrm{OWD DL/UL}_j)$}. Equivalently, in the throughput analysis, the probability that the TP of the technology in row $i$ is greater than that of the technology in row $j$ is determined, i.e., \mbox{$P(\textrm{TP DL/UL}_i >\textrm{TP DL/UL}_j)$}. Given this representation, submatrix $\mathbf{A}_{1:2,1:2}$ shows an intra-comparison between satellite technologies, submatrix $\mathbf{A}_{3:4,3:4}$ shows an intra-comparison between cellular technologies, and submatrices $\mathbf{A}_{1:2,3:4}$ and $\mathbf{A}_{3:4,1:2}$ show an inter-comparison between terrestrial and non-terrestrial technologies. Visually, blue tones indicate that the technology in row $i$ is superior to that in column $j$. Conversely, red tones indicate that the technology in row $i$ is worse to that in column $j$. In general terms, and in line with the results outlined in the statistical analysis, it can be concluded that both mobile operators are clearly superior in all scenarios under all metrics (see blue tones in submatrix $\mathbf{A}_{3:4,1:2}$ versus red tones in submatrix $\mathbf{A}_{1:2,3:4}$). Thus, satellite networks should not be considered a substitute technology for the current terrestrial deployment in the current state of technology. Nevertheless, they can be considered as a complement in cases where cellular networks are not operating under normal conditions. This idea gives rise to the concept of multi-connectivity, in which several interfaces or technologies can be integrated to increase the resilience and reliability of the network. This concept is explored and detailed in Section~V.

\section{Multi-Connectivity Evaluation}

Taking advantage of the synchronization when forwarding packets across the network in the setup proposed in Section~III, it is possible to develop the concept of multi-connectivity. Specifically, multi-connectivity involves the packet duplication and packet forwarding through two or more independent network interfaces. This duplication of packets brings two direct advantages: (i) larger network reliability, since in the event of packet loss on one of the interfaces, the alternative interface may still be able to forward that packet, and (ii) better KPI performance, since in terms of latency it will be reduced to that of the interface with the lowest delay, and in terms of throughput, this will be lower bounded by that of the interface with the highest available throughput. In exchange, the main cost of this approach lies in the redundancy of transmitted packets, which increases linearly with the number of interfaces used. Therefore, when two interfaces are considered, the traffic volume doubles. From a hardware perspective, the additional equipment must also be taken into account, as multi-connectivity implies multiple modems and/or user terminals, which leads to higher energy consumption and economic cost of the hardware and subscriptions to terrestrial and non-terrestrial networks. From a practical point of view, duplication can be carried out at the network layer by encapsulating IP packets with a sequence number and unique identifiers associated with each of the available interfaces. Similarly, this process is reversible on reception by decapsulating the IP packet, making it a completely transparent process at the application level. Note that when the same IP packet is received by a different interface, it can be discarded if the sequence number has already been received through an alternative interface. Applications implementing the described process have been successfully developed and tested at Aalborg University (AAU). These open-source tools are publicly available for general use~\cite{MC_tool}. In practice, the IP packet encapsulation involves an overhead of approximately 2.7\% for 1500 Bytes IP packets. As indicated in Section III, multi-connectivity performance is evaluated using synchronized time traces acquired from all available interfaces during the measurement campaign. This approach enables a fair comparison between single-connectivity and multi-connectivity cases under identical radio and physical conditions, which is not possible with multiple drive tests using live multi-connectivity implementations, where only one solution can be assessed at a time. While this methodology does not aim to emulate the full control plane behavior of a specific live multi-connectivity protocol (e.g., packet reordering, buffering, or timeout mechanisms), it allows us to evaluate the performance gains achieved by multi-connectivity due to interface diversity and packet duplication. Experimentation in controlled and real environments has shown that the latency and throughput distributions obtained through post-processing synchronization closely match those observed in live multi-connectivity implementations, with performance overhead due to additional processing remaining negligible compared to \mbox{end-to-end} network delays~\cite{procesado_MC}.

\renewcommand{\arraystretch}{0.9}

\begin{table}[!b]
\centering
\caption{One-way delay (ms) percentiles (downlink/uplink) for multi-connectivity modes in urban, suburban, and forest scenarios}
\label{tab:percentiles_OWD_MC}
\resizebox{\columnwidth}{!}{%
\begin{tabular}{|c|c|c||c|c|c|c|c|c|}
\hline\hline
Scenario                                     & Link                                     & Technology       & Min       & P50         & P90        & P99        & P99.9      & P99.99      \\ \hline\hline
\multirow{6}{*}{\rotatebox{50}{Urban}}       & \multirow{3}{*}{DL}                      & Cell./Cell. MC   & \ccl{7}   & \ccl{13}    & \ccl{15}   & \ccl{17}   & \ccl{21}   & \ccl{52}    \\ \cline{3-9} 
                                             &                                          & Sat./Sat. MC     & \ccl{17}  & \ccl{36}    & \ccl{62}   & \ccl{Out.} & \ccl{Out.} & \ccl{Out.}  \\ \cline{3-9} 
                                             &                                          & Cell./Sat. MC    & \ccl{7}   & \ccl{13}    & \ccl{18}   & \ccl{40}   & \ccl{76}   & \ccl{134}   \\ \cline{2-9} 
                                             & \multirow{3}{*}{UL}                      & Cell./Cell. MC   & \ccl{7}   & \ccl{16}    & \ccl{22}   & \ccl{30}   & \ccl{105}  & \ccl{156}   \\ \cline{3-9} 
                                             &                                          & Sat./Sat. MC     & \ccl{15}  & \ccl{51}    & \ccl{108}  & \ccl{Out.} & \ccl{Out.} & \ccl{Out.}  \\ \cline{3-9}  
                                             &                                          & Cell./Sat. MC    & \ccl{7}   & \ccl{17}    & \ccl{28}   & \ccl{122}  & \ccl{160}  & \ccl{296}   \\ \hline\hline
\multirow{6}{*}{\rotatebox{50}{Suburban}}    & \multirow{3}{*}{DL}                      & Cell./Cell. MC   & \ccl{9}   & \ccl{13}    & \ccl{15}   & \ccl{17}   & \ccl{19}   & \ccl{27}    \\ \cline{3-9} 
                                             &                                          & Sat./Sat. MC     & \ccl{18}  & \ccl{25}    & \ccl{43}   & \ccl{79}   & \ccl{Out.} & \ccl{Out.}  \\ \cline{3-9} 
                                             &                                          & Cell./Sat. MC    & \ccl{9}   & \ccl{13}    & \ccl{17}   & \ccl{21}   & \ccl{33}   & \ccl{72}    \\ \cline{2-9} 
                                             & \multirow{3}{*}{UL}                      & Cell./Cell. MC   & \ccl{7}   & \ccl{15}    & \ccl{20}   & \ccl{25}   & \ccl{45}   & \ccl{79}    \\ \cline{3-9} 
                                             &                                          & Sat./Sat. MC     & \ccl{16}  & \ccl{32}    & \ccl{63}   & \ccl{125}  & \ccl{Out.} & \ccl{Out.}  \\ \cline{3-9} 
                                             &                                          & Cell./Sat. MC    & \ccl{7}   & \ccl{15}    & \ccl{20}   & \ccl{26}   & \ccl{47}   & \ccl{78}    \\ \hline\hline
\multirow{6}{*}{\rotatebox{50}{Forest}}       & \multirow{3}{*}{DL}                     & Cell./Cell. MC   & \ccl{11}  & \ccl{16}    & \ccl{19}   & \ccl{52}   & \ccl{173}  & \ccl{202}   \\ \cline{3-9} 
                                             &                                          & Sat./Sat. MC     & \ccl{19}  & \ccl{45}    & \ccl{Out.} & \ccl{Out.} & \ccl{Out.} & \ccl{Out.}  \\ \cline{3-9} 
                                             &                                          & Cell./Sat. MC    & \ccl{11}  & \ccl{16}    & \ccl{25}   & \ccl{72}   & \ccl{174}  & \ccl{207}   \\ \cline{2-9} 
                                             & \multirow{3}{*}{UL}                      & Cell./Cell. MC   & \ccl{7}   & \ccl{21}    & \ccl{118}  & \ccl{Out.} & \ccl{Out.} & \ccl{Out.}  \\ \cline{3-9} 
                                             &                                          & Sat./Sat. MC     & \ccl{15}  & \ccl{227}   & \ccl{Out.} & \ccl{Out.} & \ccl{Out.} & \ccl{Out.}  \\ \cline{3-9} 
                                             &                                          & Cell./Sat. MC    & \ccl{7}   & \ccl{21}    & \ccl{121}  & \ccl{423}  & \ccl{Out.} & \ccl{Out.}  \\ \hline\hline
\end{tabular}%
}
\end{table}

\begin{figure}[t]
	\centering
	\subfigure[]{\includegraphics[width=1\columnwidth]{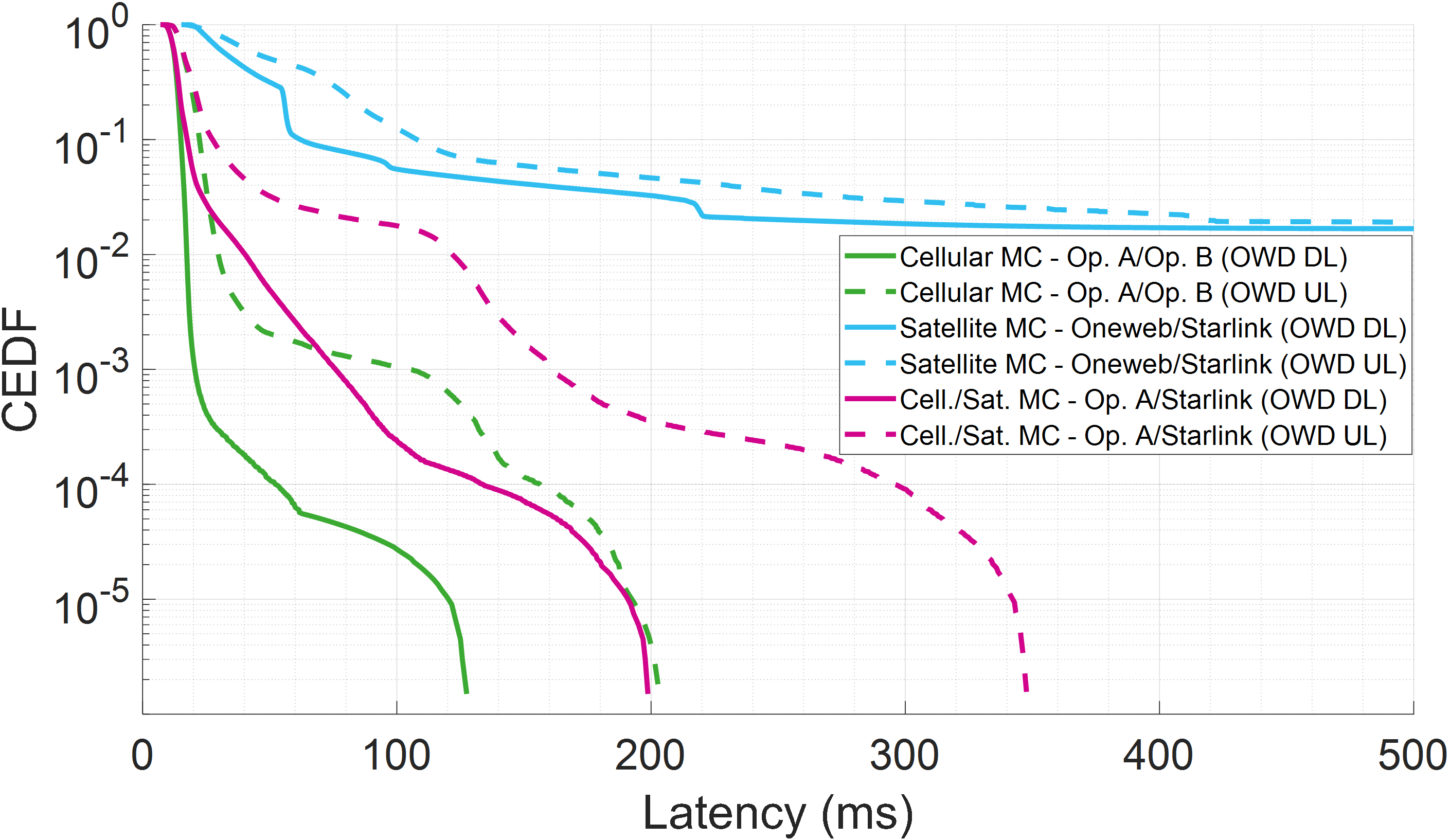}
	} 
    \subfigure[]{\includegraphics[width=1\columnwidth]{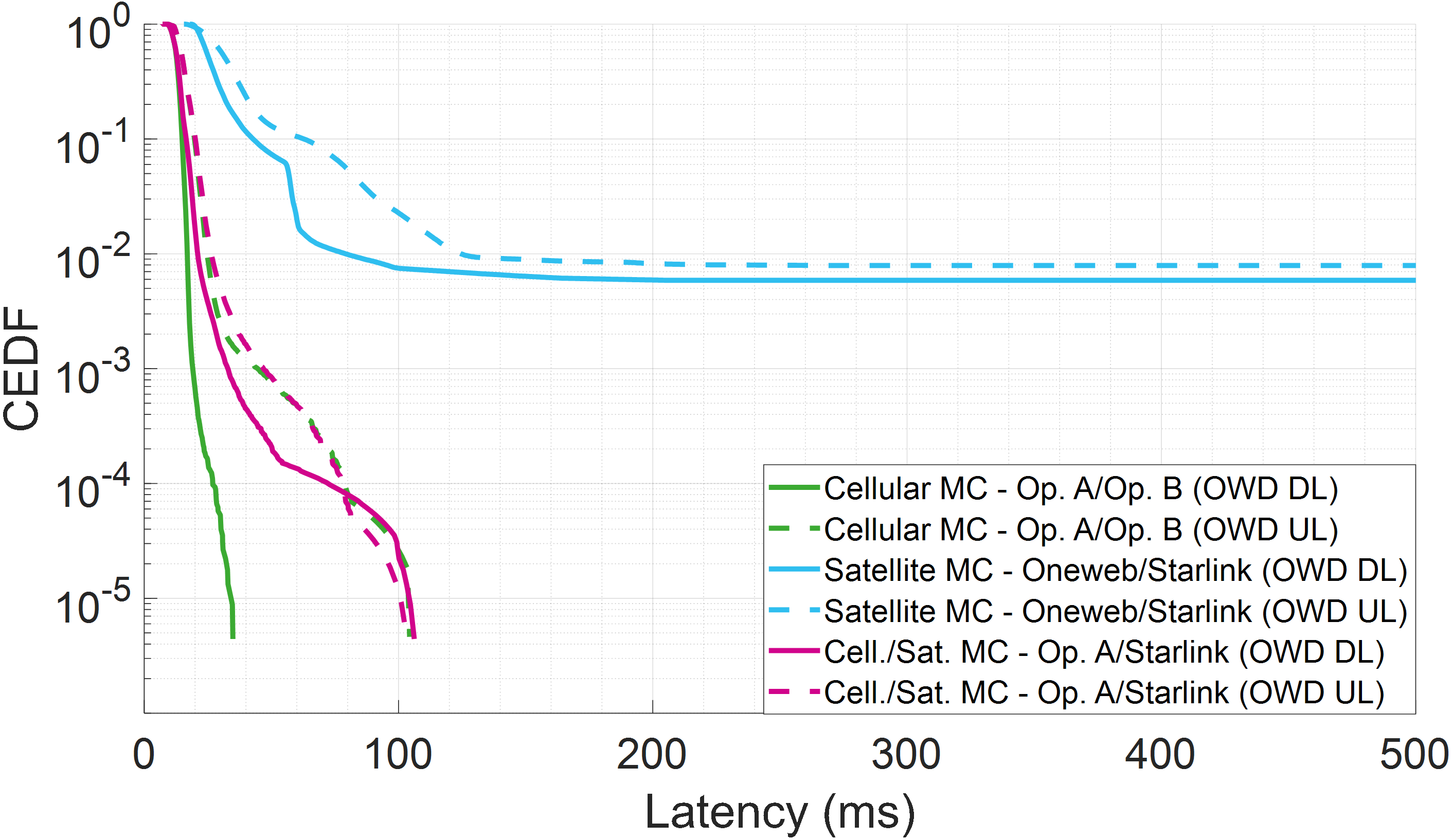}
	}
	\subfigure[]{\includegraphics[width=1\columnwidth]{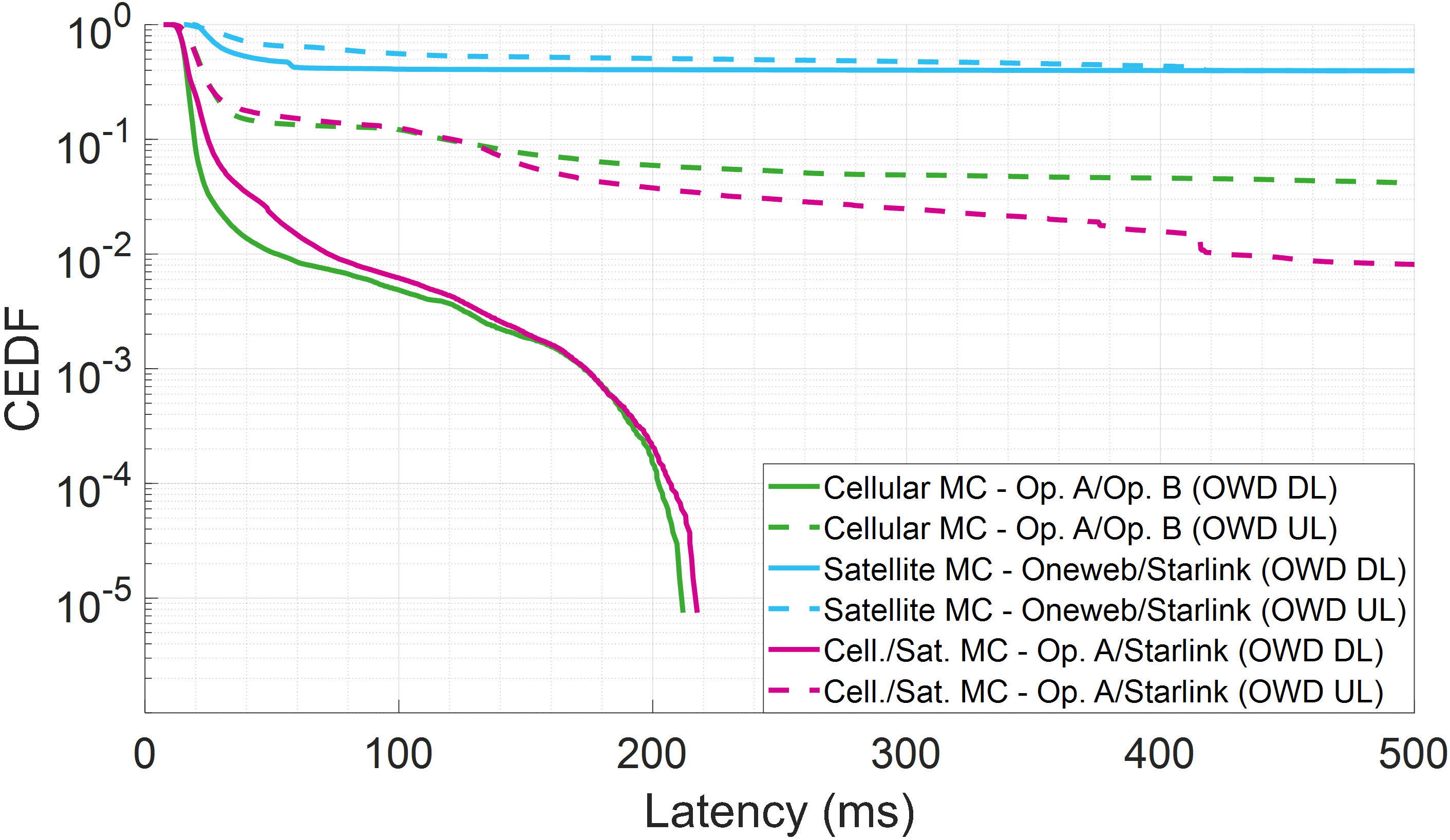}
	}
    \caption{Complementary empirical distribution function for one-way delay in downlink (solid line) and uplink (dashed line) for cellular (\scalebox{0.8}{\textcolor[rgb]{0.2314,0.6667,0.1961}{\ding{108}}}), satellite (\scalebox{0.8}{\textcolor[rgb]{0.1843,0.7451,0.9373}{\ding{108}}}) and cellular/satellite (\scalebox{0.8}{\textcolor[rgb]{0.8196,0.0157,0.5451}{\ding{108}}}) multi-connectivity in (a) urban, (b) suburban, and (c) forest environments.} 
	\label{fig10}
\end{figure}

\begin{figure}[t]
	\centering
	\subfigure[]{\includegraphics[width=0.977\columnwidth]{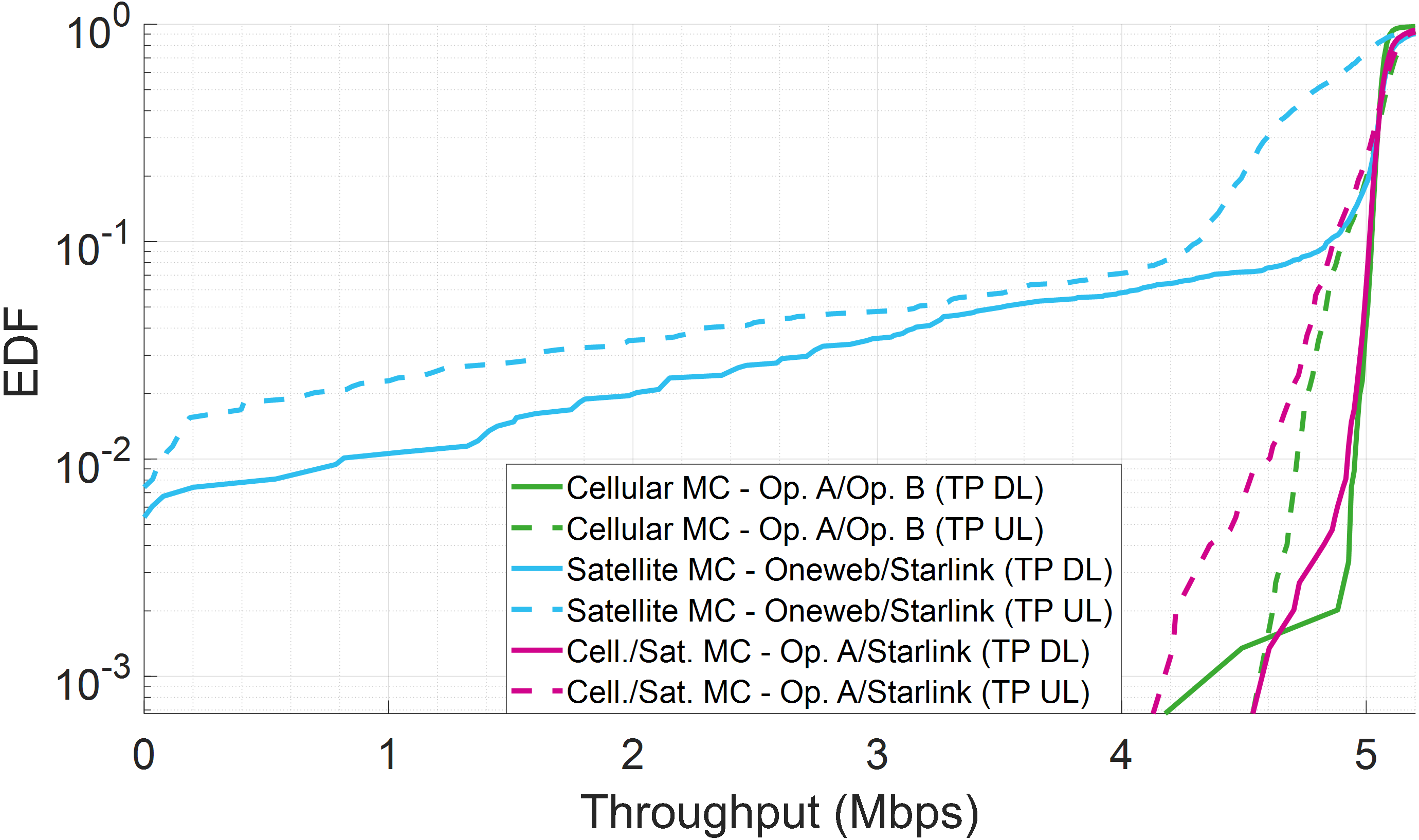}
	} 
    \subfigure[]{\includegraphics[width=0.98\columnwidth]{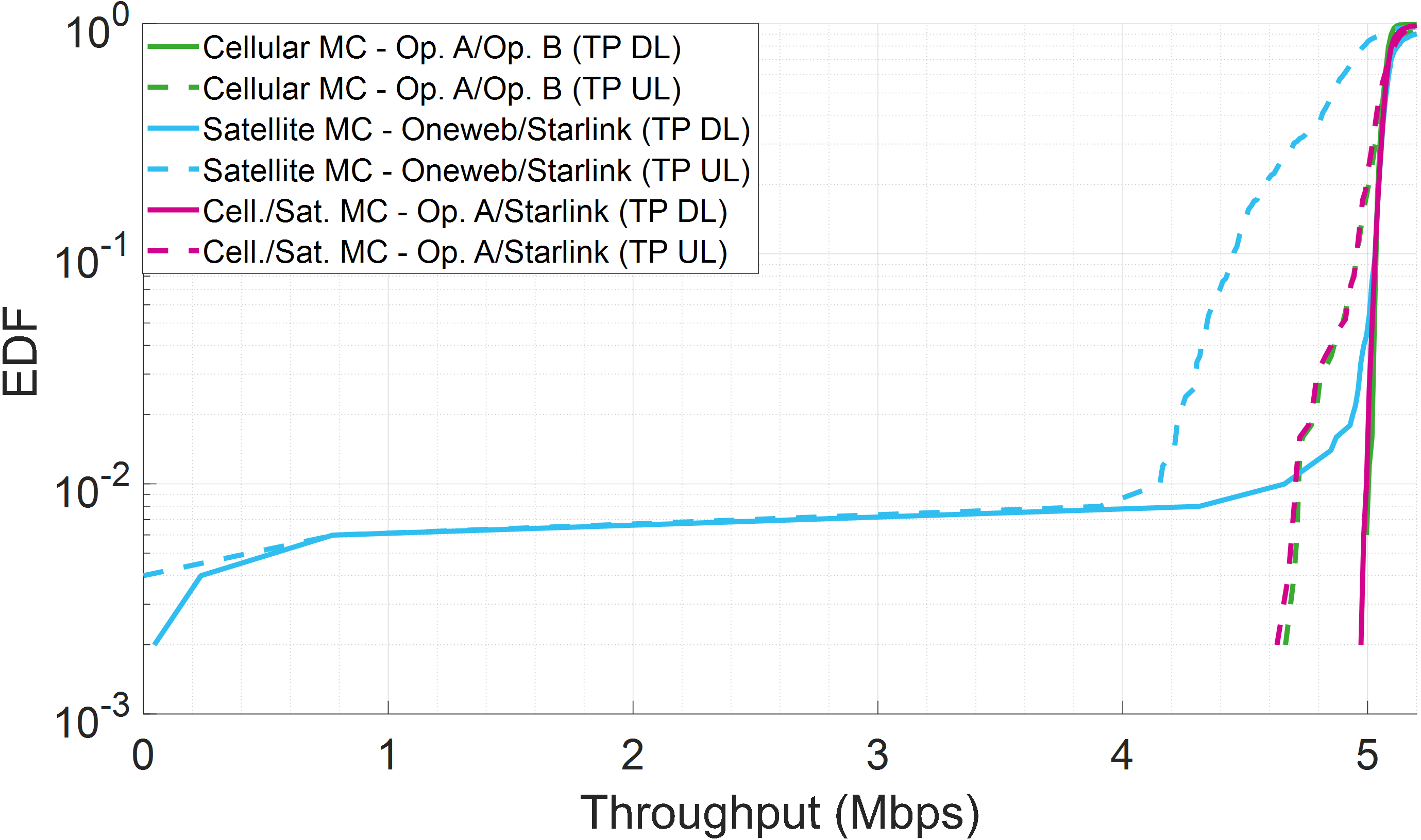}
	}
	\subfigure[]{\includegraphics[width=0.975\columnwidth]{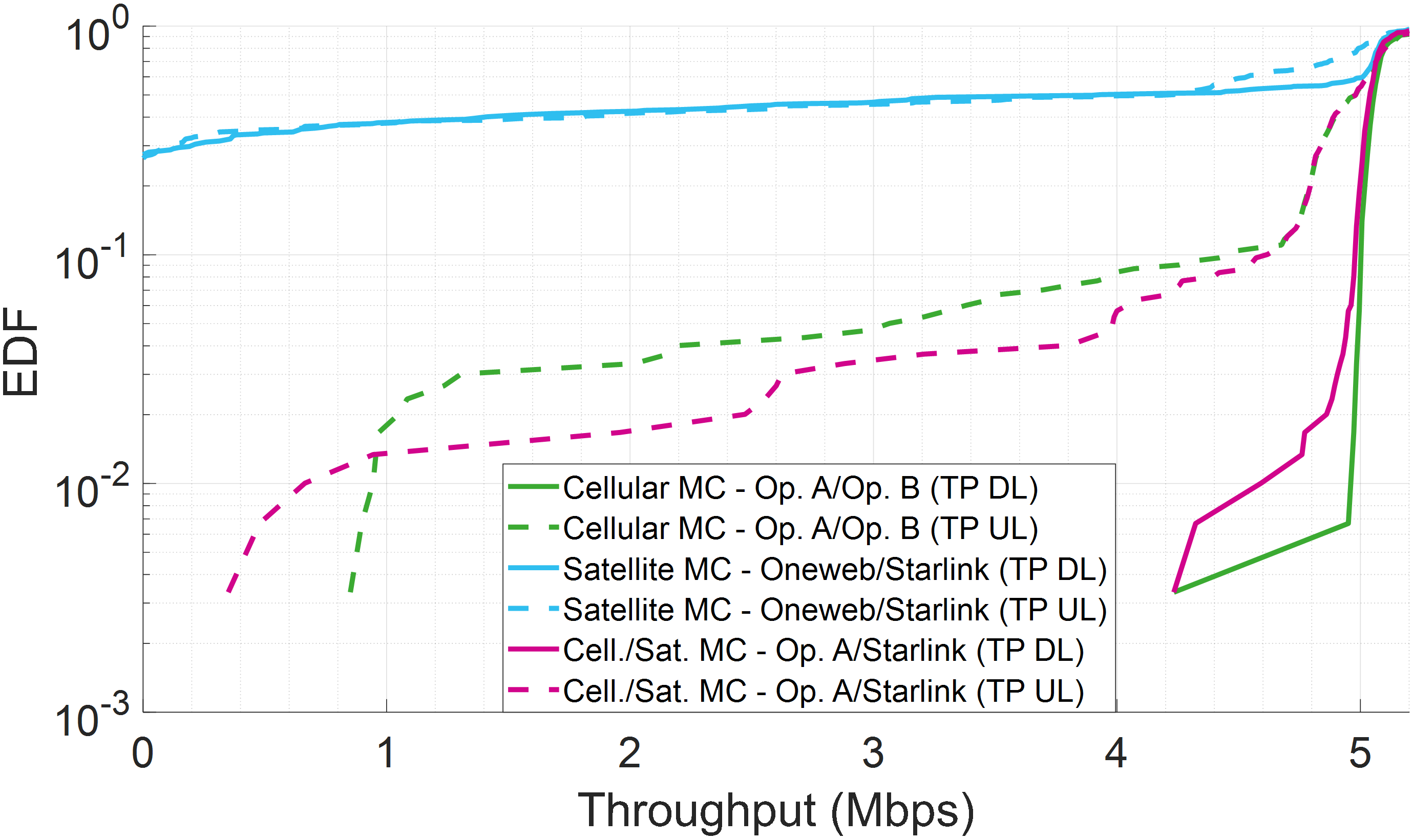}
	}
    \caption{Complementary empirical distribution function for throughput in downlink (solid line) and uplink (dashed line) for cellular (\scalebox{0.8}{\textcolor[rgb]{0.2314,0.6667,0.1961}{\ding{108}}}), satellite (\scalebox{0.8}{\textcolor[rgb]{0.1843,0.7451,0.9373}{\ding{108}}}) and cellular/satellite (\scalebox{0.8}{\textcolor[rgb]{0.8196,0.0157,0.5451}{\ding{108}}}) multi-connectivity in (a) urban, (b) suburban, and (c) forest environments.} 
	\label{fig11}
\end{figure}

Taking into account the connectivity options available in the measurement campaign presented throughout the work, three possible types of multi-connectivity are proposed: (i) cellular-cellular (Cell./Cell.) multi-connectivity between Op. A and Op. B, (ii) satellite-satellite (Sat./Sat.) multi-connectivity between OneWeb and Starlink, and (iii) cellular-satellite (Cell./Sat.) multi-connectivity between Op. A and Starlink. The first case (Cell./Cell.) takes advantage of the already dense deployment of cellular network base stations to maximize the joint coverage area of the infrastructure of Op.~A and Op.~B. The second case (Sat./Sat.) relies exclusively on satellite technology in the event of total failure of the terrestrial infrastructure. The combination of both satellite networks aims to take advantage of the satellites in both constellations to minimize spots without ground coverage. The third case (Cell./Sat.) simultaneously combines TN-NTN technology, thereby leveraging the complete independence of both communication systems to enhance the resilience of the communication channel. Tables~\ref{tab:percentiles_OWD_MC} and~\ref{tab:percentiles_TP_MC}, as well as Figs.~\ref{fig10} and~\ref{fig11}, show the percentile, CEDF, and EDF statistics for the one-way delay and throughput KPIs in the three scenarios, i.e., urban, suburban and forest, for the three multi-connectivity cases described.

Focusing on the one-way delay, a direct comparison can be performed between single-connectivity and multi-connectivity modes through Table~\ref{tab:percentiles_OWD} and Fig.~\ref{fig7} (single-connectivity), and Table~\ref{tab:percentiles_OWD_MC} and Fig.~\ref{fig10} (multi-connectivity), respectively. In particular, cellular multi-connectivity significantly reduces high latency peaks in the CEDF tails (see Fig.~\ref{fig10}(a)), reaching a maximum of 52/156~ms (DL/UL) in the urban scenario and 27/79~ms (DL/UL) in the suburban scenario for the P99 percentile, in contrast to the values above 300 ms observed for single-connectivity cases. Similarly, in the forest environment, the downlink OWD is limited to below 202 ms for the P99 percentile. However, the uplink remains poor due to link budget limitations in both operators.

\renewcommand{\arraystretch}{0.9}

\begin{table}[!b]
\centering
\caption{Throughput (Mbps) percentiles (downlink/uplink) for multi-connectivity modes in urban, suburban, and forest scenarios}
\label{tab:percentiles_TP_MC}
\resizebox{0.961\columnwidth}{!}{%
\begin{tabular}{|c|c|c||c|c|c|c|c|c|}
\hline\hline
Scenario                                     & Link                                     & Technology     & Min        & P0.5       & P1         & P5         & P10        & P50       \\ \hline\hline
\multirow{6}{*}{\rotatebox{50}{Urban}}       & \multirow{3}{*}{DL}                      & Cell./Cell. MC & \cct{4.2}  & \cct{4.9}  & \cct{5.0}  & \cct{5.0}  & \cct{5.0}  & \cct{5.0} \\ \cline{3-9} 
                                             &                                          & Sat./Sat. MC   & \cct{Out.} & \cct{Out.} & \cct{0.8}  & \cct{3.5}  & \cct{4.8}  & \cct{5.0} \\ \cline{3-9} 
                                             &                                          & Cell./Sat. MC  & \cct{4.5}  & \cct{4.9}  & \cct{4.9}  & \cct{5.0}  & \cct{5.0}  & \cct{5.0} \\ \cline{2-9} 
                                             & \multirow{3}{*}{UL}                      & Cell./Cell. MC & \cct{4.5}  & \cct{4.7}  & \cct{4.7}  & \cct{4.8}  & \cct{4.9}  & \cct{5.0} \\ \cline{3-9} 
                                             &                                          & Sat./Sat. MC   & \cct{Out.} & \cct{Out.} & \cct{0.1}  & \cct{3.2}  & \cct{4.3}  & \cct{4.8} \\ \cline{3-9} 
                                             &                                          & Cell./Sat. MC  & \cct{4.1}  & \cct{4.4}  & \cct{4.6}  & \cct{4.8}  & \cct{4.9}  & \cct{5.0} \\ \hline\hline
\multirow{6}{*}{\rotatebox{50}{Suburban}}    & \multirow{3}{*}{DL}                      & Cell./Cell. MC & \cct{5.0}  & \cct{5.0}  & \cct{5.0}  & \cct{5.0}  & \cct{5.0}  & \cct{5.0} \\ \cline{3-9} 
                                             &                                          & Sat./Sat. MC   & \cct{0.1}  & \cct{0.5}  & \cct{4.7}  & \cct{5.0}  & \cct{5.0}  & \cct{5.0} \\ \cline{3-9} 
                                             &                                          & Cell./Sat. MC  & \cct{5.0}  & \cct{5.0}  & \cct{5.0}  & \cct{5.0}  & \cct{5.0}  & \cct{5.0} \\ \cline{2-9} 
                                             & \multirow{3}{*}{UL}                      & Cell./Cell. MC & \cct{4.7}  & \cct{4.7}  & \cct{4.7}  & \cct{4.9}  & \cct{5.0}  & \cct{5.0} \\ \cline{3-9} 
                                             &                                          & Sat./Sat. MC   & \cct{Out.} & \cct{0.4}  & \cct{4.2}  & \cct{4.3}  & \cct{4.5}  & \cct{4.9} \\ \cline{3-9} 
                                             &                                          & Cell./Sat. MC  & \cct{4.6}  & \cct{4.7}  & \cct{4.7}  & \cct{4.9}  & \cct{5.0}  & \cct{5.0} \\ \hline\hline
\multirow{6}{*}{\rotatebox{50}{Forest}}       & \multirow{3}{*}{DL}                     & Cell./Cell. MC & \cct{4.2}  & \cct{4.6}  & \cct{5.0}  & \cct{5.0}  & \cct{5.0}  & \cct{5.0} \\ \cline{3-9} 
                                             &                                          & Sat./Sat. MC   & \cct{Out.} & \cct{Out.} & \cct{Out.} & \cct{Out.} & \cct{Out.} & \cct{3.9} \\ \cline{3-9} 
                                             &                                          & Cell./Sat. MC  & \cct{4.2}  & \cct{4.3}  & \cct{4.6}  & \cct{4.9}  & \cct{5.0}  & \cct{5.0} \\ \cline{2-9} 
                                             & \multirow{3}{*}{UL}                      & Cell./Cell. MC & \cct{0.8}  & \cct{0.9}  & \cct{0.9}  & \cct{3.1}  & \cct{4.5}  & \cct{5.0} \\ \cline{3-9} 
                                             &                                          & Sat./Sat. MC   & \cct{Out.} & \cct{Out.} & \cct{Out.} & \cct{Out.} & \cct{Out.} & \cct{4.1} \\ \cline{3-9} 
                                             &                                          & Cell./Sat. MC  & \cct{0.3}  & \cct{0.4}  & \cct{0.7}   & \cct{4.0} & \cct{4.6}  & \cct{5.0} \\ \hline\hline
\end{tabular}%
}
\end{table}

With regard to satellite multi-connectivity, it experiences a notable improvement in reliability compared to the use of single-connectivity with Starlink or OneWeb. In the urban scenario, the probability of an out-of-coverage situation goes from 12\%/17\% (DL/UL) for Starlink and 19\%/21\% (DL/UL) for OneWeb to 2\%/2\% (DL/UL) with satellite multi-connectivity. These results are even better in suburban areas, where the probability drops to 0.6\%/0.8\% (DL/UL) with multi-satellite connectivity. Finally, in the forest, the probability of signal loss is 40\%/40\% (DL/UL) when multi-connectivity between OneWeb and Starlink is applied, which improves performance compared to single-connectivity, although not substantially, mainly due to the low performance of each solution separately in this scenario. In general terms, it is noteworthy how the independence of both systems and the simultaneous use of two different satellite constellations significantly improve connectivity in both urban and suburban environments.

Finally, the case of cellular-satellite multi-connectivity achieves performance results very similar to those observed for cellular multi-connectivity, but with the advantage of providing the flexibility of two communication systems with completely independent architectures. In terms of reliability, this option achieves coverage percentages similar to those observed for cellular multi-connectivity with slightly higher latencies due to the complexity of the satellite system architecture. In fact, it is noteworthy that in the case of uplink OWD in the forest, where both single-connectivity and other multi-connectivity options are unable to establish connectivity with 99\% reliability, cellular-satellite multi-connectivity is the only case that can reach this target.

Focusing on the throughput, Table~\ref{tab:percentiles_TP} and Fig.~\ref{fig8} (single-connectivity), and Table~\ref{tab:percentiles_TP_MC} and Fig.~\ref{fig11} (multi-connectivity) show a direct comparison between both connectivity modes. As in the case of one-way delay, both cellular multi-connectivity and satellite-cellular multi-connectivity obtain very similar results. Both ensure a minimum throughput exceeding 4~Mbps in the urban scenario, which surpasses the minimum performance obtained in single-connectivity mode, remaining close to the target of 5 Mbps. In the suburban scenario, performance is excellent, with a minimum downlink throughput of 5 Mbps, i.e., the target throughput, and 4.6~Mbps in uplink. In the forest, the minimum DL throughput found is 4.2~Mbps, although in the uplink this degrades to values below 1 Mbps. In any case, performance is better than with single-connectivity. For instance, the P5 percentile was only capable of reaching 1.5 Mbps for cellular operators, and in this case it is at least 3.1 Mbps for cellular multi-connectivity and 4 Mbps for cellular-satellite multi-connectivity. For satellite multi-connectivity, the probability of losing connectivity in the one-second intervals over which throughput is measured decreases to 0.5\%/0.7\% (DL/UL) in the urban scenario, which is significantly lower compared to the individual use of OneWeb or Starlink (see Table \ref{tab:percentiles_TP}). Finally, in the forest scenario, the distribution tails are not significantly improved compared to the single-connectivity case, although the median does improve to around 4 Mbps bidirectional compared to values of 2 Mbps in single-connectivity mode.

In summary, cellular and cellular-satellite multi-connectivity does not offer a notable improvement in terms of connectivity compared to cellular single-connectivity, since in terms of coverage, the latter was more than satisfactory with virtually ubiquitous connectivity. However, there is a general improvement in KPI statistics for multi-connectivity compared to single-connectivity, highlighted in the minimization of latency and maximization of throughput in the statistical distribution tails. On the other hand, purely satellite multi-connectivity offers a leap in quality in terms of connectivity compared to single-connectivity, minimizing the connectivity gaps that exist in the scenarios assessed. Thus, this section has empirically demonstrated the benefits that can be achieved by integrating terrestrial and non-terrestrial networks, as well as hybrid combinations of both in real scenarios.

\section{Satellite Static Performance}

\begin{figure}[!t]
	\centering
	\subfigure[{\footnotesize Scenario 1}]{\includegraphics[width=1\columnwidth]{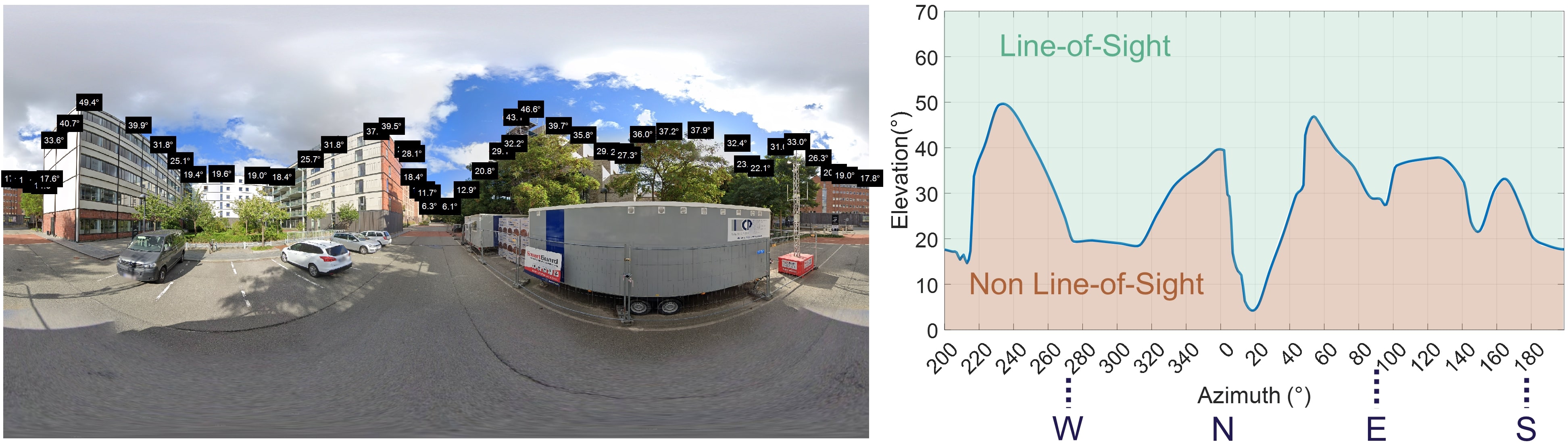}
	} 
    \subfigure[{\footnotesize Scenario 2}]{\includegraphics[width=1\columnwidth]{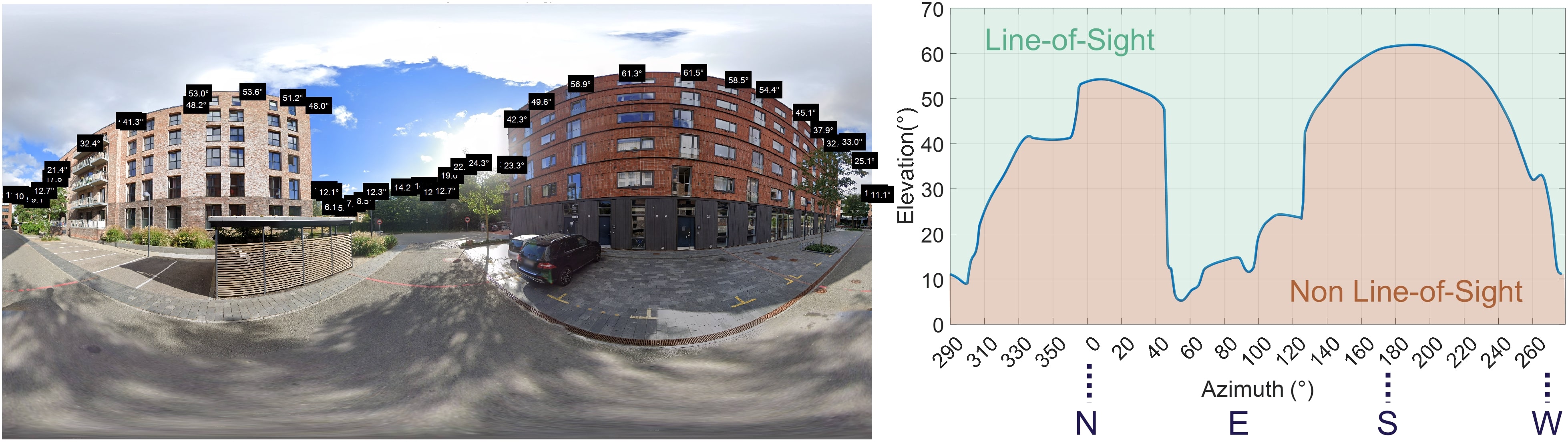}
	}
	\subfigure[{\footnotesize Scenario 3}]{\includegraphics[width=1\columnwidth]{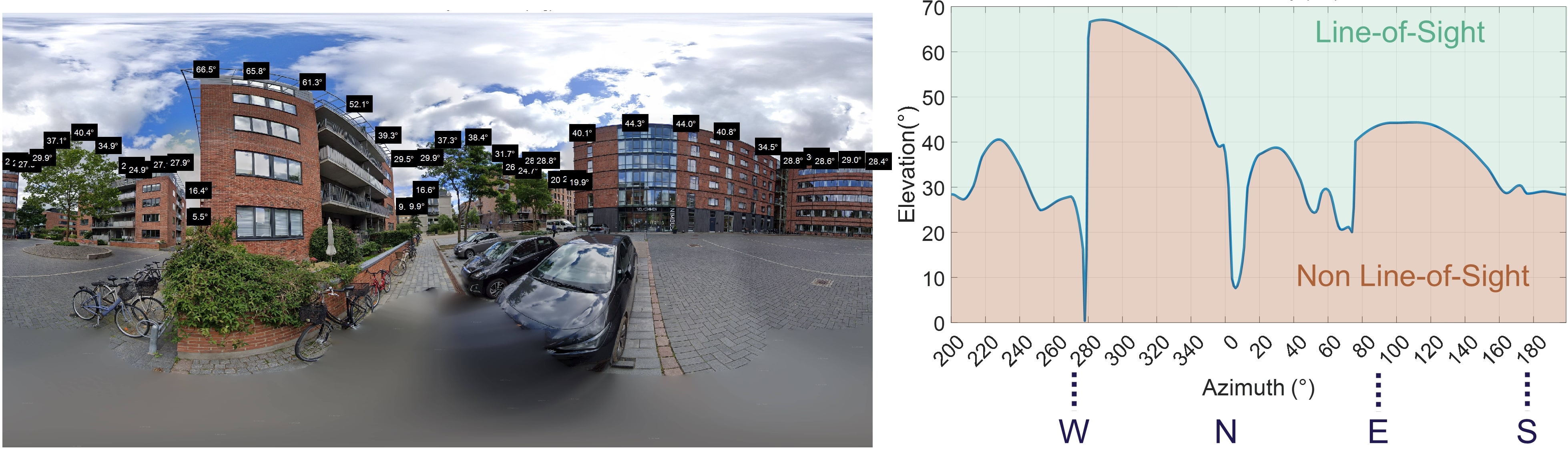}
	}
    \caption{360º photographs of the locations where static connectivity analysis and visibility elevation profiles were carried out for scenarios with (a) $\theta_{\max} = 49\degree$, $\theta_{\mu} = 29\degree$, (b) $\theta_{\max} = 61\degree$, $\theta_{\mu} = 39\degree$, and (c) $\theta_{\max} = 67\degree$, $\theta_{\mu} = 39\degree$. Images extracted from Google Street View.} 
	\label{fig12}
\end{figure}

While the analysis has thus far focused on satellite connectivity under mobility, it is equally critical to evaluate performance under static conditions in obstacle-rich settings, such as urban environments. Satellite operators are aware of this issue and recommend positioning ground user equipment in areas with full visibility towards the sky. Notwithstanding, this may not always be possible for logistical reasons. Therefore, in addition to the mobility study, we propose studying the satellite connectivity of Starlink and OneWeb operators in three static locations in the urban area under analysis. Fig.~\ref{fig12} shows a $360\degree$ panoramic view for the three locations, as well as the elevation profile of the obstacles blocking the horizon of the user equipment, where $\theta = 0\degree$ corresponds to the horizon and $\theta = 90\degree$ to the zenith. The elevation profile was obtained by computing the elevation angle of the boundary curve separating street-level obstacles (NLoS) from the sky (LoS). Each area has been characterized based on two parameters: (i)~$\theta_{\max}$, which is the maximum elevation angle blocked on the horizon by some element in the environment, e.g., a building or a tree. Therefore, $\theta_{max}$ is the maximum elevation angle attained over the full azimuthal range  $\phi \in [0\degree, 360\degree)$. (ii)~$\theta_{\mu}$, which is the average blocked elevation angle along the $360\degree$ azimuth profile. It can be formally defined as $\theta_\mu=\frac{1}{N} \sum_{i=1}^N \theta_i$, where $N$ denotes the number of points used to discretize the elevation profile along the azimuthal range.

\begin{figure}[!t]
	\centering
	\includegraphics[width= 1\columnwidth]{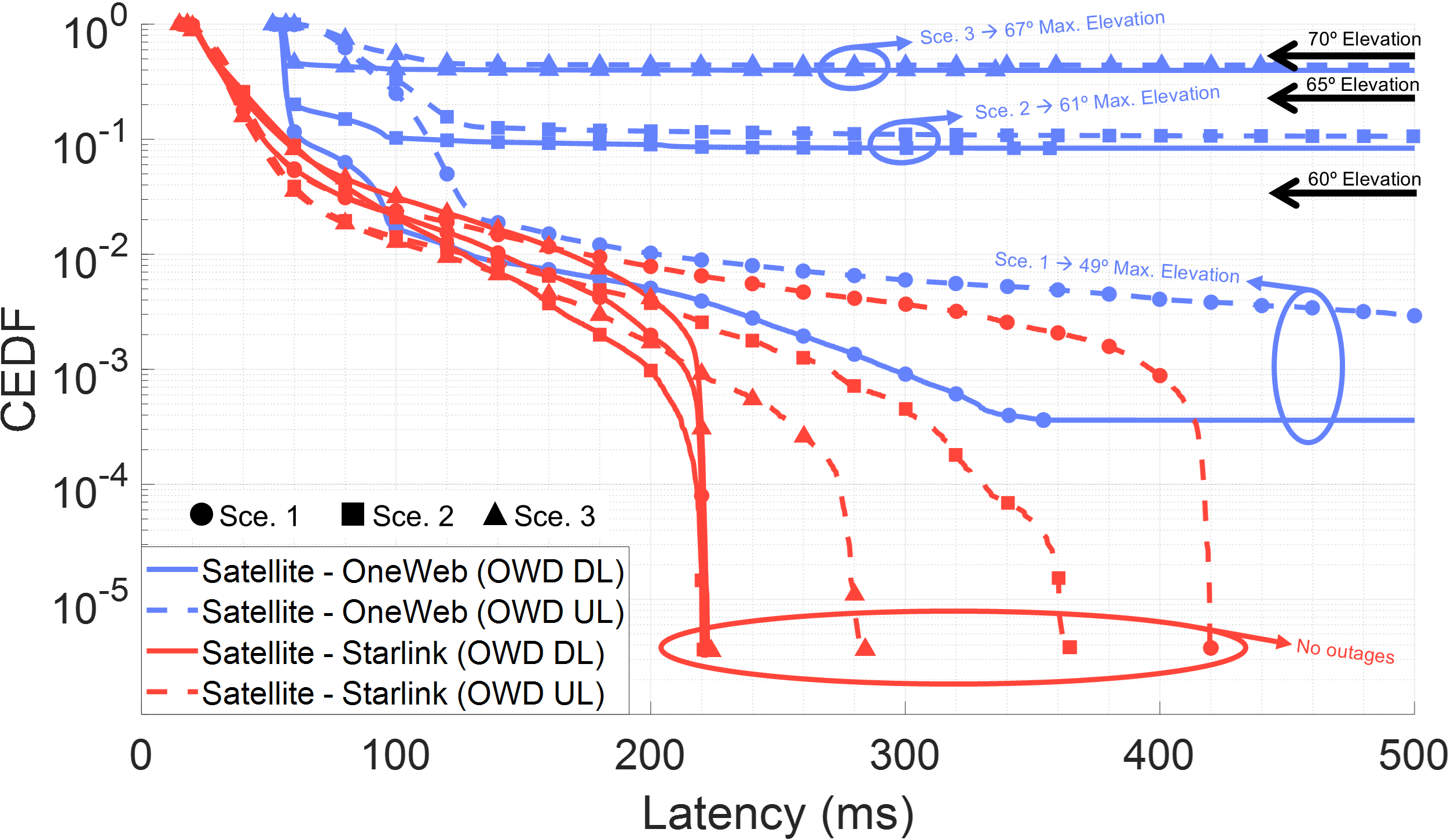}
	\caption{Complementary empirical distribution function for one-way delay in downlink (solid line) and uplink (dashed line) for satellite operators OneWeb (\scalebox{0.8}{\textcolor[rgb]{0.3961,0.5098,0.9922}{\ding{108}}}) and Starlink (\scalebox{0.8}{\textcolor[rgb]{1,0.2706,0.2275}{\ding{108}}}) under static conditions in three different urban area locations. The black horizontal arrows show the simulated probability of null connectivity for OneWeb constellation under different required elevation angles.} 
	\label{fig13}
\end{figure}

Fig.~\ref{fig13} shows the one-way delay CEDF in DL/UL for each static scenario in both satellite technologies. In each scenario, a time window is considered during which the user equipment has visibility of at least 10 satellites from the OneWeb constellation and several tens from the Starlink constellation over time, thus acquiring 600k UDP packet one-way delay values to estimate the visibility and connectivity conditions of the satellite link for each OneWeb and Starlink technologies. For Starlink, it can be observed that regardless of the scenario studied, it is always possible to transmit UDP packets with latencies below 500~ms, indicating no network outages and stable coverage despite the presence of obstacles with maximum elevations of up to~$67\degree$. Note that the divergence observed at the tail of the CEDF in the uplink is attributable to the global network reconfiguration that Starlink performs every 15 seconds, leading to latency spikes when handovers occur between the ground station and satellite, and between the satellite and user terminal~\cite{Reconfiguration_Starlink}. These spikes every 15 seconds have been confirmed by the time traces, thus ruling out the lack of visibility due to the presence of obstacles. Therefore, the behavior of the network in the three scenarios is stable according to Starlink's procedures, regardless of the environment. However, the same effect is not observed with OneWeb connectivity, where latency spikes exceeding a 500 ms threshold are experienced with some regularity, suggesting a temporary loss of visibility with the satellites in the constellation. In particular, in the first scenario (see~Fig.~\ref{fig12}(a)) with $\theta_{\max} = 49\degree$ and $\theta_{\mu} = 29\degree$, the threshold is exceeded with a probability of 0.04\%/0.2\% (DL/UL). In the second scenario (see Fig.~\ref{fig12}(b)), where the presence of obstacles increases with $\theta_{\max} = 61\degree$ and $\theta_{\mu} = 39\degree$, the probability of exceeding the latency threshold also increases to 8.4\%/10.6\% (DL/UL). Finally, in the third scenario (see Fig.~\ref{fig12}(c)), where the antenna is placed particularly close to a building with $\theta_{\max} = 67\degree$ and $\theta_{\mu} = 39\degree$, the probability of exceeding the threshold shoots up to 39.8\%/44.1\% (DL/UL). These results suggest a direct relation between the probability of signal blocking, denoted by the high latencies found in communications, and the $\theta_{\max}$ present in our measurement scenario. While Starlink is able to overcome signal blocking through satellite redundancy on the horizon, OneWeb cannot do it due to a much smaller constellation.

To validate the previous measurements, visibility simulations have been carried out for user equipment located at the latitude and longitude of the scenarios with the satellites of both constellations simulated through a Simplified General Perturbations-4 (SGP4) propagator~\cite{SGP4}. To establish visibility between the satellites in the constellation and the user equipment, a variable minimum elevation angle has been set as a constraint, corresponding to the presence of obstacles on the horizon. In the case of Starlink simulations, it is possible to establish LoS visibility between the satellite and the ground even for elevation angles larger than the maximum angle of $67\degree$ in the third scenario, which is consistent with the continuous coverage found in the empirical analysis. For OneWeb, simulations indicate that visibility is total up to an elevation angle of $50\degree$, beginning to degrade for higher angles to visibilities of 98.8\% ($\theta = 55\degree$), 97.8\% ($\theta = 60\degree$), 77.7\% ($\theta = 65\degree$), and 47.1\% ($\theta = 70\degree$), respectively. The probabilities of simulated visibility in OneWeb for different elevation angles are shown in Fig.~\ref{fig13} for the sake of clarity. Therefore, these simulations indicate that visibility declines significantly at elevations larger than $60\degree$, which is also corroborated empirically in the measurements of the three scenarios. In conclusion, it is advisable to avoid obstacles with elevations equal to or larger than $60\degree$ when the user equipment is fixed statically, even more so in commercial constellations that do not have satellite redundancy, as observed in the case of OneWeb.

\section{Conclusion}

This work has assessed an experimental measurement campaign conducted under mobility conditions in three different scenarios (urban, suburban, and forest) covering four commercial non-terrestrial satellite networks and terrestrial cellular networks. For the analysis, four KPIs of the network related to unidirectional link latency were considered, namely one-way delay in uplink and downlink, as well as data rates in both uplink and downlink.

The use of 5G cellular connectivity in urban and suburban areas proves to be satisfactory, with network reliability exceeding 99.9\% due to the dense infrastructure deployments by mobile network operators. However, this performance is degraded in forested areas, especially in the uplink, due to limitations in the link budget of user equipment and the sparse existing infrastructure. As an alternative to well-established cellular solutions, the use of emerging commercial satellite networks such as Starlink or OneWeb is proposed in scenarios where there is lack or vulnerability of existing infrastructure. The results show that in urban areas, which are challenging in terms of radio propagation due to the high NLoS likelihood with the satellite link, the outage of Starlink is around 12-17\% and that of OneWeb around 19-21\%. These technologies are particularly sensitive to mobility due to changes in orientation at urban intersections that require satellite re-acquisition from the constellation to reestablish connectivity.

To improve both KPI performance and Internet connectivity, up to three multi-connectivity options have been considered given the four available interfaces: (i)~cellular multi-connectivity, (ii)~satellite multi-connectivity, and (iii)~cellular-satellite multi-connectivity. The results indicate that cellular and cellular-satellite multi-connectivity improve latency and throughput due to the redundancy of the cellular infrastructure (Cell./Cell.) and the independence of the satellite network (Cell./Sat.), respectively. On the other hand, satellite multi-connectivity improves reliability rates in terms of connectivity up to 98\% in urban areas, thus taking advantage of satellite redundancy between both constellations with different orbits. Therefore, this approach enables this solution as a reliable and fully independent alternative to cellular networks in scenarios where the latter may be compromised. Finally, the impact of urban obstacles on establishing satellite connectivity has also been quantified through tests in static conditions, showing that in constellations with a limited number of satellites, where satellite redundancy might not be available, connectivity can be severely compromised when obstacles are blocking the horizon at elevations of $60\degree$ or higher.

In conclusion, with the emergence of new satellite communication systems and the 6G vision technology of enabling hybrid communications through the integration of terrestrial and non-terrestrial networks, we consider the characterization of the performance of these solutions across multiple environments and connectivity combinations to be fundamental for the efficient and reliable future deployment of mobile communication systems. Thus, this work has provided data measured in real-world environments that serve as a comprehensive assessment of the current state of these technologies. Future work considers complementing this study by analyzing variable and high-demand traffic patterns, including congested scenarios, to further improve the understanding of performance in terrestrial and non-terrestrial networks.

\section*{Acknowledgment}

The authors would like to thank the Center for Beredskabskommunikation (CFB) for providing the OneWeb service plan and hardware, and for the help provided during the measurement campaign. The work of O. S. Peñaherrera-Pulla has been partially supported by Junta de Andalucía through Secretaría General de Universidades, Investigación y Tecnología under predoctoral Grant PREDOC\_01712.


\begin{thebibliography}{50}

\bibitem{5G_society}
G. Fettweis and S. Alamouti, ``5G: Personal mobile internet beyond what cellular did to telephony,'' \textit{IEEE Communications Magazine}, vol. 52, no. 2, pp. 140-145, 2014.

\bibitem{5G_defacto}
M. Vaezi \textit{et al.}, ``Cellular, Wide-Area, and Non-Terrestrial IoT: A Survey on 5G Advances and the Road Toward 6G,'' \textit{IEEE Communications Surveys \& Tutorials}, vol. 24, no. 2, pp. 1117-1174, 2022.


\bibitem{5G_urban}
J. Navarro-Ortiz, P. Romero-Diaz, S. Sendra, P. Ameigeiras, J. J. Ramos-Munoz, and J. M. Lopez-Soler, ``A Survey on 5G Usage Scenarios and Traffic Models,'' \textit{IEEE Communications Surveys \& Tutorials}, vol. 22, no. 2, pp. 905-929, 2020.

\bibitem{5G_ericsson}
Ericsson, \textit{Ericsson Mobility Report, June 2025}, Stockholm, Sweden: Ericsson AB, 2025. [Online]. Available: \href{https://www.ericsson.com/en/reports-and-papers/mobility-report/reports/june-2025}{https://www.ericsson.com/en/reports-and-papers/mobility-report/ reports/june-2025}.

\bibitem{5G_disaster_1}
Q. Shen \textit{et al.}, ``Fair Communications in UAV Networks for Rescue Applications,'' \textit{IEEE Internet of Things Journal}, vol. 10, no. 23, pp. 21013-21025, 2023.

\bibitem{5G_disaster_2}
M. Matracia, N. Saeed, M. A. Kishk, and M. -S. Alouini, ``Post-Disaster Communications: Enabling Technologies, Architectures, and Open Challenges,'' \textit{IEEE Open Journal of the Communications Society}, vol. 3, pp. 1177-1205, 2022.

\bibitem{5G_disaster_3}
D. G.C., A. Ladas, Y. A. Sambo, H. Pervaiz, C. Politis, and M. A. Imran, ``An Overview of Post-Disaster Emergency Communication Systems in the Future Networks,'' \textit{IEEE Wireless Communications}, vol. 26, no. 6, pp. 132-139, 2019.

\bibitem{NTN_SOTA_1}
O. Kodheli \textit{et al.}, ``Satellite Communications in the New Space Era: A Survey and Future Challenges,'' \textit{IEEE Communications Surveys} \& Tutorials, vol. 23, no. 1, pp. 70-109, 2021.

\bibitem{NTN_SOTA_2}
M. Giordani and M. Zorzi, ``Non-Terrestrial Networks in the 6G Era: Challenges and Opportunities,'' \textit{IEEE Network}, vol. 35, no. 2, pp. 244-251, 2021.

\bibitem{NTN_SOTA_3}
C. T. Nguyen \textit{et al.}, ``Emerging Technologies for 6G Non-Terrestrial-Networks: From Academia to Industrial Applications,'' \textit{IEEE Open Journal of the Communications Society}, vol. 5, pp. 3852-3885, 2024.

\bibitem{NTN_P2P}
N. Saeed, H. Almorad, H. Dahrouj, T. Y. Al-Naffouri, J. S. Shamma, and M. -S. Alouini, ``Point-to-Point Communication in Integrated Satellite-Aerial 6G Networks: State-of-the-Art and Future Challenges,'' \textit{IEEE Open Journal of the Communications Society}, vol. 2, pp. 1505-1525, 2021.

\bibitem{3GPP_NTN_1}
X. Lin, S. Rommer, S. Euler, E. A. Yavuz, and R. S. Karlsson, ``5G from Space: An Overview of 3GPP Non-Terrestrial Networks,'' \textit{IEEE Communications Standards Magazine}, vol. 5, no. 4, pp. 147-153, 2021.

\bibitem{3GPP_NTN_2}
M. M. Saad, M. A. Tariq, M. T. R. Khan, and D. Kim, ``Non-Terrestrial Networks: An Overview of 3GPP Release 17 \& 18,'' \textit{IEEE Internet of Things Magazine}, vol. 7, no. 1, pp. 20-26, 2024.

\bibitem{3GPP_NTN_3}
G. A. Medina-Acosta, R. K. Mungara, S. G. Eriksson, and T. Khan, ``3GPP Release-18 Physical Layer Enhancements for IoT-NTN,'' \textit{IEEE Communications Standards Magazine}, vol. 8, no. 3, pp. 18-24, 2024.

\bibitem{LEO_1}
Y. Su, Y. Liu, Y. Zhou, J. Yuan, H. Cao, and J. Shi, ``Broadband LEO Satellite Communications: Architectures and Key Technologies,'' \textit{IEEE Wireless Communications}, vol. 26, no. 2, pp. 55-61, 2019.

\bibitem{LEO_2}
I. Leyva-Mayorga \textit{et al.}, ``LEO Small-Satellite Constellations for 5G and Beyond-5G Communications,'' \textit{IEEE Access}, vol. 8, pp. 184955-184964, 2020.

\bibitem{LEO_examples_1}
I. del Portillo, B. G. Cameron, and E. F. Crawley, ``A Technical Comparison of Three Low Earth Orbit Satellite Constellation Systems to Provide Global Broadband,'' \textit{Acta Astronautica}, vol. 159, pp. 123–135, 2019.

\bibitem{LEO_examples_2}
Z. M. Kassas, N. Khairallah, and S. Kozhaya, ``Ad Astra: Simultaneous Tracking and Navigation With Megaconstellation LEO Satellites,'' \textit{IEEE Aerospace and Electronic Systems Magazine}, vol. 39, no. 9, pp. 46-71, Sept. 2024

\bibitem{challenges_1}
M. Hosseinian, J. P. Choi, S. -H. Chang, and J. Lee, ``Review of 5G NTN Standards Development and Technical Challenges for Satellite Integration With the 5G Network,'' \textit{IEEE Aerospace and Electronic Systems Magazine}, vol. 36, no. 8, pp. 22-31, 2021.

\bibitem{link_budget_1}
M. Conti, A. Guidotti, C. Amatetti, and A. Vanelli-Coralli, ``NB-IoT over Non-Terrestrial Networks: Link Budget Analysis,'' in \textit{GLOBECOM 2020 - 2020 IEEE Global Communications Conference}, Taipei, Taiwan, pp. 1-6, 2020.

\bibitem{TN_NLoS_1}
V. Kristem, C. U. Bas, R. Wang, and A. F. Molisch, ``Outdoor Wideband Channel Measurements and Modeling in the 3–18 GHz Band,'' \textit{IEEE Transactions on Wireless Communications}, vol. 17, no. 7, pp. 4620-4633, 2018.

\bibitem{TN_NLoS_2}
J. Huang, C. -X. Wang, Y. Yang, Y. Liu, J. Sun, and W. Zhang, ``Channel Measurements and Modeling for 400–600-MHz Bands in Urban and Suburban Scenarios,'' \textit{IEEE Internet of Things Journal}, vol. 8, no. 7, pp. 5531-5543, 2021.

\bibitem{TN_NLoS_3}
H. Miao \textit{et al.}, ``Sub-6 GHz to mmWave for 5G-Advanced and Beyond: Channel Measurements, Characteristics and Impact on System Performance,'' \textit{IEEE Journal on Selected Areas in Communications}, vol. 41, no. 6, pp. 1945-1960, 2023.

\bibitem{SOTA_THE_5G_1}
S. Kang \textit{et al.}, ``Cellular Wireless Networks in the Upper Mid-Band,'' \textit{IEEE Open Journal of the Communications Society}, vol. 5, pp. 2058-2075, 2024.

\bibitem{SOTA_THE_5G_2}
S. Iranmanesh, F. S. Abkenar, R. Raad, and A. Jamalipour, ``Improving Throughput of 5G Cellular Networks via 3D Placement Optimization of Logistics Drones,'' \textit{IEEE Transactions on Vehicular Technology}, vol. 70, no. 2, pp. 1448-1460, 2021.

\bibitem{SOTA_THE_5G_3}
C. L. Vielhaus \textit{et al.}, "vBerlinV2N: Recreating a Cellular Network Measurement Campaign With Simulations," \textit{IEEE Access}, vol. 13, pp. 127023-127044, 2025.

\bibitem{SOTA_PRA_5G_1}
S. Aerts \textit{et al.}, ``In-situ Measurement Methodology for the Assessment of 5G NR Massive MIMO Base Station Exposure at Sub-6 GHz Frequencies,'' \textit{IEEE Access}, vol. 7, pp. 184658-184667, 2019.

\bibitem{SOTA_PRA_5G_2}
L. Chiaraviglio \textit{et al.}, ``EMF Exposure in 5G Standalone mm-Wave Deployments: What Is the Impact of Downlink Traffic?,'' \textit{IEEE Open Journal of the Communications Society}, vol. 3, pp. 1445-1465, 2022.

\bibitem{SOTA_PRA_5G_3}
S. Mohebi, F. Michelinakis, A. Elmokashfi, O. Grøndalen, K. Mahmood, and A. Zanella, ``Sectors, Beams and Environmental Impact on the Performance of Commercial 5G mmWave Cells: An Empirical Study,'' \textit{IEEE Access}, vol. 10, pp. 133309-133323, 2022.

\bibitem{SOTA_THE_SAT_1}
D. -H. Jung, J. -G. Ryu, W. -J. Byun, and J. Choi, ``Performance Analysis of Satellite Communication System Under the Shadowed-Rician Fading: A Stochastic Geometry Approach,'' \textit{IEEE Transactions on Communications}, vol. 70, no. 4, pp. 2707-2721, 2022.

\bibitem{SOTA_THE_SAT_2}
M. Y. Abdelsadek, G. Karabulut-Kurt, H. Yanikomeroglu, P. Hu, G. Lamontagne, and K. Ahmed, ``Broadband Connectivity for Handheld Devices via LEO Satellites: Is Distributed Massive MIMO the Answer?,'' \textit{IEEE Open Journal of the Communications Society}, vol. 4, pp. 713-726, 2023.

\bibitem{SOTA_PRA_SAT_1}
D. Laniewski, E. Lanfer, S. Beginn, J. Dunker, M. Dückers, and N. Aschenbruck, ``Starlink on the Road: A First Look at Mobile Starlink Performance in Central Europe,'' in \textit{2024 8th Network Traffic Measurement and Analysis Conference (TMA)}, Dresden, Germany, pp. 1-8, 2024.

\bibitem{SOTA_PRA_SAT_2}
D. Laniewski, E. Lanfer, and N. Aschenbruck, ``Measuring Mobile Starlink Performance: A Comprehensive Look,'' \textit{IEEE Open Journal of the Communications Society}, vol. 6, pp. 1266-1283, 2025.

\bibitem{SOTA_PAN_1}
J. Pan, J. Zhao, and L. Cai, ``Measuring a Low-Earth-Orbit Satellite Network,'' in \textit{2023 IEEE 34th Annual International Symposium on Personal, Indoor and Mobile Radio Communications (PIMRC)}, Toronto, ON, Canada, pp. 1-6, 2023.

\bibitem{SOTA_PAN_2}
J. Pan, J. Zhao, and L. Cai, ``Measuring the Satellite Links of a LEO Network,'' in \textit{ICC 2024 - IEEE International Conference on Communications}, Denver, CO, USA, 2024, pp. 4439-4444.

\bibitem{SOTA_PAN_3}
J. Zhao, O. Perrin, A. Ahangarpour, and J. Pan, ``Measuring the OneWeb Satellite Network,'' in \textit{2025 9th Network Traffic Measurement and Analysis Conference (TMA)}, Copenhagen, Denmark, pp. 1-10, 2025.

\bibitem{SOTA_TCPUDP_1}
C. Careau, E. Fredriksson, R. Olsson, P. Sjödin, and C. Beckman, ``Throughput Analysis of Starlink Satellite Internet: Study on the Effects of Precipitation and Hourly Variability with TCP and UDP,'' in \textit{The Seventeenth International Conference on Advances in Satellite and Space Communications (SPACOMM)}, Athens, Greece,  pp.~1-4, 2025.

\bibitem{SOTA_TCPUDP_2}
J. Garcia, S. Sundberg, and A. Brunstrom, ``TCP Congestion Control Performance over Starlink,'' in \textit{Proceedings of the 2025 Applied Networking Research Workshop}, Madrid, Spain, pp.~70–77, 2025.

\bibitem{PW_1}
A. Ramírez-Arroyo, M. López, I. Rodríguez, S. B. Damsgaard, and P. Mogensen, ``Multi-Connectivity Solutions for Rural Areas: Integrating Terrestrial 5G and Satellite Networks to Support Innovative IoT Use Cases,'' \textit{Smart Agricultural Technology}, vol. 12, p. 101260, 2025.

\bibitem{PW_2}
A. Ramírez-Arroyo, T. B. Sørensen, and P. Mogensen, ``Terrestrial 5G and Starlink NTN Multi-Connectivity Toward 6G Communications Integration Era: An Empirical Assessment,'' \textit{IEEE Open Journal of the Communications Society}, vol. 6, pp. 5269-5283, 2025.

\bibitem{mastedatabasen}
Danish Agency for Digital Government: Base Station Database [Online]. Available: \url{https://www.mastedatabasen.dk/viskort/PageMap.aspx}.

\bibitem{number_starlink}
P. Gomez-del-Hoyo and P. Samczynski, ``Starlink-Based Passive Radar for Earth's Surface Imaging: First Experimental Results,'' \textit{IEEE Journal of Selected Topics in Applied Earth Observations and Remote Sensing}, vol. 17, pp. 13949-13965, 2024.

\bibitem{number_oneweb}
S. Kozhaya and Z. M. Kassas, ``A First Look at the OneWeb LEO Constellation: Beacons, Beams, and Positioning,'' \textit{IEEE Transactions on Aerospace and Electronic Systems}, vol. 60, no. 5, pp. 7528-7534, 2024.

\bibitem{NT_Starlink}
B. Wang, X. Zhang, S. Wang, L. Chen, J. Zhao, D. Li, and Y. Jiang, ``A Large-Scale IPv6-Based Measurement of the Starlink Network,'' 	arXiv:2412.18243 [cs.NI], 2026.

\bibitem{SS_Starlink}
S. Kozhaya, J. Saroufim, and Z. M. Kassas, ``Unveiling Starlink for PNT,'' \textit{NAVIGATION: Journal of the Institute of Navigation}, vol. 72, no. 1, 2025.

\bibitem{SS_OneWeb}
Z. M. Komodromos and T. E. Humphreys, ``Signal parameter estimation and demodulation of the OneWeb Ku-band downlink,'' \textit{npj Wireless Technology}, vol. 2, no. 1, p. 7, 2026.

\bibitem{Reconfiguration_Starlink}
N. Mohan \textit{et al.}, ``A Multifaceted Look at Starlink Performance,'' in \textit{Proceedings of the ACM Web Conference 2024}, Singapore, Singapore, pp. 2723–2734, 2024.

\bibitem{MC_tool}
\textit{multi-connect - The open source multi-path connectivity tool}, S.~B.~Damsgaard, 2024. [Online]. Available: \url{https://github.com/drblah/multi-connect/}.

\bibitem{procesado_MC}
I. Rodriguez \textit{et al.}, ``An Experimental Framework for 5G Wireless System Integration into Industry 4.0 Applications,'' \textit{Energies}, vol. 14, no. 15, 2021.


\bibitem{SGP4}
D. Vallado, P. Crawford, R. Hujsak, and T. S. Kelso, ``Revisiting Spacetrack Report \#3: Rev 1,'' in \textit{AIAA/AAS Astrodynamics Specialist Conference and Exhibit}, Keystone, USA, pp. 1-92, 2006.

\end{thebibliography}
\end{document}